\documentclass[review, 12pt, authoryear]{elsarticle}
\usepackage[english]{babel}
\usepackage{lineno}
\graphicspath{Figures/}
\usepackage{graphicx}
\usepackage{booktabs} 
\usepackage{enumerate}
\usepackage{subcaption}
\usepackage{times}
\usepackage[colorinlistoftodos]{todonotes}
\usepackage{natbib}
\usepackage{amsmath}
\usepackage{url}
\usepackage{array}
\usepackage{acro}
\usepackage[document]{ragged2e}

\DeclareAcronym{c1}{
	short = \textit{Inside},
	long = Inside, 
    tag = abbrev
}
\DeclareAcronym{c2}{
	short = \textit{Edge},
	long = Edge,  
	tag = abbrev
}
\DeclareAcronym{c3}{
	short = \textit{Outside},
	long = Outside, 
	tag = abbrev
}

\DeclareAcronym{p1}{
	short = \textit{Prototypical},
	long = Prototypical, 
	tag = abbrev
}
\DeclareAcronym{p2}{
	short = \textit{Positive},
	long = Positive, 
	tag = abbrev
}
\DeclareAcronym{p3}{
	short = \textit{Negative},
	long = Negative, 
	tag = abbrev
}
\usepackage[left=1in, right=1in, top=1in, bottom=1in]{geometry}
\newcolumntype{L}[1]{>{\raggedright\let\newline\\\arraybackslash\hspace{0pt}}m{#1}}
\newcolumntype{C}[1]{>{\centering\let\newline\\\arraybackslash\hspace{0pt}}m{#1}}
\newcolumntype{R}[1]{>{\raggedleft\let\newline\\\arraybackslash\hspace{0pt}}m{#1}}
\makeatletter
\newcommand\xLongLeftRightArrow[2][]{%
	\ext@arrow 0099{\LongLeftRightArrowfill@}{#1}{#2}}
\def\LongLeftRightArrowfill@{%
	\arrowfill@\Leftarrow\Relbar\Rightarrow}
\makeatother
\usepackage{rotating}
\usepackage{tablefootnote}
\usepackage{multirow}
\usepackage{textcomp}
\usepackage{color}
\usepackage{soul}
\usepackage{booktabs, makecell}
\usepackage{lscape}

\journal{International Journal of Human Computer Studies}

\begin{document}
	\acsetup{first-style=long}
	\begin{frontmatter}
		
	\title{Exploration Strategies for Tactile Graphics Displayed by Electrovibration on a Touchscreen}
	\author[mymainaddress]{Bushra Sadia}
	\author[mymainaddress]{Ayberk Sadic}
	\author[mysecondaryaddress]{Mehmet Ayyildiz}
	\author[mymainaddress]{Cagatay Basdogan\corref{CorrespondingAuthor}} 
	\address[mymainaddress]{College of Engineering, Ko\c{c} University, 34450, Istanbul, Turkey}
	\address[mysecondaryaddress]{Faculty of Engineering and Natural Sciences, Istanbul Bilgi University, 34060, Istanbul, Turkey}

	\cortext[CorrespondingAuthor]{Email: cbasdogan@ku.edu.tr (Cagatay Basdogan), \\ College of Engineering,  Ko\c{c} University, 34450, Istanbul, Turkey.}
		
	\begin{abstract}
		\justify 
		Advancements in surface haptics technology have given rise to the development of  interactive applications displaying tactile content on touch surfaces such as images, signs, diagrams, plots, charts, graphs, maps, networks, and tables. In those applications, users manually explore the touch surface to interact with the tactile data using some intuitive strategies. The user's exploration strategy, tactile data's complexity, and tactile rendering method all affect the user's haptic perception, which plays a critical role in design and prototyping of those applications. In this study, we conducted experiments with human participants to investigate the recognition rate and time of five tactile shapes (i.e., triangle, square, pentagon, hexagon, and octagon) rendered by electrovibration on a touchscreen using three different methods and displayed in prototypical orientation and non-prototypical orientations (i.e., 15 degrees CW and CCW to the prototypical orientation). The results showed that the correct recognition rate of the shapes was higher when the haptically active area (area where electrovibration was on) was larger. However, as the number of edges was increased, the recognition time increased and the recognition rate dropped significantly, arriving to a value slightly higher than the chance rate of 20\% for non-prototypical octagon. Moreover, the recognition time for inside rendering condition was significantly shorter in comparison with edge and outside rendering conditions and edge rendering condition led to the longest recognition time. We also recorded the participants' finger movements on the touchscreen to examine their haptic exploration strategies. Based on our temporal analysis, we classified six exploration strategies adopted by participants to identify the shapes, which were different for the prototypical and non-prototypical shapes. Moreover, our spatial analysis revealed that the participants first used global scanning to extract the coarse features of the displayed shapes, and then they applied local scanning to identify finer details, but needed another global scan for final confirmation in the case of non-prototypical shapes, possibly due to the current limitations of electrovibration technology in displaying tactile stimuli to a user. We observed that it was highly difficult to follow the edges of shapes and recognize shapes with more than five edges under electrovibration when a single finger was used for exploration.
			
	\end{abstract}

	\acuseall
	\end{frontmatter}
	\section{Introduction}
	
	\justify
	
	Today, there is already an extensive adoption of touch-input surfaces in our daily devices, such as in mobile phones, tablets, navigation interfaces in cars, HMIs in factories, and entertainment systems in homes.  It is also anticipated that many physical surfaces such as refrigerator doors, bathroom mirrors, tabletops,  and home windows will be converted to touch-input surfaces in the future. Haptics offers a new form of engagement with such surfaces by presenting compelling touch experiences. The interest in touch surfaces displaying tactile feedback to user has risen over the last decade, and a new research field called surface haptics has emerged.  Researchers have presented various ways of rebuilding the feel of natural sensations and interactions by displaying programmable haptic effects in the form of vibrations, pulses, and variable friction forces to fingertips of users (see a review of surface haptics in \citealp{basdogan2020review}).
	
	The most common actuation methods for displaying tactile feedback on touch surfaces are vibrotactile, ultrasonic, and electrostatic. In vibrotactile actuation, a vibration signal at a frequency that is detectable by the fingertip of user (typically below 500 Hz) is displayed through a touch surface (\citealp{sadia2020data, kim2021development}). The simplest and most prevalent example for this technology is the touch surface of a mobile phone, which is typically actuated by a low-cost vibration motor in order to inform the user about the incoming calls and to provide conformation for button clicks. In ultrasonic actuation, touch surface is vibrated mechanically at an ultrasonic resonance frequency, which produces a thin layer of air gap between the user's finger and the surface (\citealp{watanabe1995method, biet2007squeeze, chubb2010shiverpad, saleem2018psychophysical}), resulting in a reduction in sliding friction. In order to render a desired tactile effect, the vibration amplitude on the touch surface can be modulated accordingly. Electrostatic actuation also modulates the friction between the user's fingertip and a touch surface. Unlike the other actuation technologies, there is no mechanical vibration on the touch surface, but an electrostatic attraction force is generated between the surface and finger when an AC voltage is applied to the conductive layer of a capacitive touchscreen. This additional force in the normal direction increases the frictional force between the finger and the touch surface in the tangential direction (called as the electrovibration effect), which can be modulated by altering the amplitude, frequency, and waveform of the applied voltage (\citealp{bau2010teslatouch, vardar2016effect, emgin2018haptable}). The physics behind the electrovibration effect is still under investigation (\citealp{ayyildiz2018contact, basdogan2020modeling})
	
	Displaying tactile feedback through a touch surface (i.e., surface haptics) has many potential and exciting applications in various domains (see \citealp{basdogan2020review}).	In applications targeted to general users, surface haptics can enrich the user experience by providing additional sensory feedback through haptic channel (e.g., feeling more frictional resistance while dragging a larger size folder on a touchscreen). Users can benefit from these sensory augmentations especially when they cannot focus their full attention on digital information displayed by visual feedback (e.g., controlling the digital buttons of a navigation interface with the help of tactile feedback while driving a car). On the other hand, users with visual impairments can take advantage of surface haptics to perceive the information provided as visual data such as pictures, diagrams, plots, charts, graphs, maps, networks, and tables.
	
    Users apply intuitive strategies to explore tactile graphics on touch surfaces. Not only the users’ exploration strategies, but also tactile data’s complexity and tactile rendering methods affect the haptic perception of the users. Therefore, it is critical to understand how users perceive tactile graphics displayed to them on these touch surfaces and what exploratory procedures are followed to perceive them. This study investigates the tactile exploration strategies followed by the users for basic geometrical shapes displayed by electrovibration on a touchscreen. We first investigated the effects of (a) number of edges, (b) rendering conditions, and (c) angular orientation on the recognition rate and time of 2D equilateral tactile shapes to understand the effect of the size of haptically active area and the complexity of geometry on the perception of tactile shapes. Afterward, we performed spatial and temporal analyses on the recorded data of finger position to understand the type of exploratory procedures followed by the participants to perceive tactile shapes. The findings of this study improve our understanding of haptic interaction with digital data displayed through a friction modulated touch surface, which can potentially guide developers in the design of future interfaces.

	\section{Background}
	
	Researchers have already utilized surface haptics technologies to render tactile graphics on touch surfaces (\citealp{,israr2012tactile, kim2013tactile, gorlewicz2014initial, mullenbach2014exploring, osgouei2016identification, osgouei2017improving, klatzky2019detection}). However, only a few studies investigated the relation between the complexity of tactile data displayed through a touch surface and haptic perception of the users (\citealp{gorlewicz2020design, costes2020towards}). The results of these studies showed that the user's haptic channel and memory should not be overloaded to deliver the desired information effectively while displaying sufficient amount of details and utilizing an established rendering methodology are both required to ensure expressiveness and efficiency. \cite{gorlewicz2020design} reported that users form a mental image of the displayed shape in memory during tactile exploration, and successful recognition depends on effective tactile rendering of vertices, lines, and particular angles of the shape. They also emphasized the importance of reducing the size of white space in applications (i.e., haptically inactive area). They recorded the user's finger position in exploring bar charts and basic geometric shapes and reported how much time is spent on different items. They concluded that appropriate tactile design (i.e., type of shapes and orientations) and suitable rendering approach (e.g., stimulation type, location, and duration) are essential to create successful surface haptics applications. 
	
	\subsection{Tactile Graphics}
	
	Tactile graphics play an essential role in the life of visually impaired and blind people. For them, it is the means to access visual information for various purposes such as education, entertainment, and navigation (\citealp{wright2008guide, smith2012role, grussenmeyer2017accessible, ducasse2018accessible, vinter2020identification}). Tactile graphics is typically displayed in the form of raised surfaces (\citealp{kennedy1993drawing,heller2002tangible,shiose2008toward, bardot2017identifying}).  It is produced by embossing on non-refreshable media such as plastic sheets, paper, or heat-sensitive swell paper. However, this approach does not support multimodal interaction, and the graphics, once created, is static and cannot be updated unless completely reproduced.  Also, the production requires special equipment and the process is relatively expensive and time-consuming. As alternatives, researchers have utilized force and tactile feedback, and sonification to convey information to blind or visually impaired users and in situations in which the visual channel is highly loaded or ineffective (\citealp{griffin2001feeling, paneels2009review, o2015designing, bateman2018user}).

	Force feedback devices, coupled with vibratory, verbal, and auditory cues, are utilized to present graphical information to the users (\citealp{petrie2002tedub, magnuson2003non, mcgookin2006multivis, crossan2008multimodal, yannier2008using}). 
	\cite{crossan2008multimodal} utilized  force feedback enriched with audio cues to guide the user's hand in outlining simple geometric shapes. \cite{yannier2008using} displayed wind forces through a force feedback device in connection with surface topography in order to educate students about cause and effect relations in climate visualization. \cite{lim2019guidance} proposed a contour rendering algorithm for 2D tactile graphics based on the haptic guidance provided by a force feedback device. They first applied this algorithm to primitive geometric shapes for testing its accuracy and then to the rendering of facial features on photographs. Readers are referred to \cite{paneels2009review} for a detailed review of the approaches implemented by force feedback devices for haptic data visualization. However, force feedback devices are often expensive, bulky, and lack portability, which hinder their usage in daily life.  
	
	For those reasons, researchers have focused on creating accessible tactile graphics on touch surfaces such as the touchscreens of mobile devices. The most common actuation approach in this regard is vibrotactile. Low-cost vibration motors embedded in mobile devices have been frequently utilized for this purpose and several studies have already reported that this approach is viable in recognizing non-visual graphics and shapes (\citealp{poppinga2011touchover, toennies2011toward, giudice2012learning, awada2013evaluation, palani2014evaluation, tennison2016toward}). However, complex tactile effects cannot be generated by such vibrators since they have a limited frequency bandwidth. Moreover, they vibrate the whole screen and do not provide localized tactile feedback. Although there are a few prototypes demonstrating multi-touch and localized tactile stimulation (\citealp{bai2011impact, hudin2015localized, emgin2018haptable, dhiab2020confinement, hwang2021light}), the techniques presented in these studies are not straightforward to implement. To overcome this problem, \cite{goncu2011gravvitas}, for instance, proposed a simple but not practical solution. They attached individual vibrators to the fingers of a user to display tactile graphics such as charts and maps.

	An alternative approach used by the researchers to render tactile graphics on a touch surface is electrovibration (\citealp{bau2010teslatouch}). Researchers have utilized electrovibration displays to render 2D tactile graphics and 3D geometrical shapes in the form of height maps (\citealp{,israr2012tactile, kim2013tactile, mullenbach2014exploring, osgouei2016identification, osgouei2017improving, tomita2019investigation, klatzky2019detection}). For example, \cite{lim2019touchphoto} prototyped a mobile application called TouchPhoto using electrovibration technology, that helps visually impaired users to not only take photographs, but also perceive the main landmarks of the person's face by touch. In this application, the geometry and texture of important face features were rendered by a gradient-based force algorithm (\citealp{osgouei2017improving}). Although these displays offer a rich touch experience, there has been insufficient research on the haptic exploratory strategies followed by the users to perceive tactile shapes. 
	
	\subsection{Haptic Exploratory Procedures}

	In their seminal studies, Lederman and Klatzky (\citealp{klatzky1987there, lederman1993extracting}) investigated how specific movements lead to extract particular features of the object's weight, texture, hardness, softness, volume, or local shape features. They classified six haptic exploratory procedures, and among those, they identified three for shapes and surface details; \textit{lateral motion} (a repetitive back and forth movement which is mainly associated with exploring textures), \textit{enclosure} (helps to extract the global information about the shape of an object), and \textit{contour following} (helps to extract the exact shape of the object). 
	
	Compared to the haptic perception of 3D objects, less number of studies have investigated the exploration strategies for real and virtual tactile shapes and graphics in 2D (\citealp{rovira2011mental,vinter2012exploratory, bara2014exploratory,guerreiro2015blind, tennison2016toward, bardot2017identifying, tekli2018evaluating, bahrin2019tactile, vinter2020identification}). \cite{rovira2011mental} conducted a study with 2D geometrical shapes (simple and composite) in the form of raised patterns to investigate the haptic exploratory procedures by comparing the performances of blind and sighted adolescents in two experiments on mental rotation (similarity judgment and recognition). They reported that blind adolescents outperformed the sighted ones in simple shapes and preferred bi-manual and multi-finger exploration. On the other hand, sighted participants explored the tactile shapes using the contour following procedure with one hand only.  \cite{vinter2012exploratory} compared the type of exploratory procedures employed by three groups of children (blind, children with low vision, and blindfolded sighted) while exploring 2D tactile patterns. Their findings suggested that children with a visual handicap employed bimanual exploration with a greater number of different procedures as compared to their sighted counterparts. They employed strategies like contour following (dynamic edge following using finger movements) and surface sweeping (dynamic and usually repetitive movement of one or more fingers or of the palm over the model’s surface) to extract the information of the pattern as a whole and strategies like pinch (holding edges in a pincer grip between the thumb and one or more other finger) and local enclosure (dynamic molding of the fingers to parts of the pattern) to get the information about the local features. \cite{bara2014exploratory} conducted a study with visually impaired children to determine how the tactile pictures were explored during joint book reading. Their findings suggested that participants preferred the lateral motion to recognize the shapes in the pictures, and contour following was preferred when finding the pictures' meaning. \cite{bardot2017identifying} investigated the role of each hand during the exploration of different raised-lines graphics (drawing, maps, and graphs) by sighted blind-folded and visually impaired participants. They concluded that visually impaired participants were faster than blind folded participants because they focused on exploring the salient areas of the displayed graphics using bimanual exploration.  \cite{bahrin2019tactile} developed a fingertip tracking system to investigate the exploration strategies employed by blind and visually impaired participants while exploring tactile graphics. The analysis of the collected data showed that  participants employed the contour following and the lateral motion strategies only. \cite{vinter2020identification} investigated the role played by the exploratory procedures employed by the children (sighted and visually impaired) to correctly identify and name the tactile images when the experimental settings were designed similar to their natural reading conditions. Their results showed that, regardless of the visual status of the children, they were successful in correctly identifying and naming the explored image when they used appropriate exploratory procedures.

	In comparison to the studies on real tactile graphics, there are less number of studies investigating the haptic exploratory procedures employed for virtual tactile graphics. 
	 \cite{tennison2016toward} investigated the exploratory procedures utilized by the participants to correctly identify the basic tactile shapes displayed by vibration on a touchscreen. They rendered three shapes (triangle, square and pentagon) in small and large sizes on the touch surface of a tablet using its built-in actuators. Vibrotactile feedback was displayed inside the shapes while stronger vibrations were provided to the participants at the edges. They reported that slow and deliberate exploration movements were preferred by the participants as a starting strategy until they became aware of the key features of the displayed graphics. They observed that some participants scanned the screen horizontally to identify the tactile shape in general and some circled around the corners to find nearby edges in particular. \cite{tekli2018evaluating} investigated how blind users utilize vibrotactile feedback displayed through a touchscreen to recognize simple shapes and graphics. They concluded that participants followed non-traditional haptic exploration strategies such as navigating the screen from left to right horizontally and then from top to bottom vertically to recognize the displayed item on the touchscreen. \cite{tennison2019non} conducted experiments with visually impaired participants to investigate how they explore virtual line profiles displayed on a touchscreen using auditory and vibratory feedback. Their results suggested that vibrating lines were preferred over auditory lines. Furthermore, participants employed similar haptic exploratory strategies to recognize the lines: zigzagging on a line to follow its profile or circling around the intersection points to determine the orientation of intersecting lines.
	
	\subsection{Our Contributions}
	
	We are on the verge of witnessing next-generation surface displays that can display tactile feedback to users. To increase the effectiveness of these displays, it is critical to understand how we perceive tactile graphics displayed on their touch surfaces and what exploratory procedures are followed by the users to perceive them. Although, earlier research has investigated this topic for vibrotactile feedback alone (\citealp{toennies2011toward, giudice2012learning, awada2013evaluation, palani2014evaluation, tennison2016toward, tekli2018evaluating}) and also the combination with auditory feedback (\citealp{poppinga2011touchover,giudice2012learning, tennison2019non, gorlewicz2020design}), to our knowledge, no previous work has investigated it for electrovibration in depth. In comparison to vibrotactile feedback which mainly stimulates the finger in the normal direction, electrovibration modulates the sliding friction between finger and the touch surface in the tangential direction. Hence, one can anticipate that haptic exploratory procedures followed by the users for virtual tactile graphics under electrovibration would be different than those employed for real tactile graphics and also the virtual ones rendered by vibrotactile actuators.

	In this study, we conducted experimental studies to investigate the tactile exploration strategies followed by the users for basic geometrical shapes displayed by electrovibration on a touchscreen. We first investigated the effects of (a) number of edges, (b) rendering conditions, and (c) angular orientation on the recognition rate and time of 2D equilateral tactile shapes to answer the following research question; \textit{How does the size of haptically active area and the complexity of geometry affect the perception of tactile shapes?} Moreover, we evaluated the strategies followed by the participants to explore these shapes by performing spatial and temporal analyses on the recorded data of finger position to tackle the following research question; \textit{What type of exploratory procedures were followed by the participants to perceive tactile shapes?} Spatial analysis  was performed quantitatively by using the heat maps generated from participants' finger positions, and temporal analysis was performed qualitatively by tagging videos of finger movements via visual inspection.  
	
	\section{Methodology}
	
	\subsection{Experimental Setup}
	
\begin{figure}[h!]
		\centering
		\subfloat[]{
			\resizebox*{12cm}{!}{\includegraphics{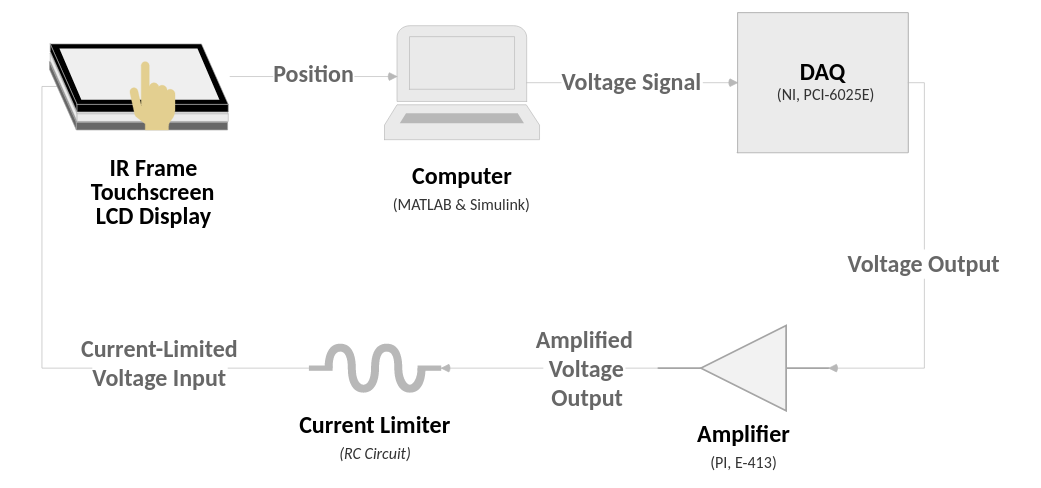}}\label{fig1a}}
		\newline
		\vspace{6pt}
		\subfloat[]{%
			\resizebox*{10cm}{!}{\includegraphics{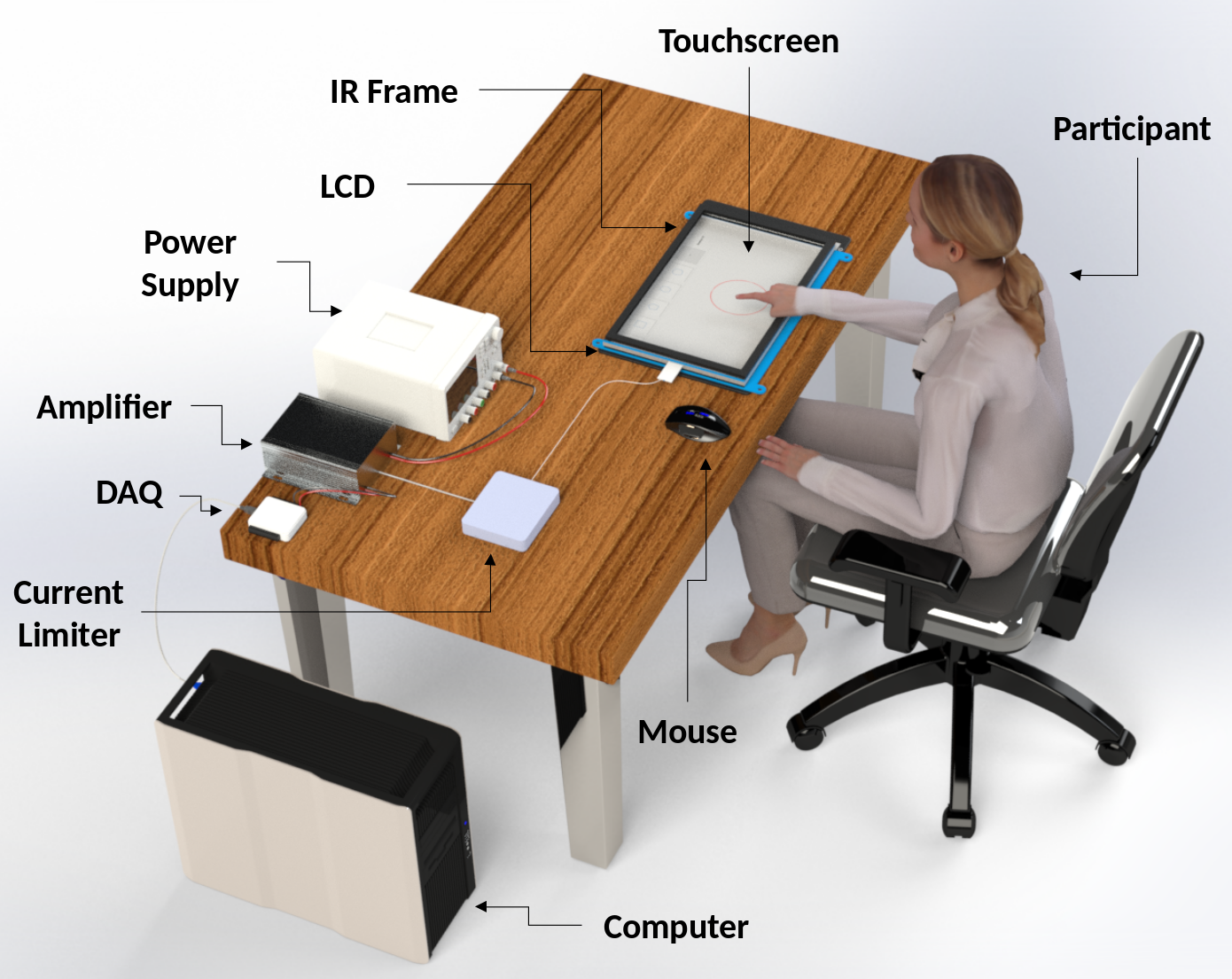}}\label{fig1b}}
		\caption{(a) Schematic showing how voltage signal is generated, processed, and transmitted to the touchscreen. (b) Illustration of the experimental setup used in our study. } \label{fig1}
		
	\end{figure}

	A 19-inch surface capacitive touchscreen (SCT3250, 3M Inc.) was used to display tactile stimuli to participants. The desired voltage signal was generated by a DAQ card (PCI-6025E, National Instruments Inc.) and boosted by an amplifier (E-413, PI Inc.) before transmitted to the touchscreen through a current limiter. Figure \ref{fig1a} depicts the process of generating voltage signal for rendering tactile shapes on the touchscreen. The waveform, amplitude, and the frequency of the voltage signal were a square wave, 200 Vpp, and 120 Hz, respectively, where the wave characteristics were selected to provide the most differentiable haptic sensation possible (\citealp{vardar2017roughness, icsleyen2019tactile}). An LCD was positioned below the touchscreen for visual feedback to participants and to let them pick the perceived tactile shape using a mouse and a GUI. An IR frame (NIB170BP, Nexio Inc.) was placed above the touchscreen to detect and record the position of participants' index finger. MATLAB and Simulink Desktop Real-Time were utilized to develop the GUI to record the participants' responses and to display electrovibration to participants in real-time based on their finger positions, respectively.  Figure \ref{fig1b} illustrates the setup used for our experiments. 
	
	\section{Experiment I}
	
	We investigated the effect of number of edges and rendering conditions on the recognition rate and time of 2D equilateral tactile shapes. Also, we examined the strategies followed by the participants while exploring these shapes on the touchscreen.
	
	\subsection{Participants}
	
	Nine participants (5 male, 4 female with an average age of 27 $\pm$ 6.7 years)  participated in this experiment. All of them were graduate students and everyday users of mobile devices. The participants read and signed a consent form before the experiment. The Ethical Committee of Ko\c{c} University approved the form for human participants.
	
	\subsection{Stimuli}
	
\begin{figure}[h!]
		\centering
		\subfloat[]{
			\resizebox*{3cm}{!}{\includegraphics{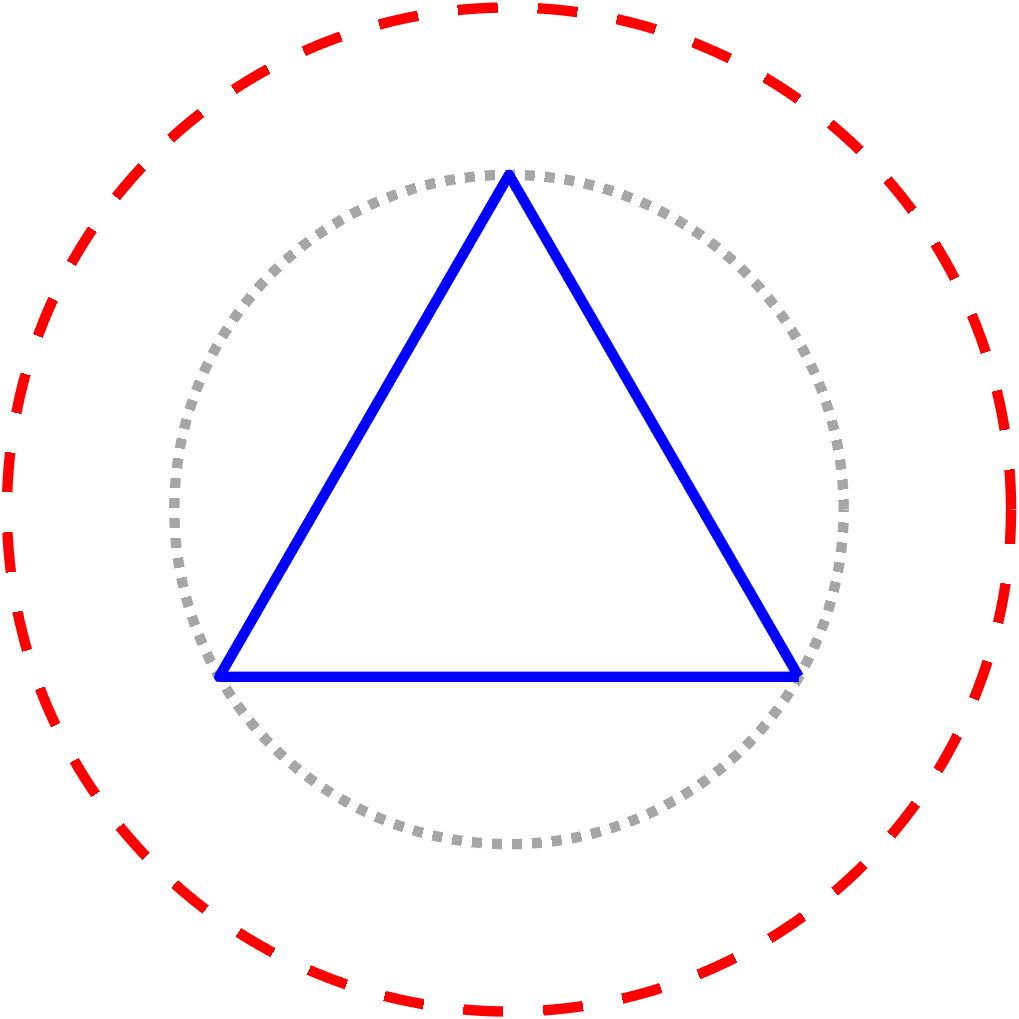}}\hspace{6pt}
			\resizebox*{3cm}{!}{\includegraphics{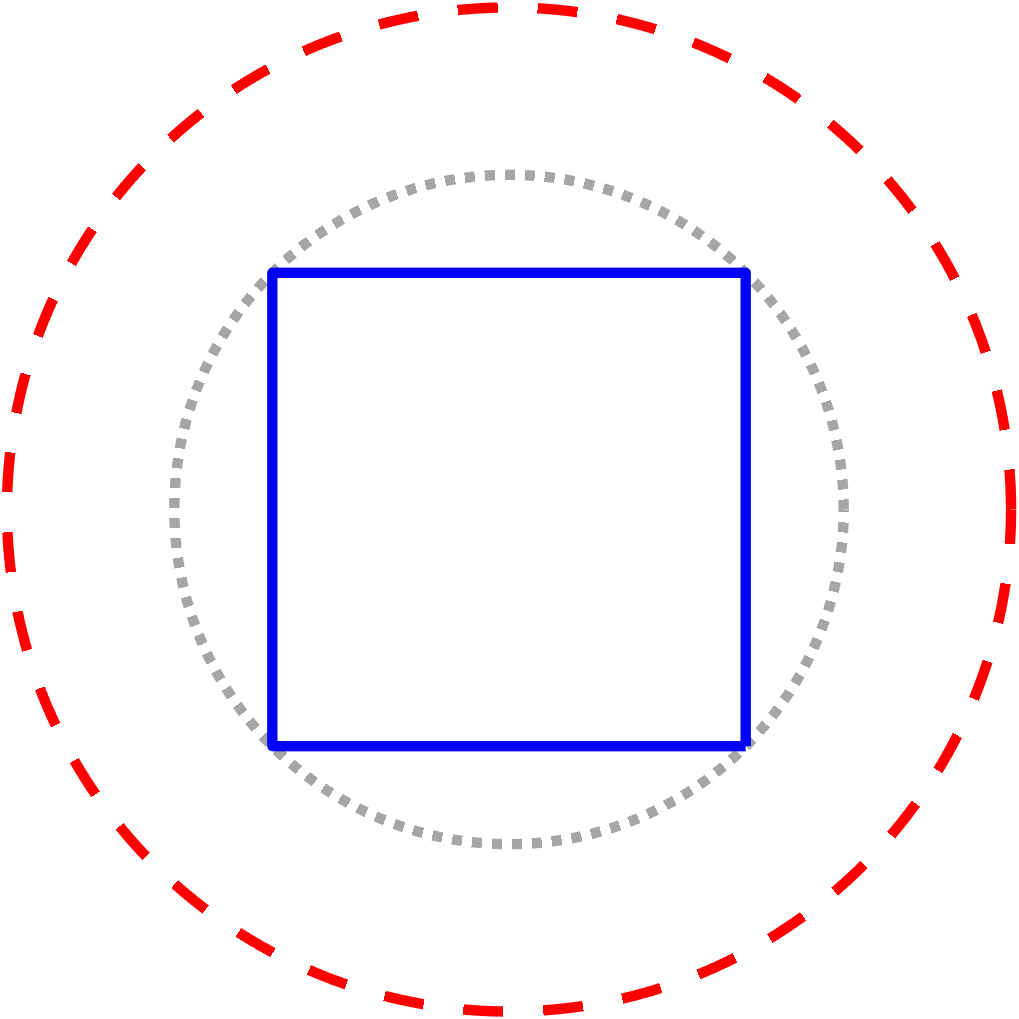}}\hspace{6pt}
			\resizebox*{3cm}{!}{\includegraphics{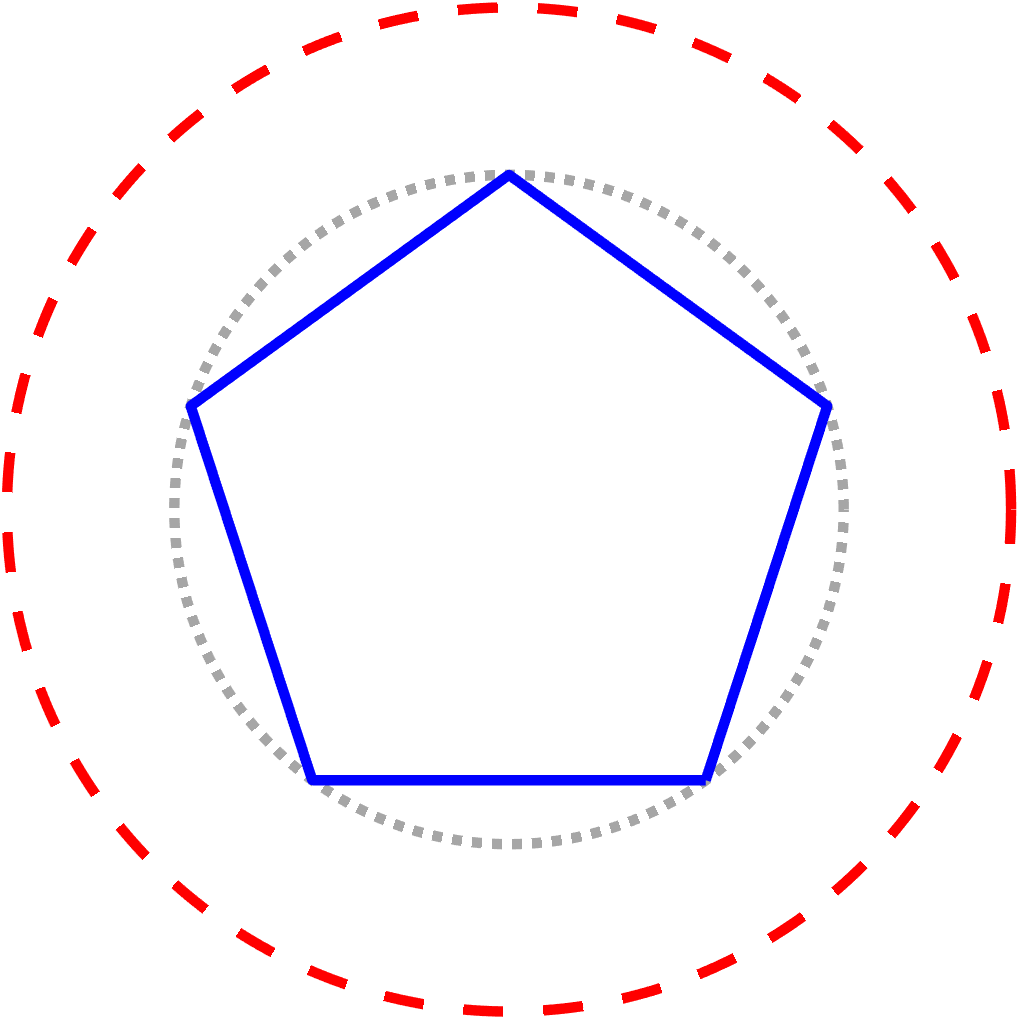}}\hspace{6pt}
			\resizebox*{3cm}{!}{\includegraphics{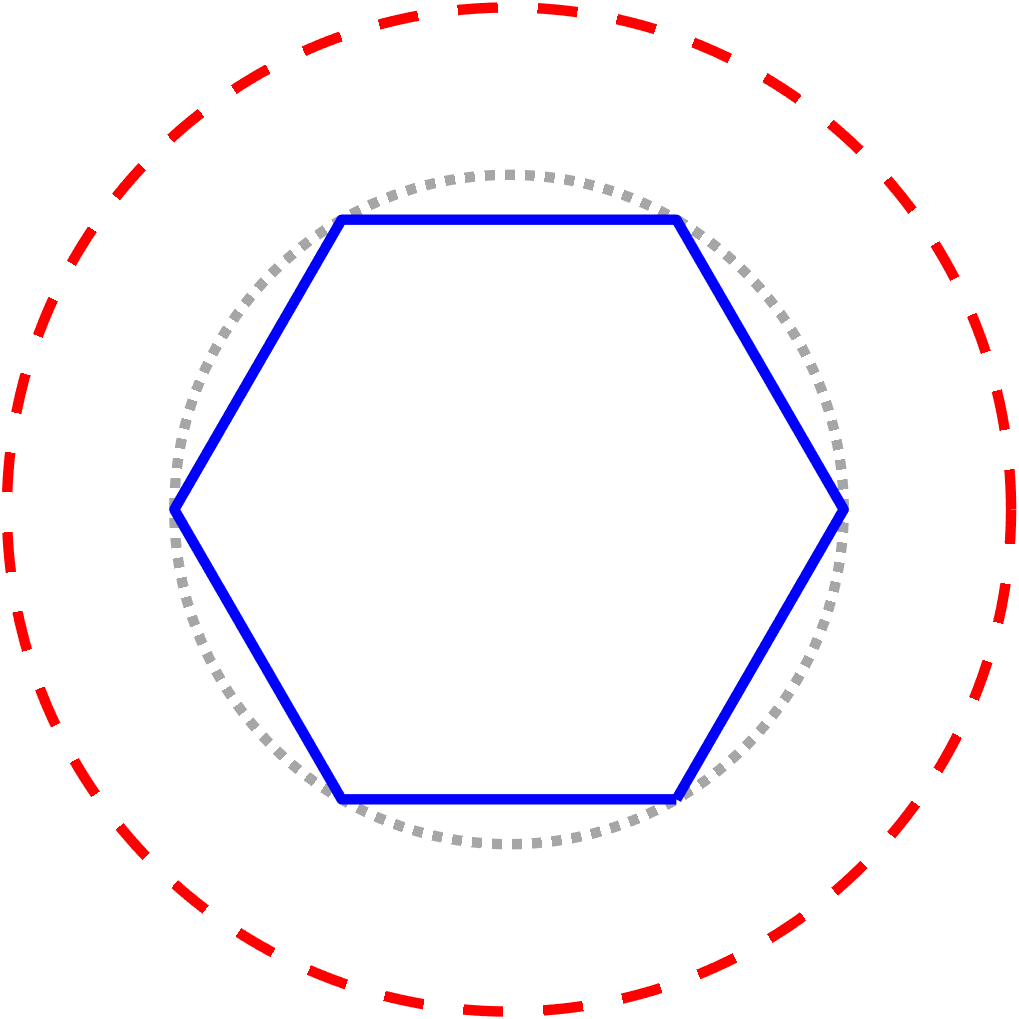}}\hspace{6pt}
			\resizebox*{3cm}{!}{\includegraphics{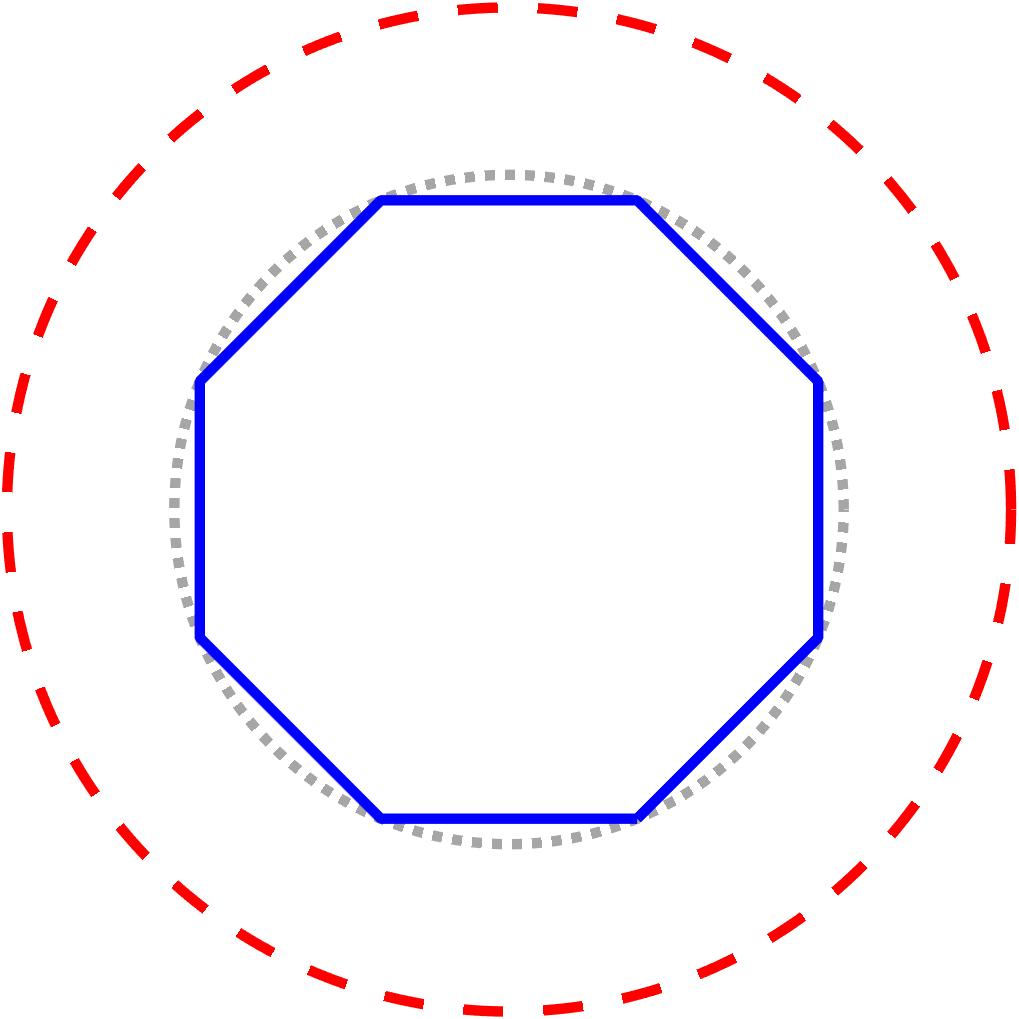}}\label{fig2a}}
		\newline
		\vspace{6pt}
		\subfloat[]{%
			\resizebox*{3cm}{!}{\includegraphics{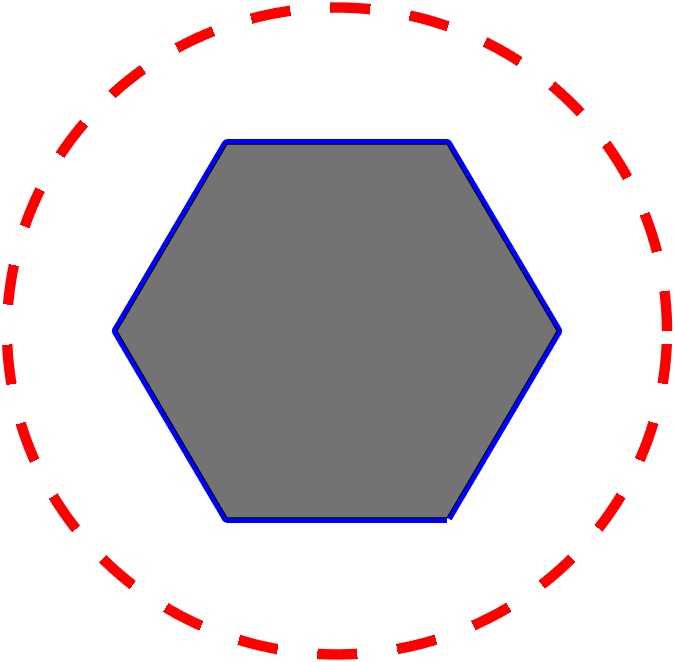}}\hspace{6pt}
			\resizebox*{3cm}{!}{\includegraphics{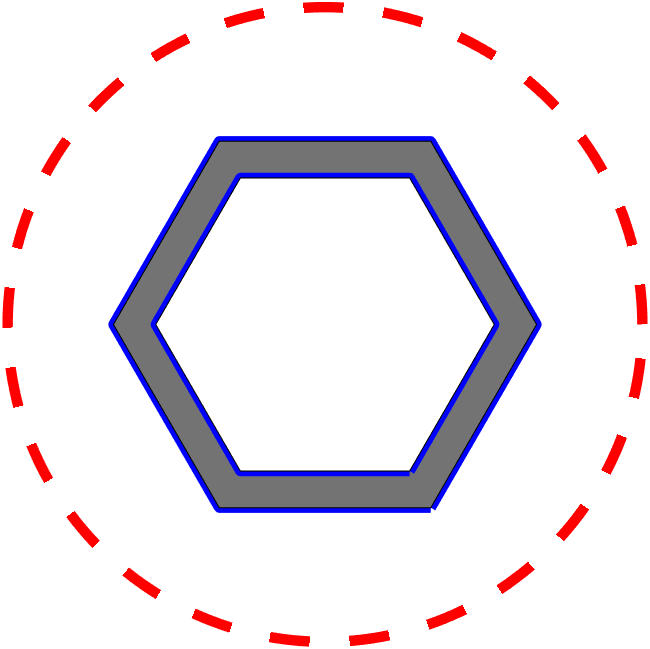}}\hspace{6pt}
			\resizebox*{3cm}{!}{\includegraphics{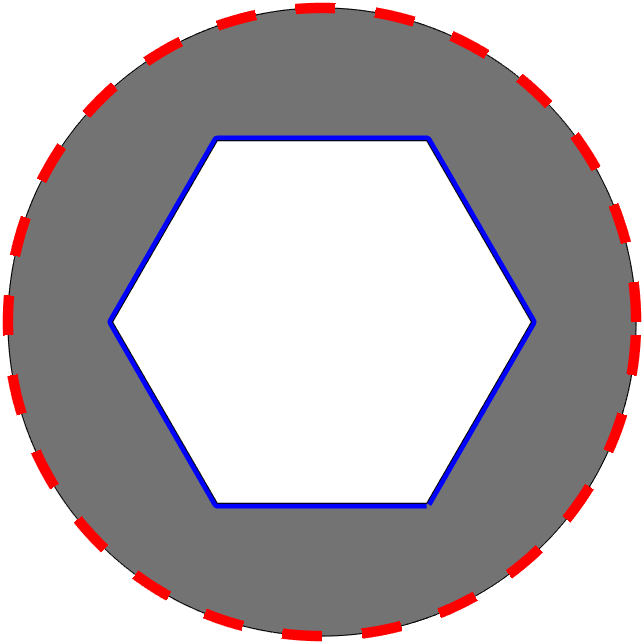}}
			\label{fig2b}}
		\caption{(a) Five tactile shapes were displayed to the participants via electrovibration under three rendering conditions (b) (left to right); electrovibration was displayed inside the shape (\ac{c1}), on the edges (\ac{c2}), and  outside the shape (\ac{c3}). The shaded areas in (b) illustrate the haptically active areas in all rendering conditions. Solid blue-colored lines in (a) represent rendered shapes, dotted gray-colored circle represents circumcircle of the shapes, and dashed red-colored circle represents the area where tactile shapes were displayed to the participants. Only red dashed circle was displayed to the participants visually during the actual experiment. }
		\label{fig2}
	\end{figure}

	The experimental stimuli consisted of five geometrical shapes (triangle, square, pentagon, hexagon, and octagon) displayed under three rendering conditions; electrovibration was displayed 1) inside the shape (\ac{c1}), 2) on the edges of the shape (\ac{c2}), and outside the shape (\ac{c3}) as shown in Figure \ref{fig2}. In edge rendering condition, the width of edges was set to 10 mm, which was chosen based on the guidelines provided by \cite{palani2018touchscreen} for vibrotactile rendering of virtual lines. The size of shapes was selected to fit in a circle with a diameter of 10 cm (see the dotted gray-colored inner circle in Figure \ref{fig2a}). This diameter was selected by trial and error such that the shapes were sufficiently large and could be detected by finger with ease. Although the shapes were not displayed visually to the participants during the actual experiment, a circle with a diameter of 15 cm (see the dashed red-colored outer circle in Figure \ref{fig2a}) was visually displayed to the participants to help them locate the rendered tactile shape more easily on the touchscreen.

	\subsection{Procedure}
	
	Before the actual experiment, participants performed a training session to familiarize themselves with the experimental setup, the tactile feedback displayed for the shapes, and rendering conditions. Each shape was displayed once to the participants in sequential order under the three rendering conditions (\ac{c1}, \ac{c2}, \ac{c3}). Participants were verbally informed about the types of shapes and the three rendering conditions. Participants were allowed to ask questions during the training session. 
	
	After the training session, participants performed the actual experiment. The task was to explore the tactile shapes on the touchscreen using the index finger and then select the perceived shape from the five choices (Triangle, Square, Pentagon, Hexagon, Octagon) presented on the GUI using a mouse. There was no time limit in the experiment. Participants were not allowed to ask any questions during the actual experiment. They were asked to put on noise-canceling headphones to prevent any external noise affecting their haptic perception. 
	
	The experiment was performed in two separate sessions with one-day interval between them. There were 30 trials in each session (5 shapes $\times$ 3 rendering conditions $\times$ 2 repetitions). Hence, the total number of trials for each participant was 60. The trials were randomized for each session, while the same randomization pattern was used for all participants.

	\subsection{Analysis} \label{sec21}
	The study was a within-subjects design with two independent variables; rendering conditions and number of edges. We analyzed the user performance based on the following metrics:
	\begin{itemize}
		\item Recognition rate: The ratio of the correctly identified shape to the total number of displayed shapes.
		\item Recognition time: The time taken by the participants to correctly identify the displayed shape, which was measured in seconds.
	\end{itemize}

	Furthermore, we recorded the participants' finger position as a function of time to investigate the strategies followed by the participants during the exploration process based on the spatial and temporal information. The collected data allowed us to generate animated video files for all the interactions made by each participant during the exploration of shapes. These video files were then used to characterize and code the exploration strategies followed by the participants.
	
\begin{figure}
		\centering
		\subfloat[]{
			\resizebox*{5cm}{!}{\includegraphics{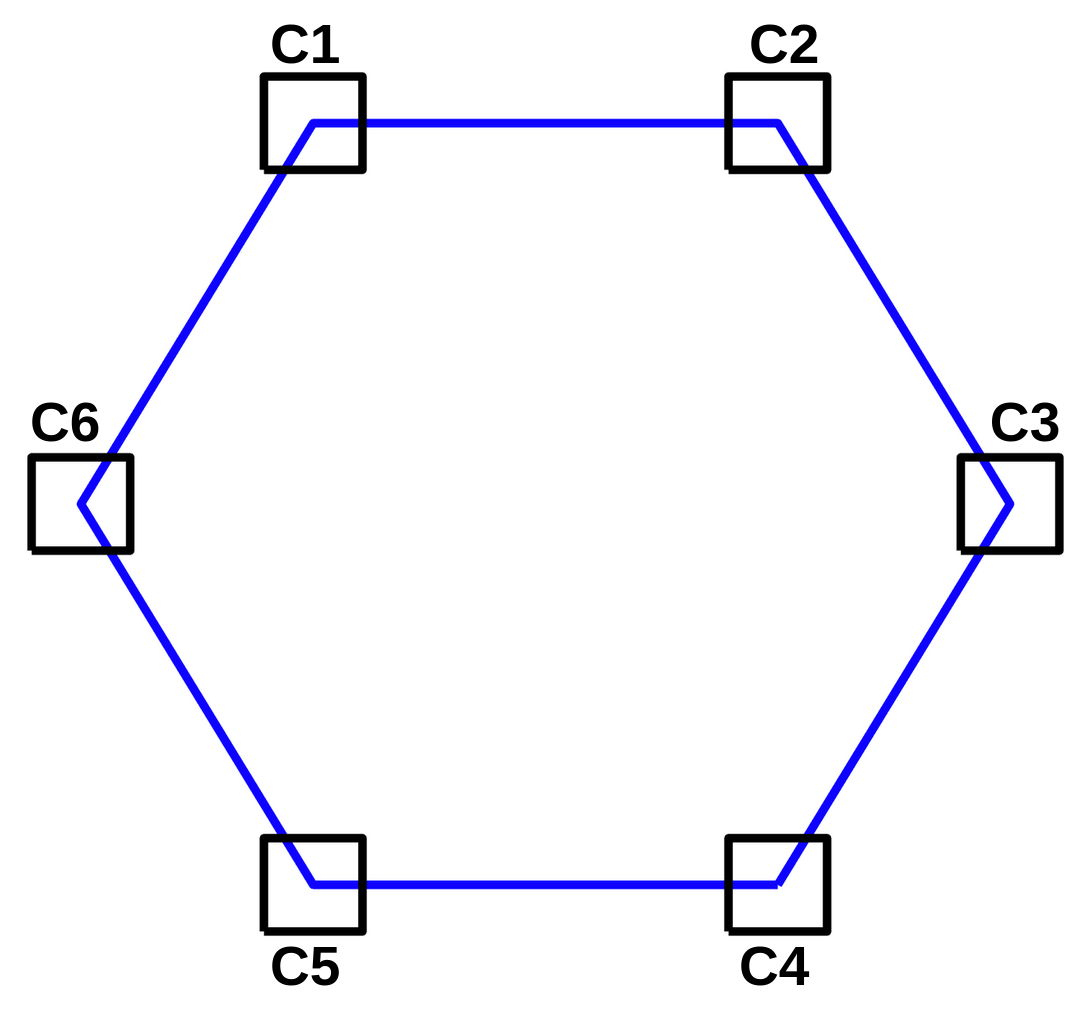}}\hspace{15pt}\label{fig3a}}
		\subfloat[]{%
			\resizebox*{5cm}{!}{\includegraphics{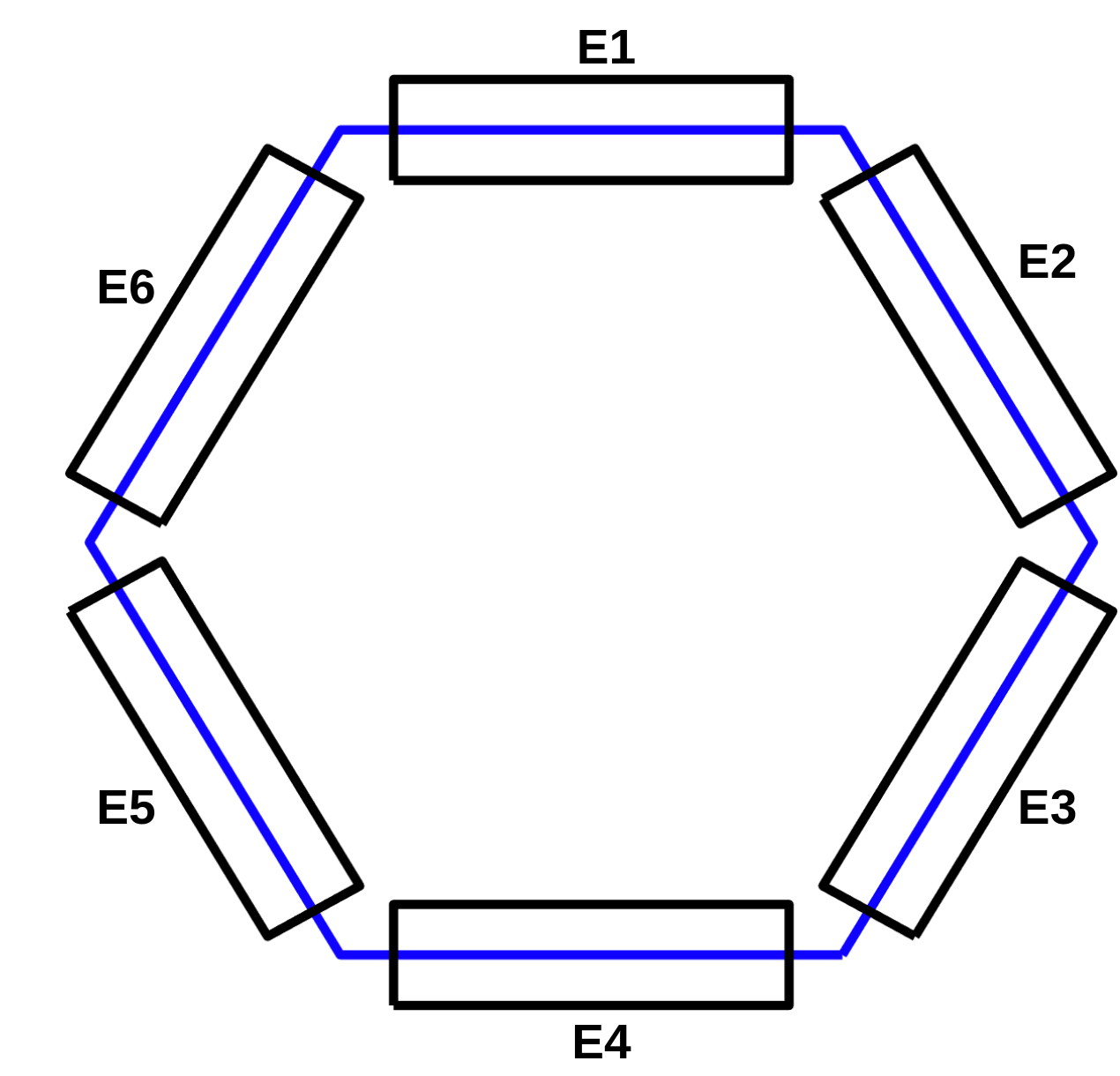}}\label{fig3b}}
		\caption{All the touches made inside the square box at each corner (a) and the rectangular box at each edge (b) were counted.}  \label{fig3}
		
	\end{figure}

	In spatial analysis, we searched for any salient spatial features that help participants to distinguish one shape from another. We used the following metrics for spatial analysis:
	
	\begin{itemize}
		\item \textbf{Number of touches to each corner} was counted by defining a small square region of 1.5 cm in size around each corner as shown in Figure \ref{fig3a}. The corners were labeled by starting from the top-left corner and moving in clockwise direction as shown in Figure \ref{fig3a}.
		\item \textbf{Number of touches to each edge} was counted  by defining a rectangular region of 3.5 cm $\times$ 1.5 cm in size along each edge as shown in Figure \ref{fig3b}. The edges were labeled by starting from the top edge and moving in clockwise direction (Figure \ref{fig3b}).
		\item \textbf{Number of touches to the haptically active area:} All the touches occurred  in the rendered area were counted. For example, if the rendering method was ``\ac{c1}", the haptically active area was inside the shape. Hence, all the touches inside the shape were counted.
		
	\end{itemize}
	
	In temporal analysis, we investigated participants' exploration strategies and identified two main strategies; global scanning and local scanning, as described below:
	\begin{itemize}
		\item \textbf{Global scanning:} If a path was traced using long and continuous movements for the identification of global features of the rendered shape (see Figure \ref{fig4}), the exploration was classified as global scanning. Global scanning was further categorized into three sub-categories based on the exploratory movements (Figure \ref{fig4}):
		
		\begin{itemize}
			\item \textbf{Global-horizontal:} If the scanning movements were along the x-axis, the exploration was classified as global-horizontal (Figure \ref{fig4a}).
			\item \textbf{Global-vertical:} If the scanning movements were along the y-axis, the exploration was classified as global-vertical (Figure \ref{fig4b}).
			\item \textbf{Global-others:} All other movement patterns such as circling and diagonal were classified as global-others (Figure \ref{fig4c}).
		\end{itemize}
		\begin{figure}[h!]
			\centering
			\subfloat[]{%
				\resizebox*{5cm}{!}{\includegraphics{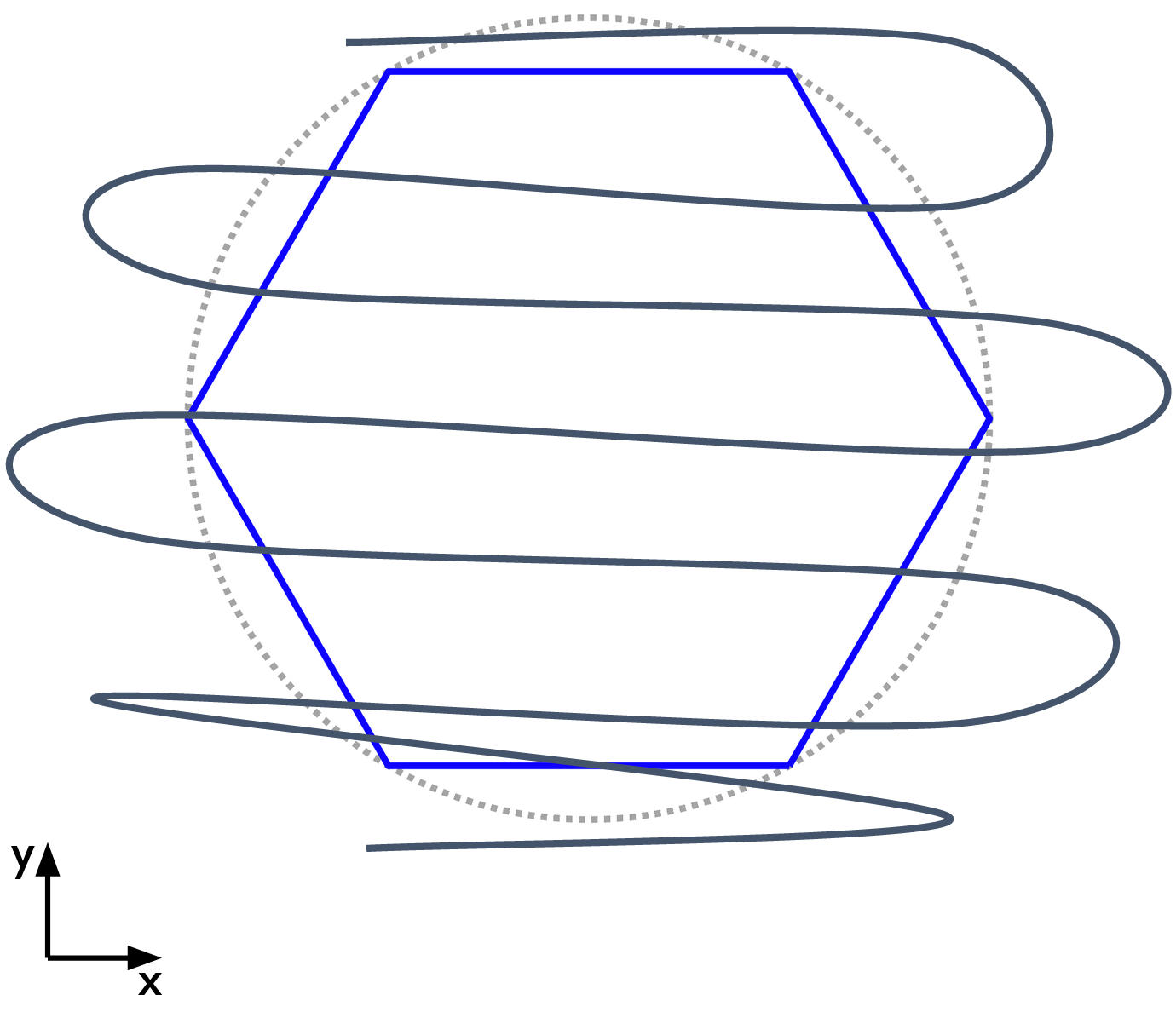}}\label{fig4a}}
			\subfloat[]{%
				\resizebox*{5cm}{!}{\includegraphics{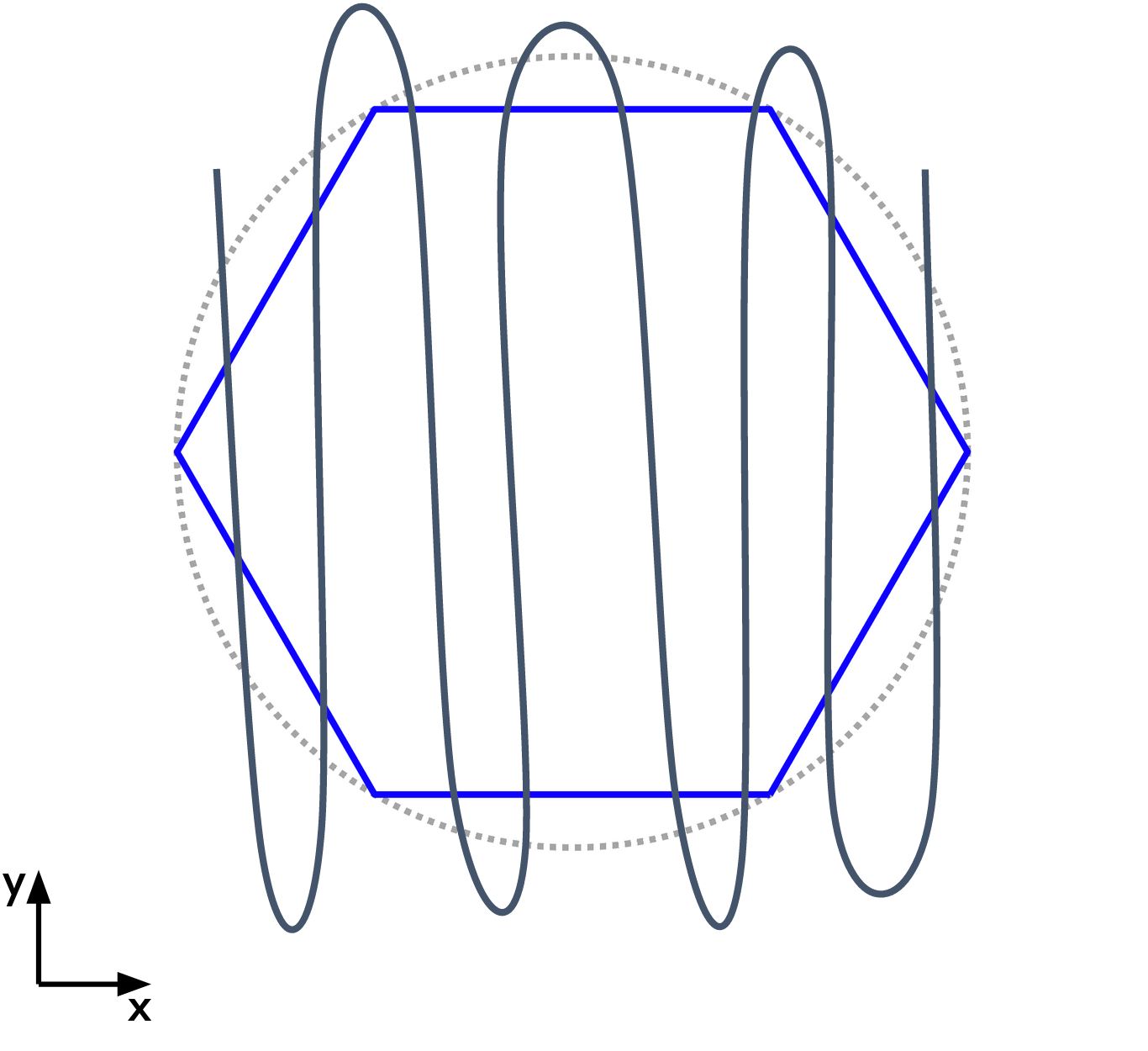}}\label{fig4b}}
			\subfloat[]{%
				\resizebox*{5cm}{!}{\includegraphics{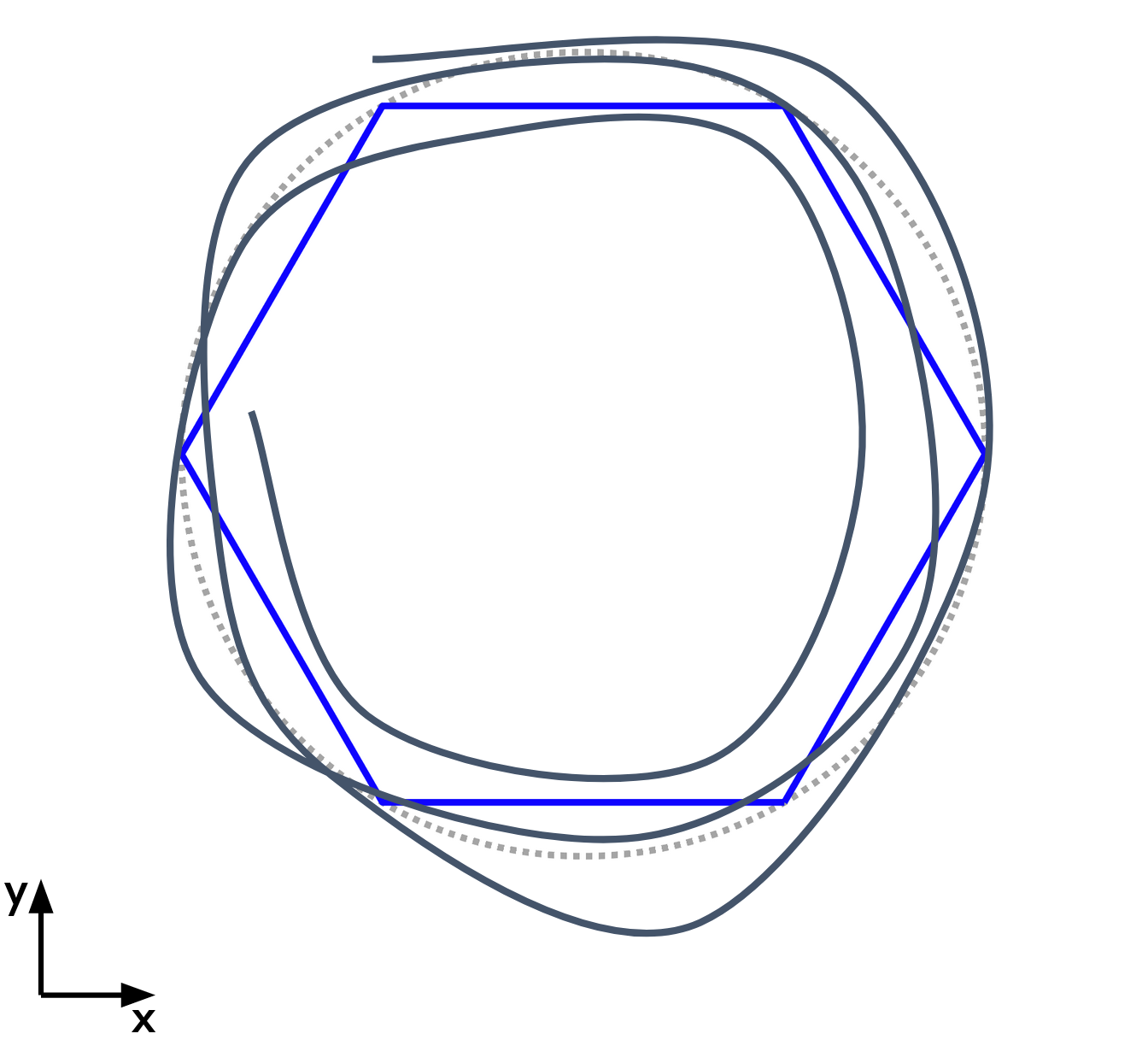}}\label{fig4c}}
			\caption{The strategies followed by the participants during global scanning: (a) global-horizontal, (b) global-vertical, and (c) global-others. } \label{fig4}
		\end{figure}
		
		\item \textbf{Local scanning:} If a path was traced using short and continuous movements for the identification of local features of the rendered shape (see Figure \ref{fig5}), the exploration was classified as local scanning. Local scanning was further categorized into three sub-categories based on the exploratory movements (Figure \ref{fig5}):
		\begin{itemize}
			\item \textbf{Local-edge-finding:} If the finger scanned an edge by moving back and forth along the edge, the exploration was classified as local-edge-finding (Figure \ref{fig5a}).
			\item \textbf{Local-edge-following:} If the finger made zig-zag movements along an edge, the exploration was classified as local-edge-following (Figure \ref{fig5b}).
			\item \textbf{ Local-corner-finding:} If the finger circled around a corner, the exploration was classified as local-corner-finding (Figure \ref{fig5c}).
		\end{itemize}
		\begin{figure}[h!]
			\centering
			\subfloat[]{%
				
				\resizebox*{5cm}{!}{\includegraphics{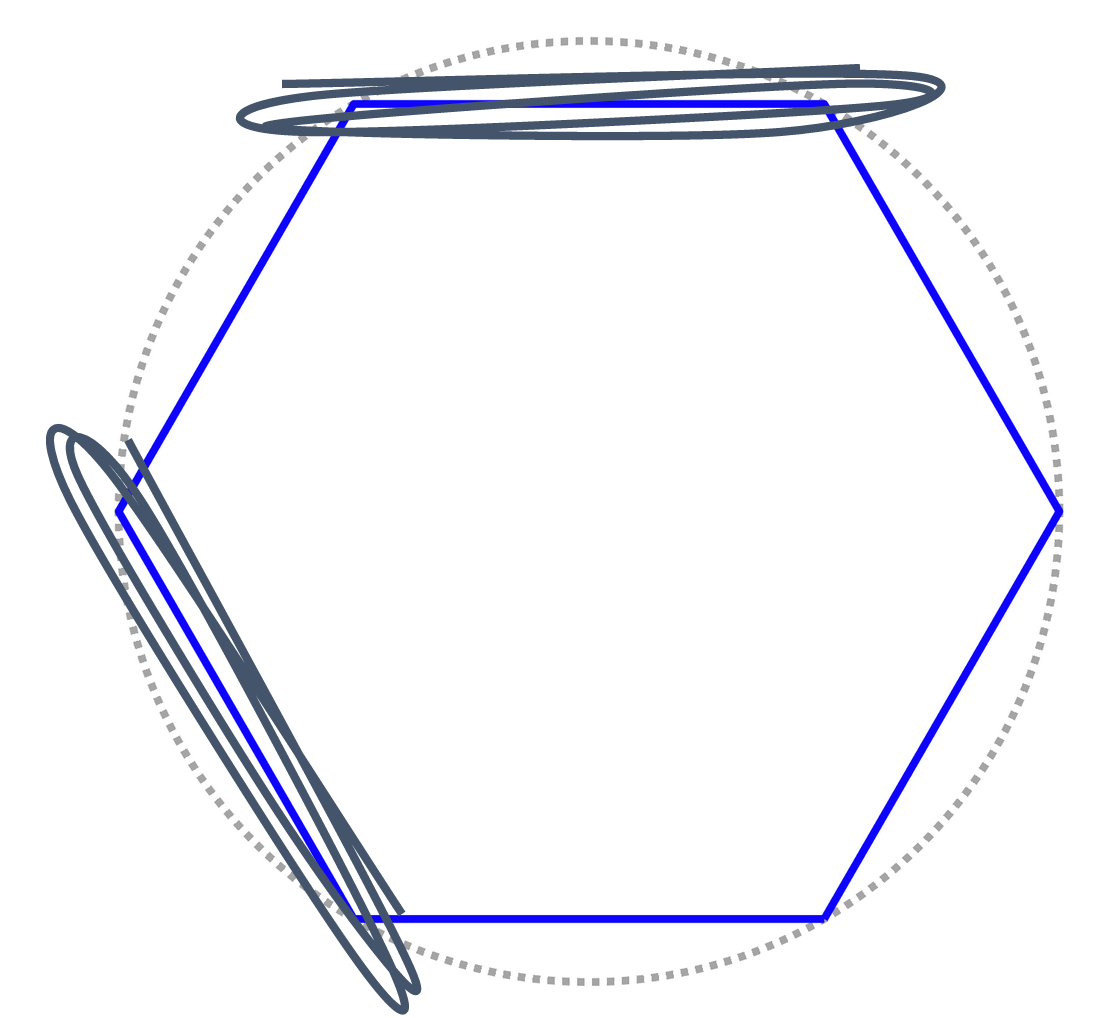}}\label{fig5a}}
			\subfloat[]{%
				
				\resizebox*{4.6cm}{!}{\includegraphics{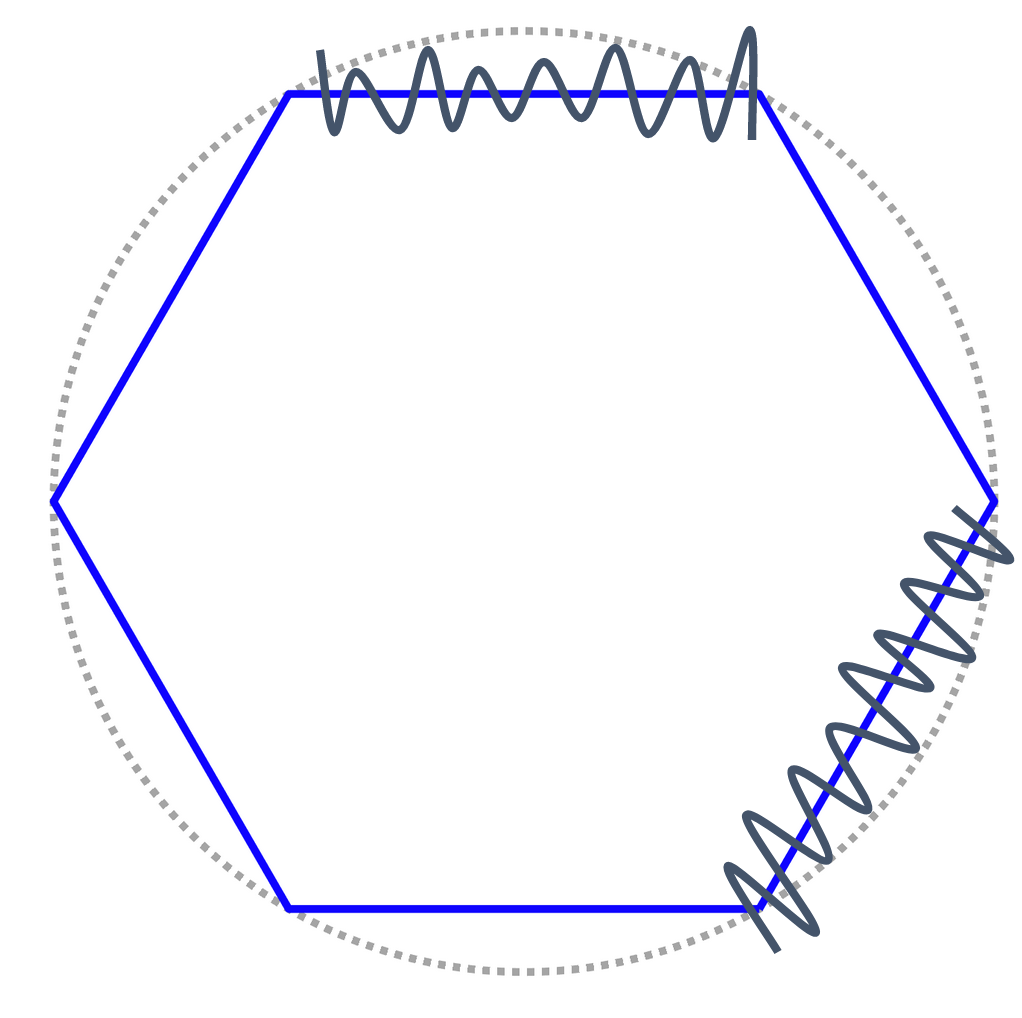}}\label{fig5b}}
			\subfloat[]{%
				
				\resizebox*{5cm}{!}{\includegraphics{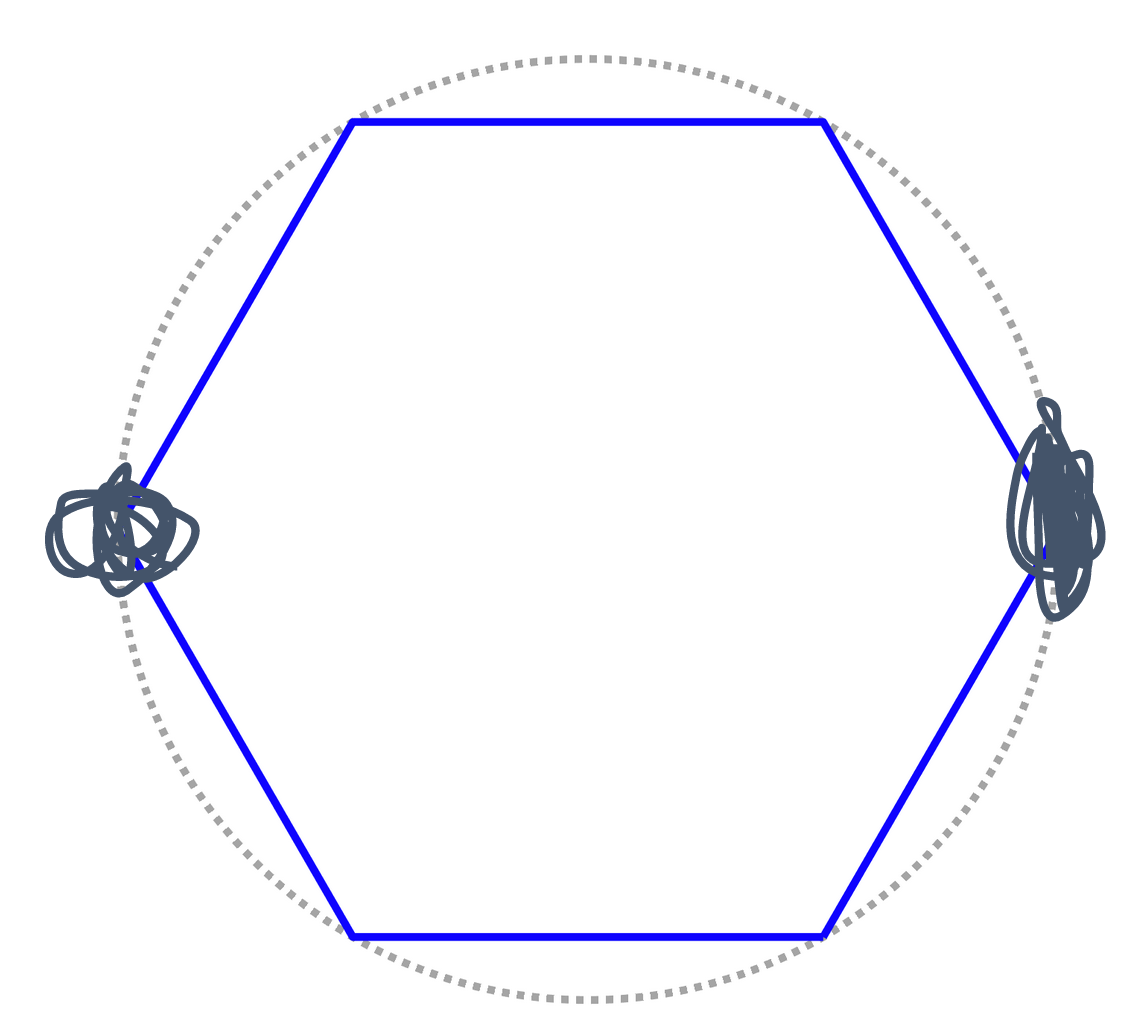}}\label{fig5c}}
			\caption{The strategies followed by the participants during local scanning: (a) local-edge-finding, (b) local-edge-following, and (c) local-corner-finding.} \label{fig5}
		\end{figure}
	\end{itemize}
	
	The time spent by the participants for each exploratory movement during local and global scanning was recorded.

	\subsection{Results}
	\subsubsection{Analysis of Recognition Rate and Time}
\begin{figure}
		\centering
		\subfloat[]{%
			\resizebox*{8cm}{!}{\includegraphics{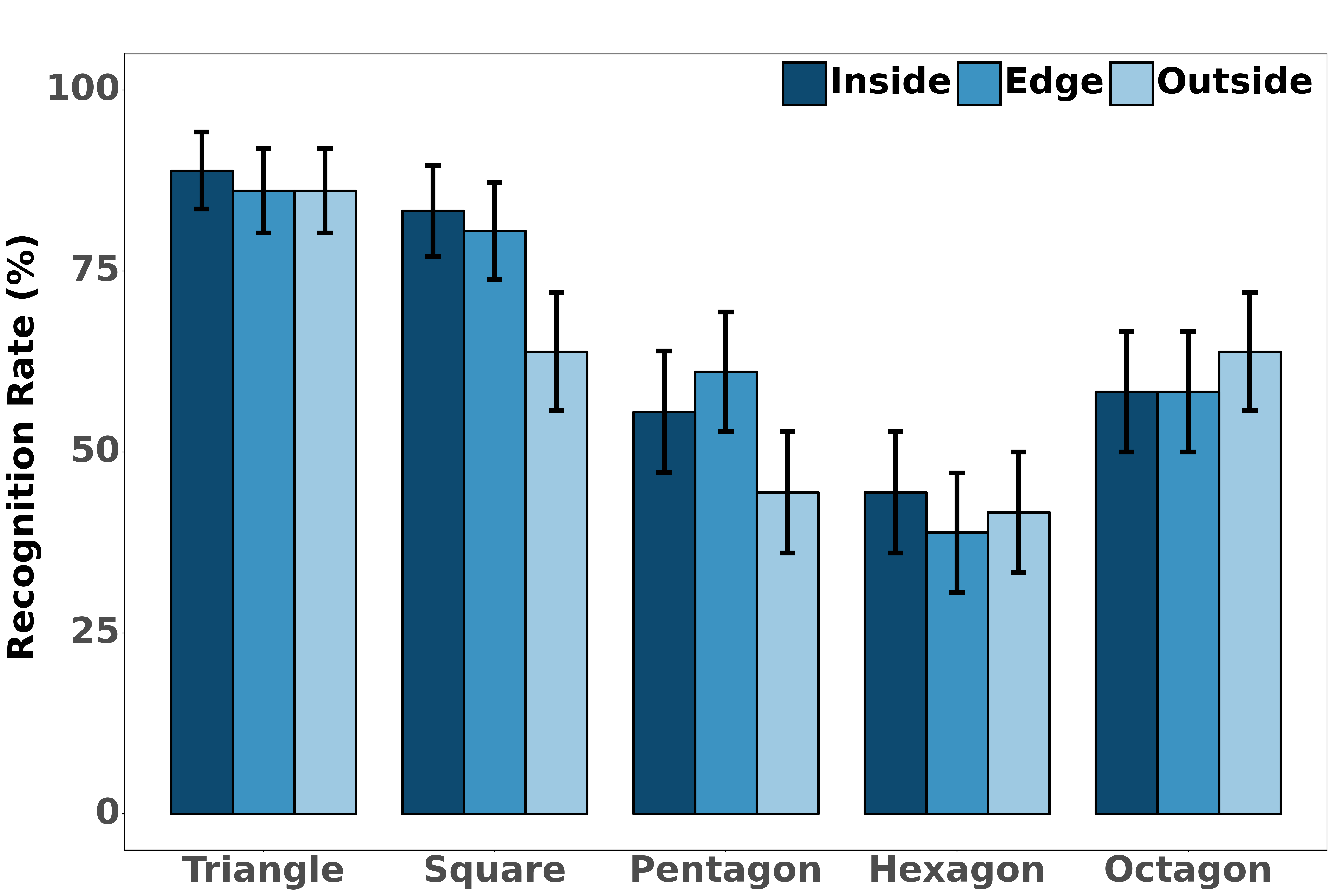}}\label{fig6a}}\hspace{5pt}
		\subfloat[]{%
			\resizebox*{8cm}{!}{\includegraphics{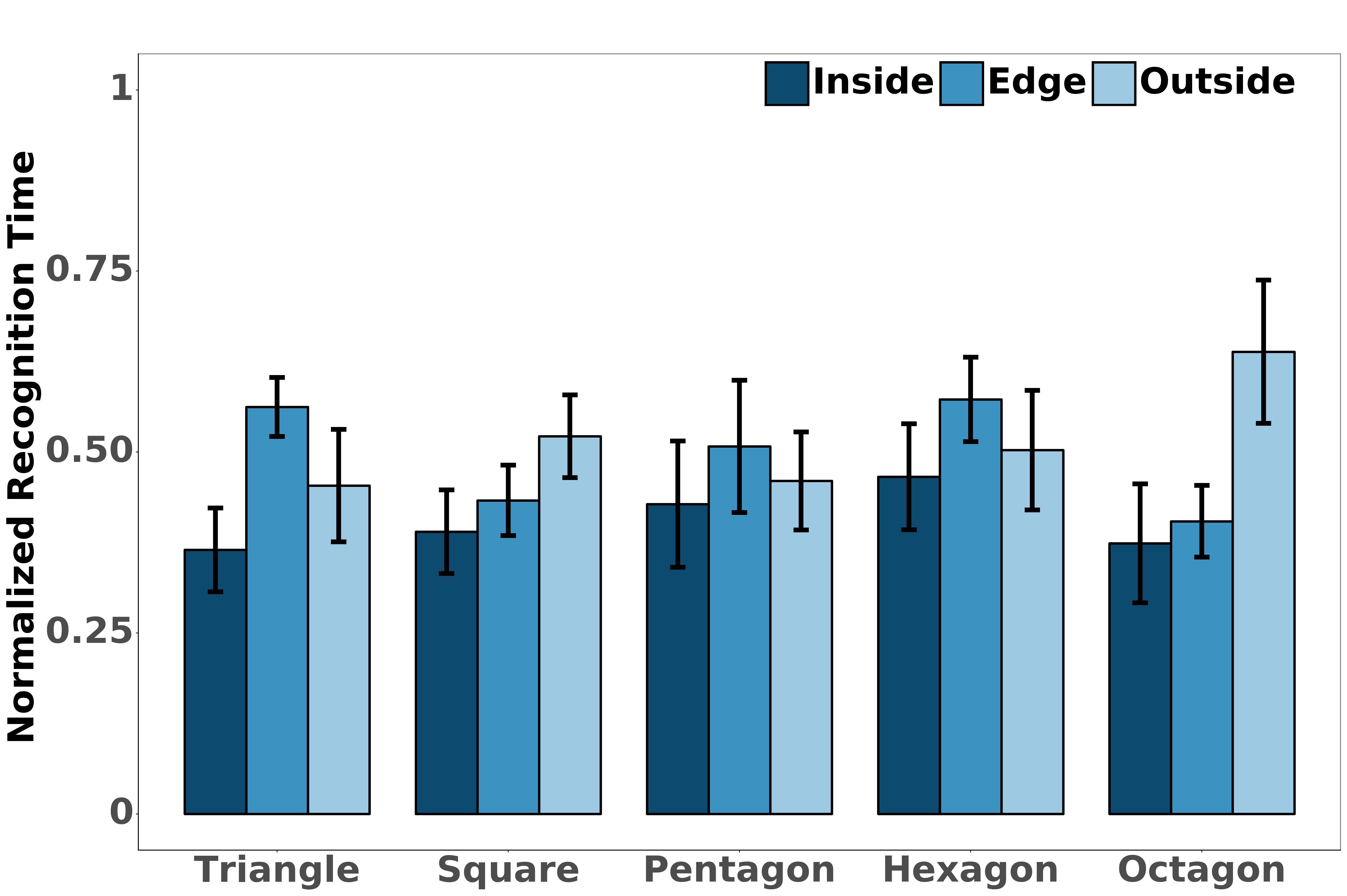}}\label{fig6b}}
		\caption{(a) Recognition rate and (b) normalized recognition time  with standard error of the means for all shapes and rendering conditions.} \label{fig6}
	\end{figure}

 	Figure \ref{fig6} shows the average recognition rate and normalized recognition time of all participants under the three rendering conditions. The normalized recognition time of each participant for each shape was computed using her/his minimum and maximum values. Since the participants were asked to recognize the tactile shapes in their own exploration speed, this normalization helps to make fair comparisons among the shapes for different participants. The mean recognition times in seconds with the standard error of the means for all shapes and rendering conditions are tabulated in Table \ref{tab1}. The results showed that irrespective of the rendering condition, as the number of edges was increased, the participants' recognition rate decreased except for octagon. The recognition rate for octagon was higher than that of  hexagon under all three rendering conditions (Figure \ref{fig6a}). Also, the participants spent less time in recognizing the shapes in the case of \ac{c1} rendering condition as compared to the \ac{c2} and \ac{c3} rendering conditions.

    \begin{table}[!ht]
    \centering
    \caption{Mean recognition times with standard error of the means for all shapes and rendering conditions.}
    \label{tab1}
    \resizebox{12cm}{!}{\begin{tabular}{|p{2.5cm}|>{\centering}m{3cm}|>{\centering}m{3cm}|>{\centering\arraybackslash}m{3.5cm}|}
    \hline
    \diaghead(5,-2){\hskip4.5cm}{\textbf{Shape}}{\textbf{Rendering}\\\textbf{Condition}} & \thead{\textit{Inside}} & \thead{\textit{Edge}} & \thead{\textit{Outside}}        \\ \hline
    Triangle & 50.97 $\pm$ 3.91 sec & 64.27 $\pm$ 2.75 sec & 56.96 $\pm$ 5.24 sec \\ \hline
    Square   & 52.66 $\pm$ 3.89 sec  & 55.58 $\pm$ 3.28 sec & 61.55 $\pm$ 3.85 sec \\ \hline
    Pentagon & 55.23 $\pm$ 5.89 sec & 60.60 $\pm$ 6.16 sec & 57.39 $\pm$ 4.57 sec \\ \hline
    Hexagon  & 57.77 $\pm$ 4.93 sec & 64.98 $\pm$ 3.94 sec & 60.25 $\pm$ 5.56 sec \\ \hline
    Octagon  & 51.58 $\pm$ 5.53 sec & 53.63 $\pm$ 3.34 sec & 69.42 $\pm$ 6.68 sec \\ \hline
    \end{tabular}}
    \end{table}

	We conducted a two-way repeated measures multivariate analysis of variance (MANOVA) with two independent variables (rendering condition: \ac{c1}, \ac{c2}, \ac{c3}; number of edges: triangle (3), square (4), pentagon (5), hexagon (6), octagon (8)) and two dependent variables (recognition rate and recognition time). The MANOVA found significant main effects of rendering condition $(F(4, 32) = 3.255, p < 0.05, \eta ^2=0.289)$ and number of edges $(F(8, 64) = 3.701, p<0.001,\eta ^2=0.316)$. The interaction effect between rendering condition and number of edges was non-significant $(F(16, 128) = 1.534, p = 0.097, \eta^2=0.161)$. Individual ANOVA found a significant main effect of rendering condition on recognition time $(F(2, 16) = 7.529, p<0.01, \eta^2=0.485)$. Also, the effect of number of edges on recognition rate was significant $(F(4, 32) = 8.848, p<0.001, \eta^2=0.525)$. Our post-hoc analysis showed that \ac{c1} rendering condition resulted in significantly shorter recognition time as compared to the \ac{c2} and \ac{c3} rendering conditions $(p_{(Inside-Edge)} = 0.003$ and $p_{(Inside-Outside)} = 0.042)$. 
	\ac{c2} rendering condition led to the longest recognition time. 
	
\begin{figure}
		\centering
		\subfloat[]{%
			\resizebox*{5cm}{!}{\includegraphics{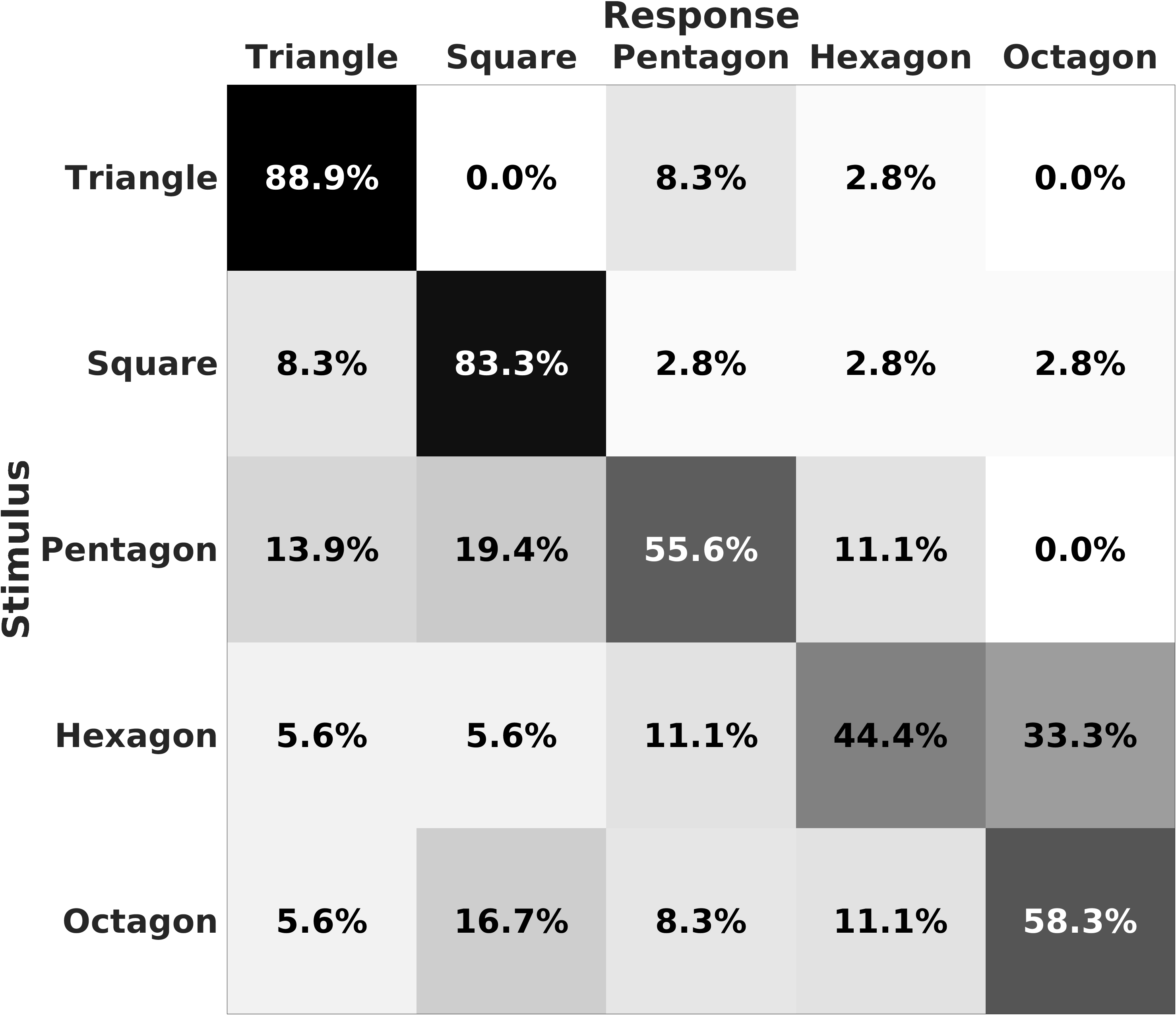}}\label{fig7a}}\hspace{5pt}
		\subfloat[]{%
			\resizebox*{5cm}{!}{\includegraphics{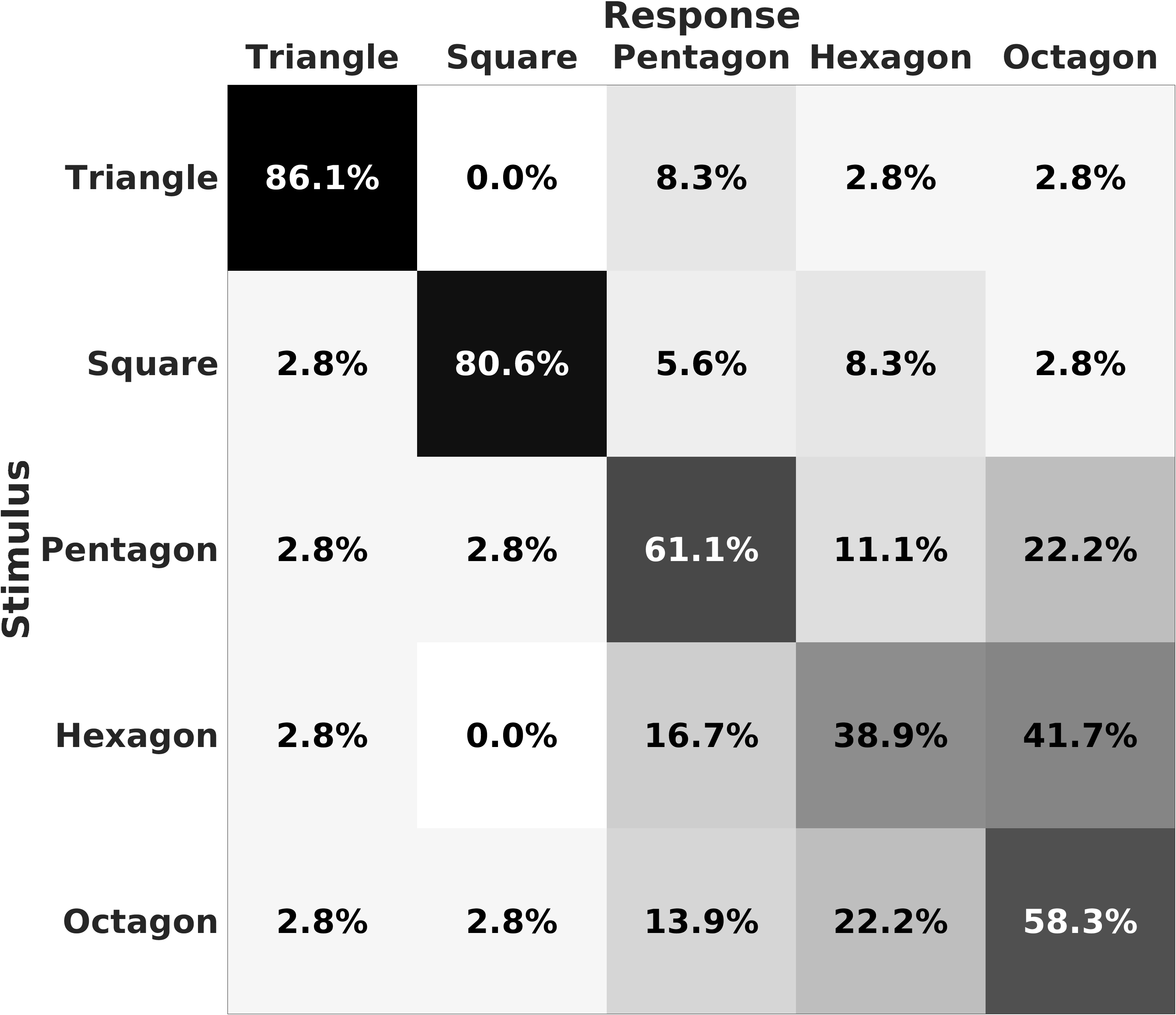}}\label{fig7b}}\hspace{5pt}
		\subfloat[]{%
			\resizebox*{5cm}{!}{\includegraphics{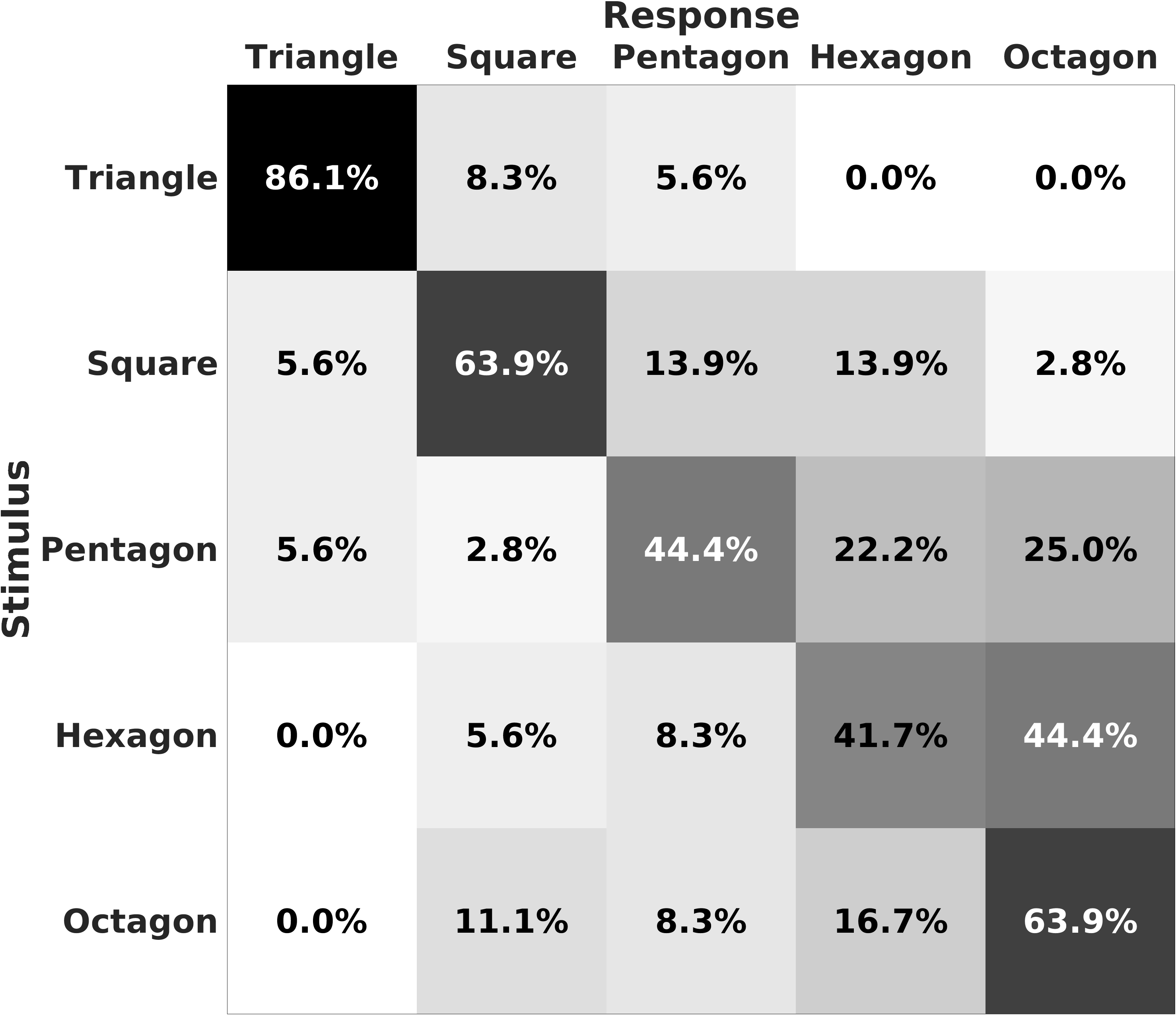}}\label{fig7c}}
		\caption{Confusion metrics of the recognition rate for (a) inside, (b) edge, and (c) outside rendering conditions.} \label{fig7}
	\end{figure}

	We constructed  confusion matrices to further investigate the recognition rate of the participants under each rendering condition (Figure \ref{fig7}).
	The confusion matrices showed that hexagon was mostly confused with octagon in all three rendering conditions; when octagon was displayed to the participants,  the confusion between hexagon and octagon was relatively low (11\% for \ac{c1}, 22\% for \ac{c2}, and 17\% for \ac{c3} rendering conditions). However, when hexagon was displayed as the stimuli, the confusion was increased by 22\%  for \ac{c1}, 20\% for  \ac{c2} and 28\% for \ac{c3} rendering conditions. Although we observed that participants identified octagon more easily than hexagon based on the confusion matrices, the mean difference between the recognition rate of hexagon and octagon was non-significant $(p = 0.815)$.

	\subsubsection{Spatial Analysis}
	
\begin{figure}[h!]
		\centering
		\subfloat[]{
			\resizebox*{3cm}{!}{\includegraphics{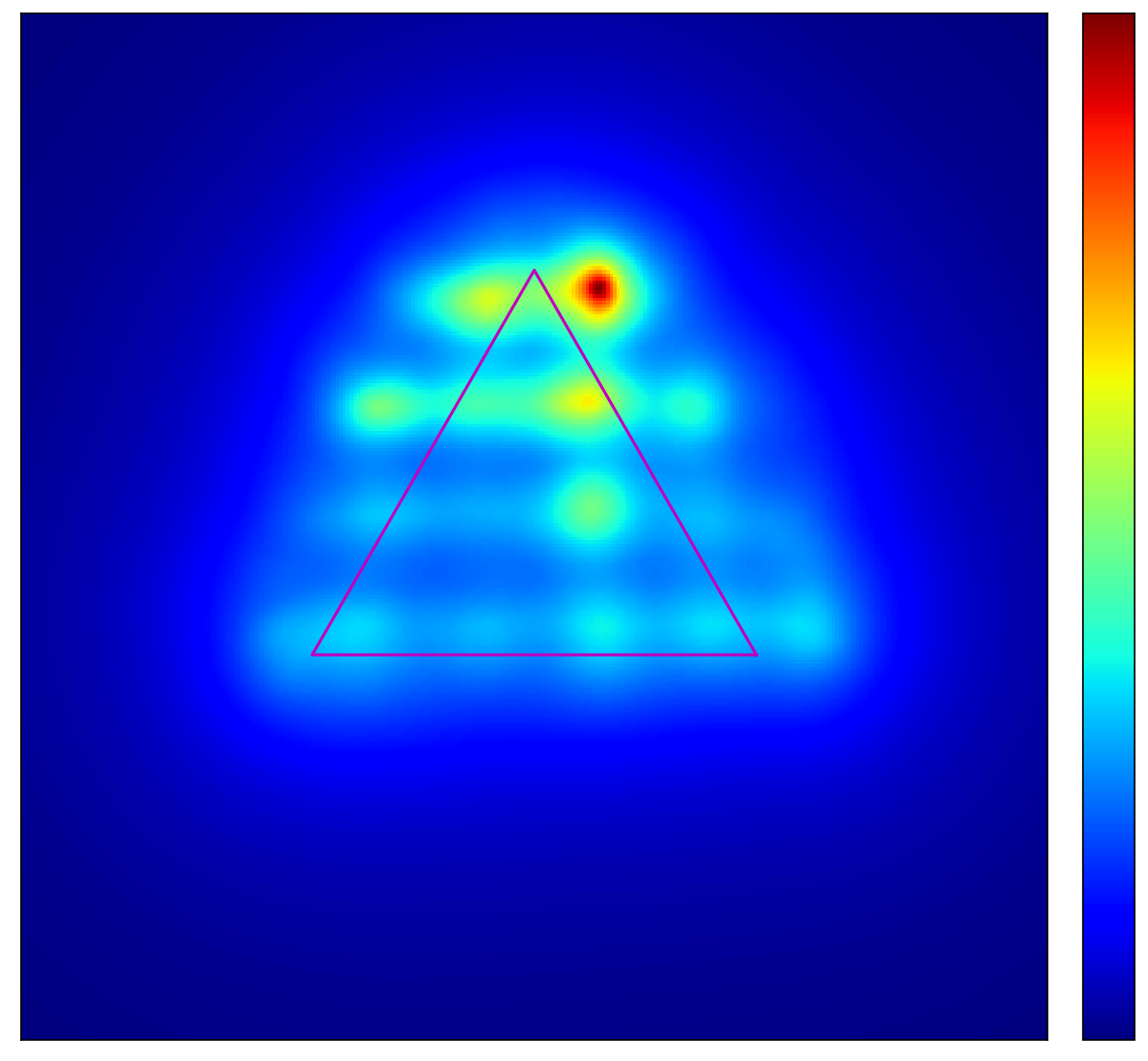}}
			\resizebox*{3cm}{!}{\includegraphics{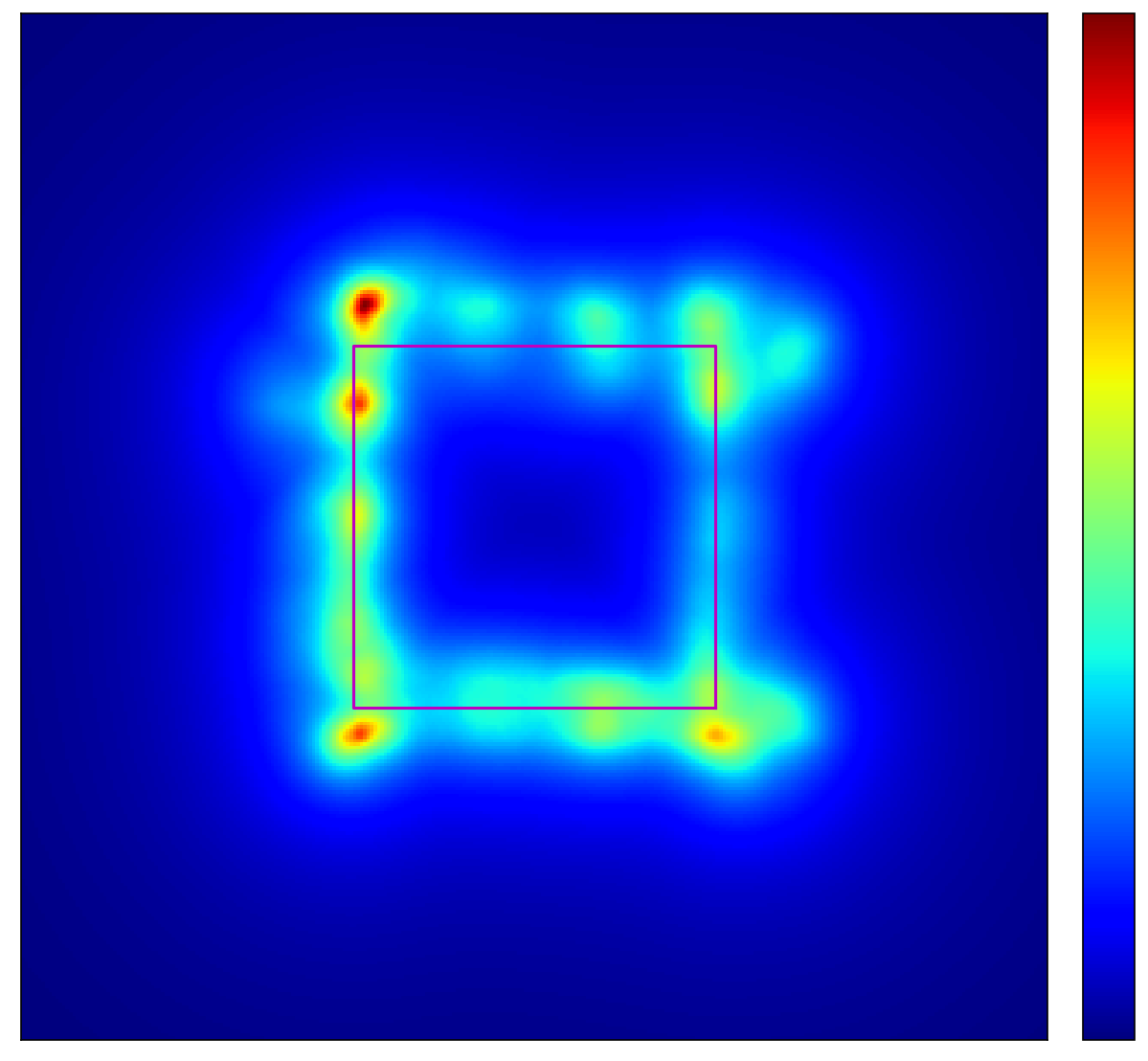}}
			\resizebox*{3cm}{!}{\includegraphics{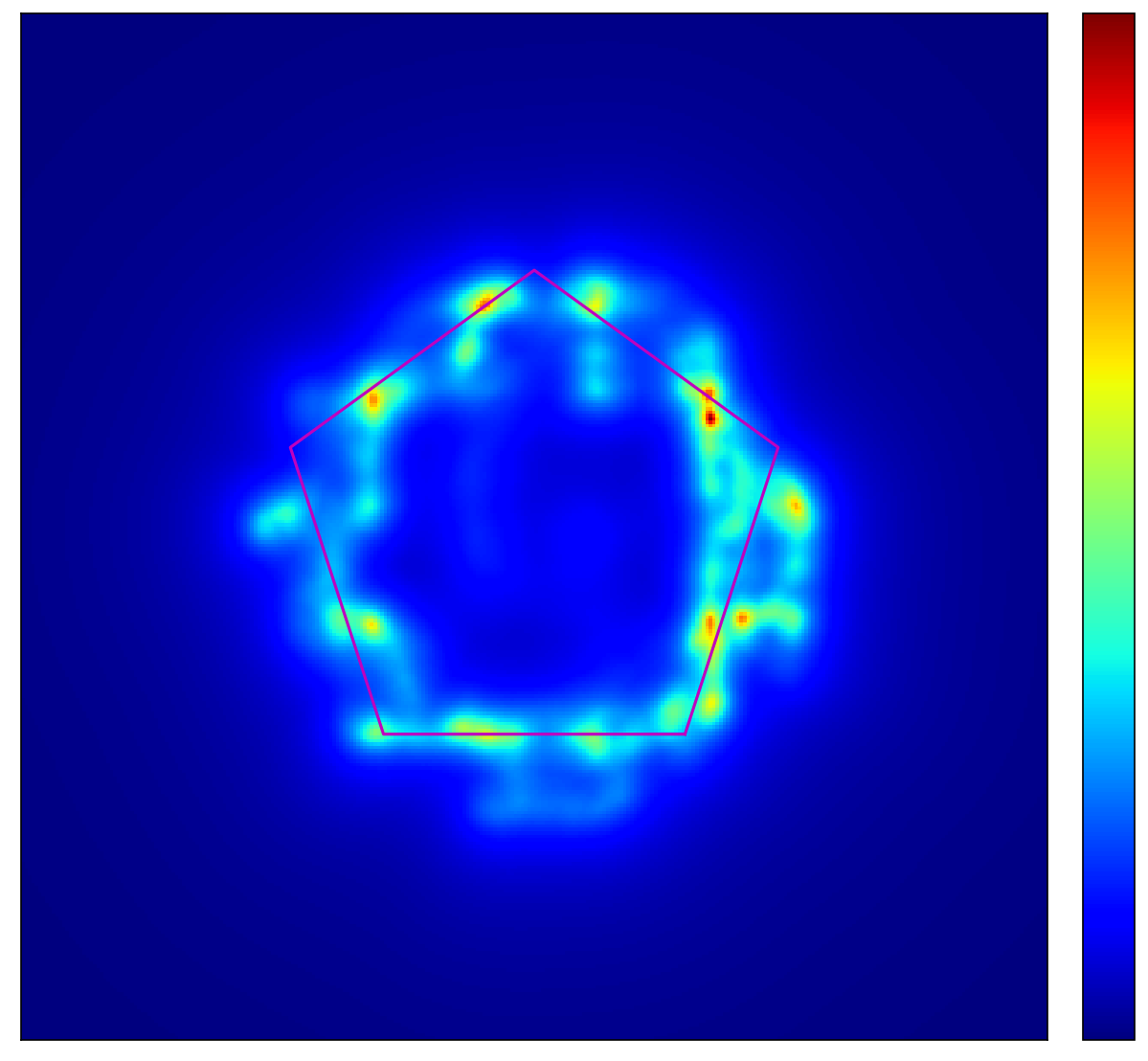}}
			\resizebox*{3cm}{!}{\includegraphics{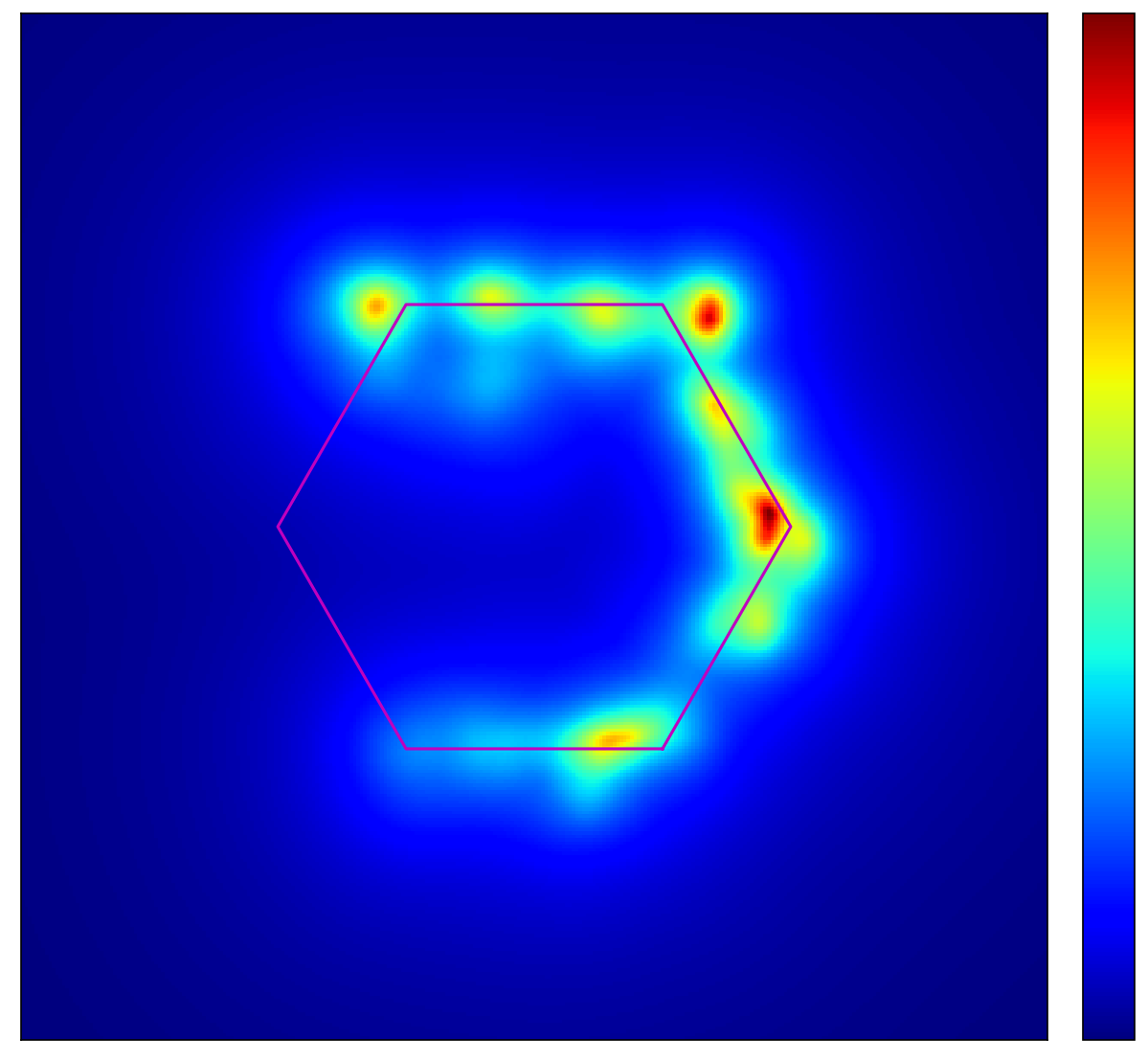}}
			\resizebox*{3cm}{!}{\includegraphics{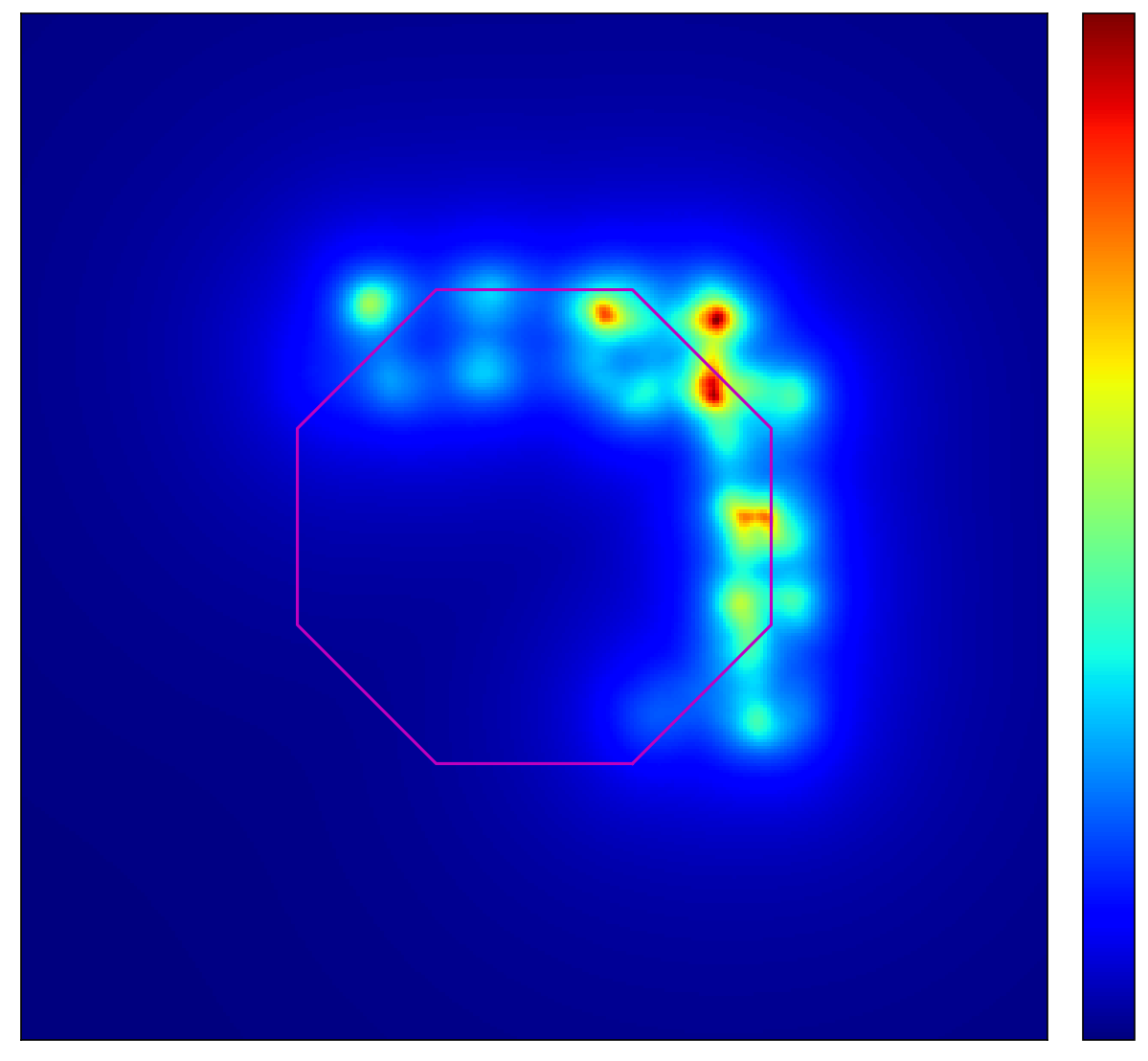}}\label{fig8a}}
		\newline
		\subfloat[]{%
			\resizebox*{3cm}{!}{\includegraphics{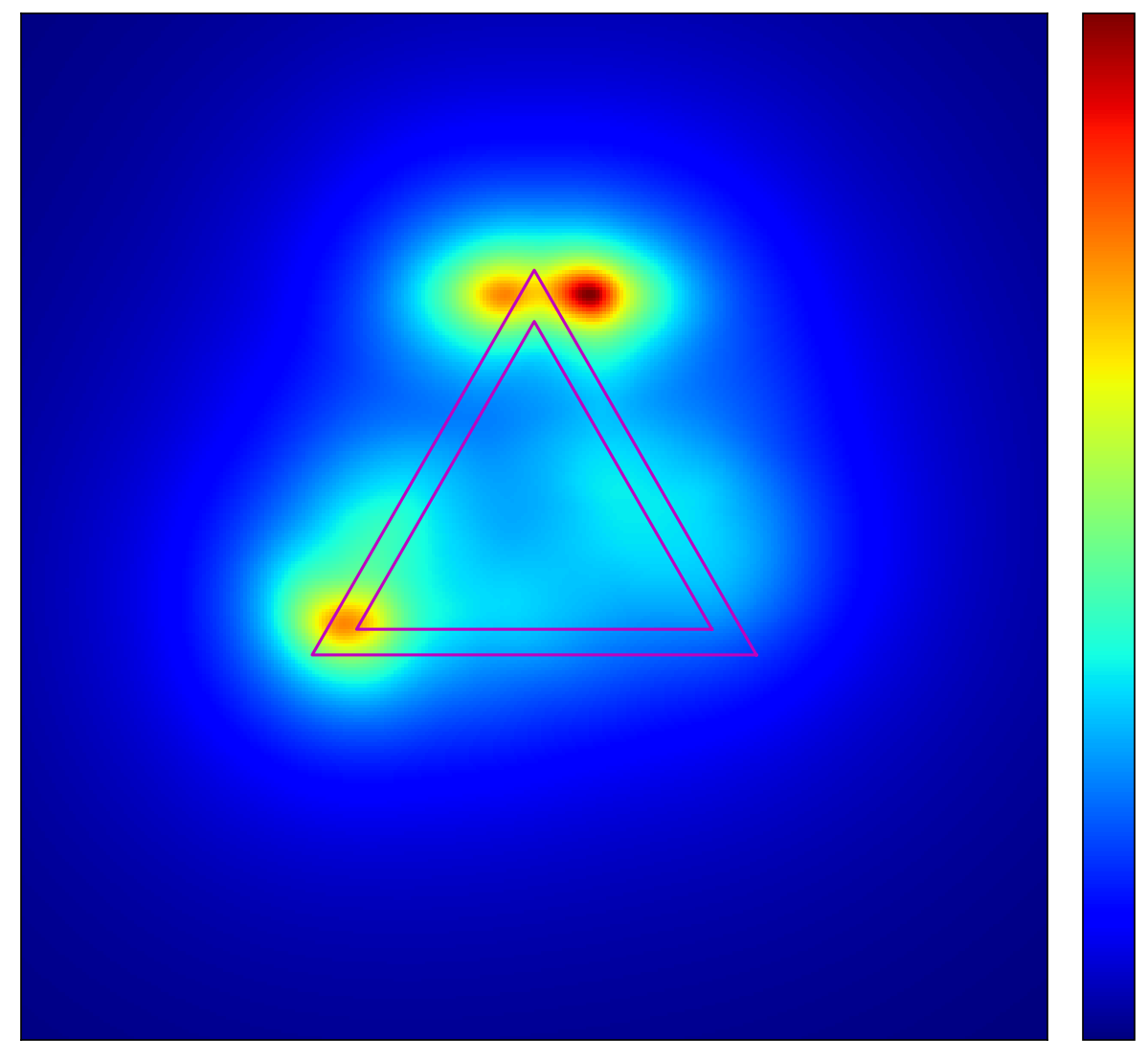}}
			\resizebox*{3cm}{!}{\includegraphics{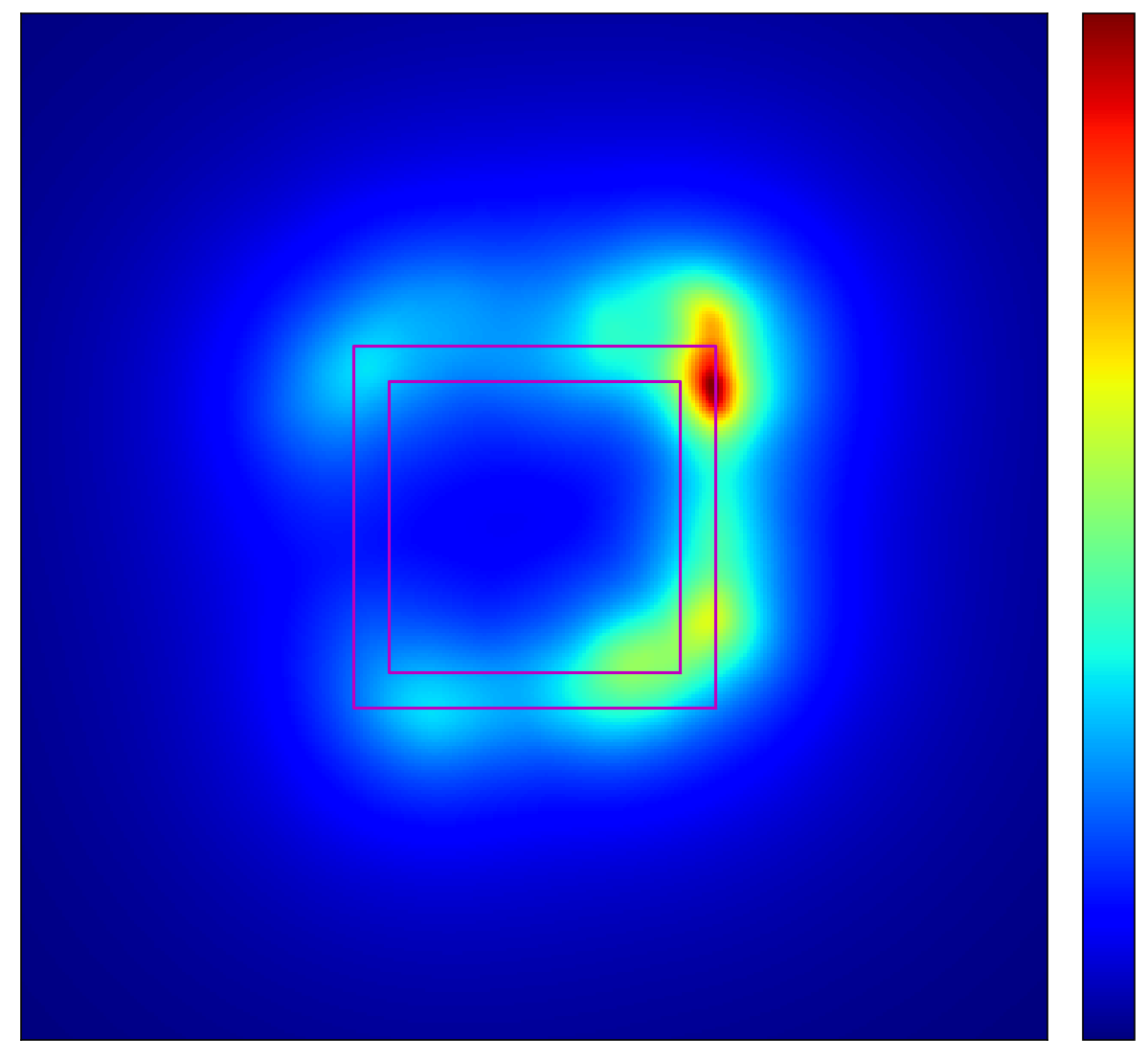}}
			\resizebox*{3cm}{!}{\includegraphics{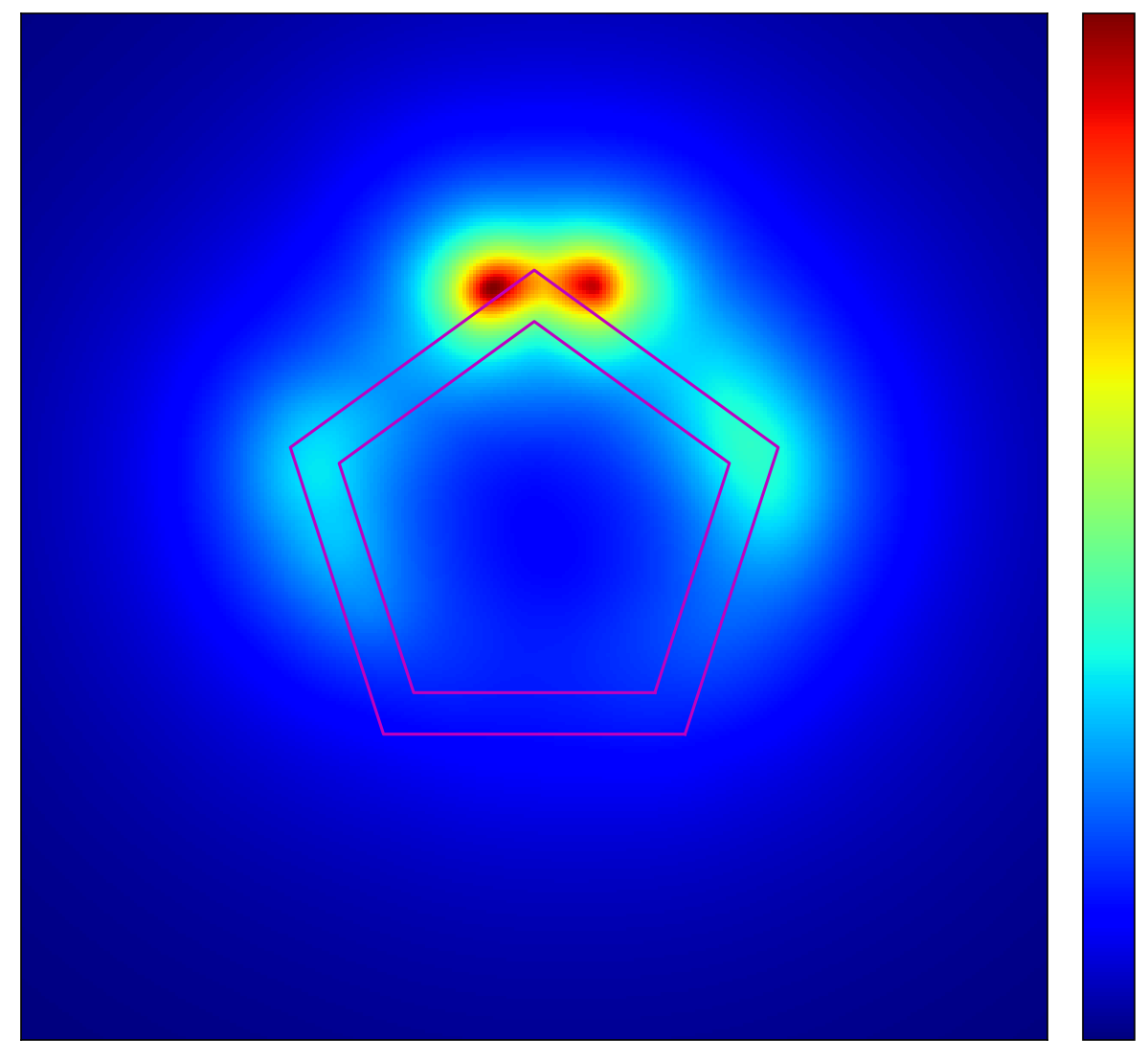}}
			\resizebox*{3cm}{!}{\includegraphics{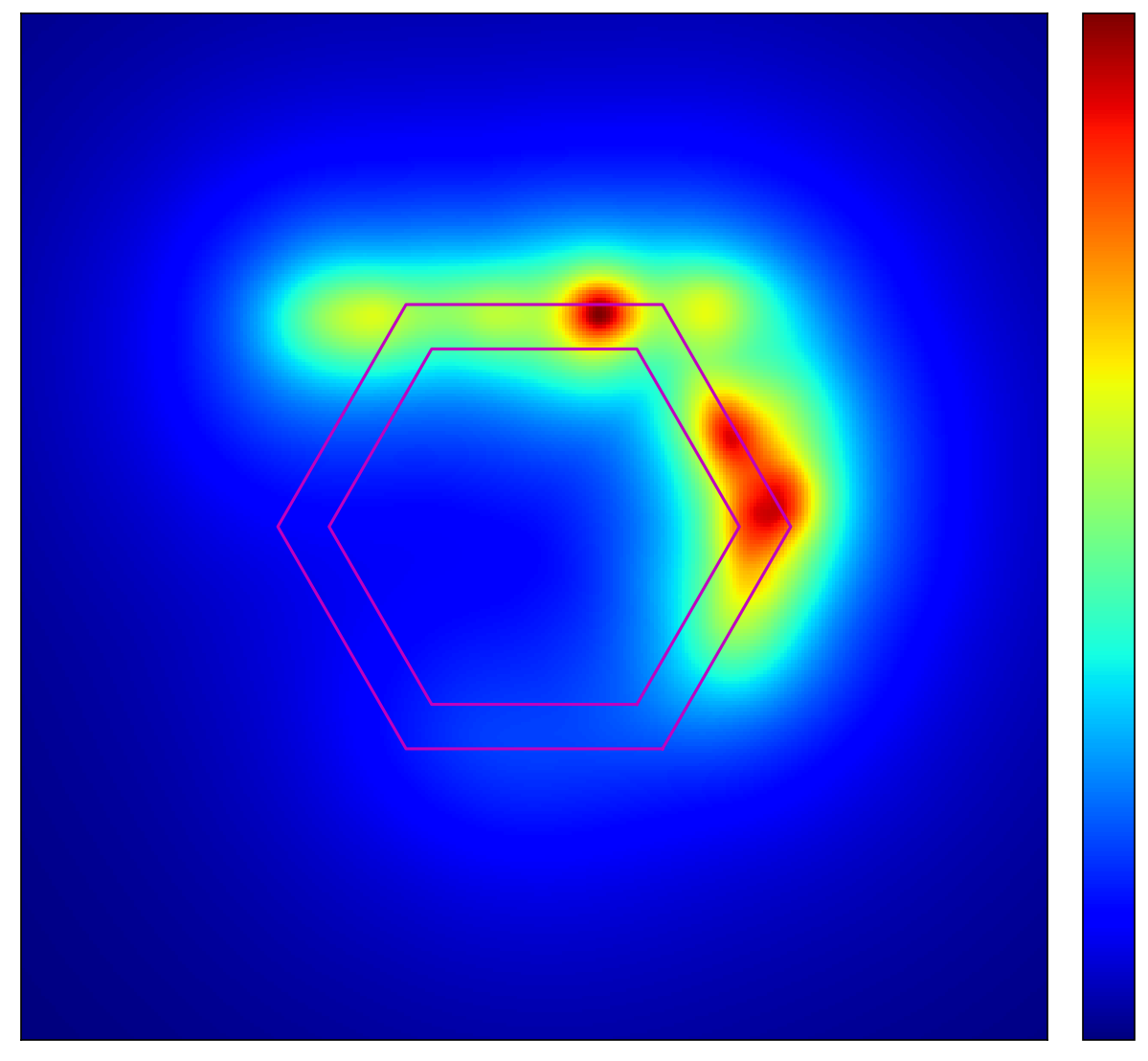}}
			\resizebox*{3cm}{!}{\includegraphics{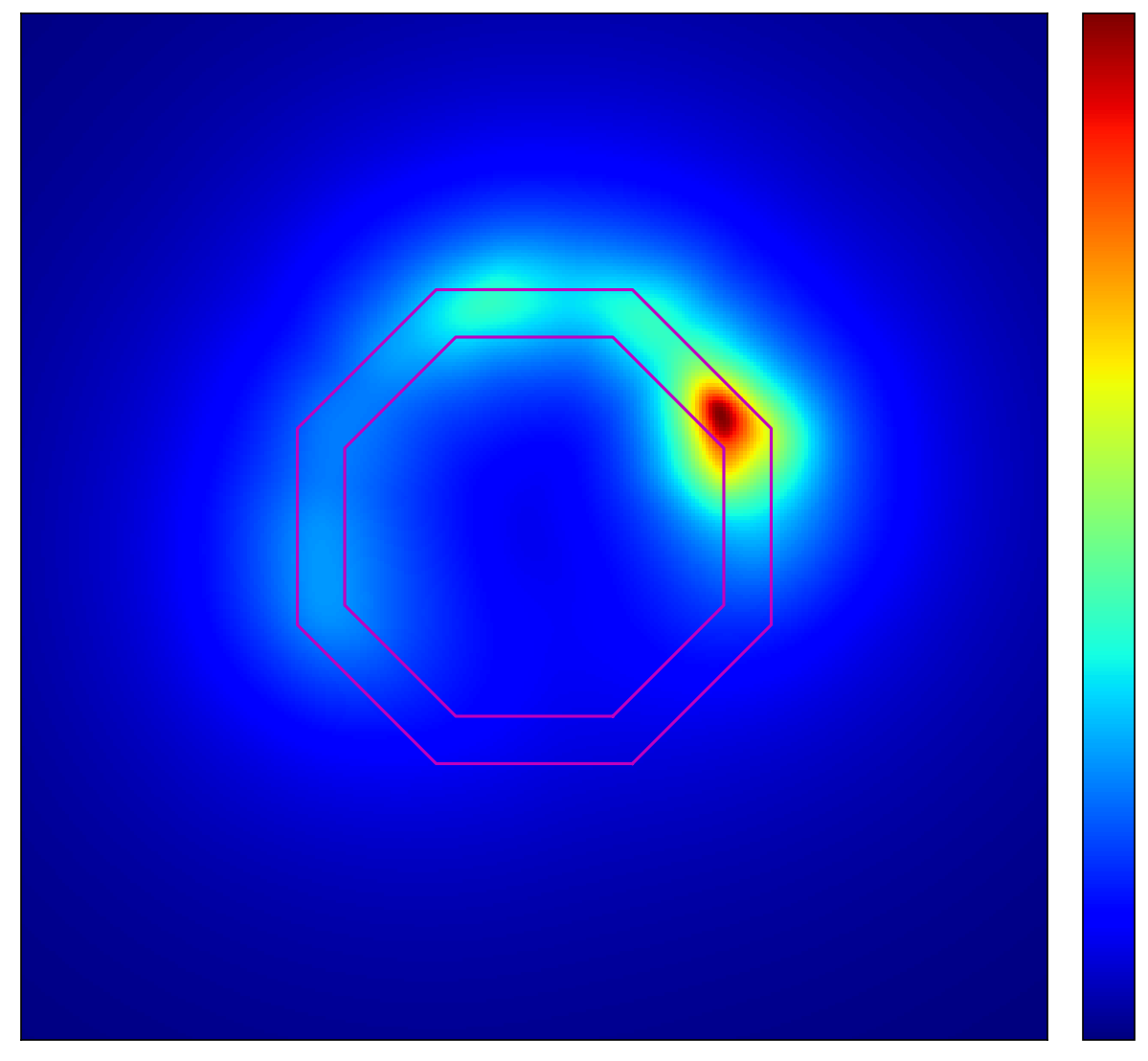}}\label{fig8b}}
		\newline
		\subfloat[]{%
			\resizebox*{3cm}{!}{\includegraphics{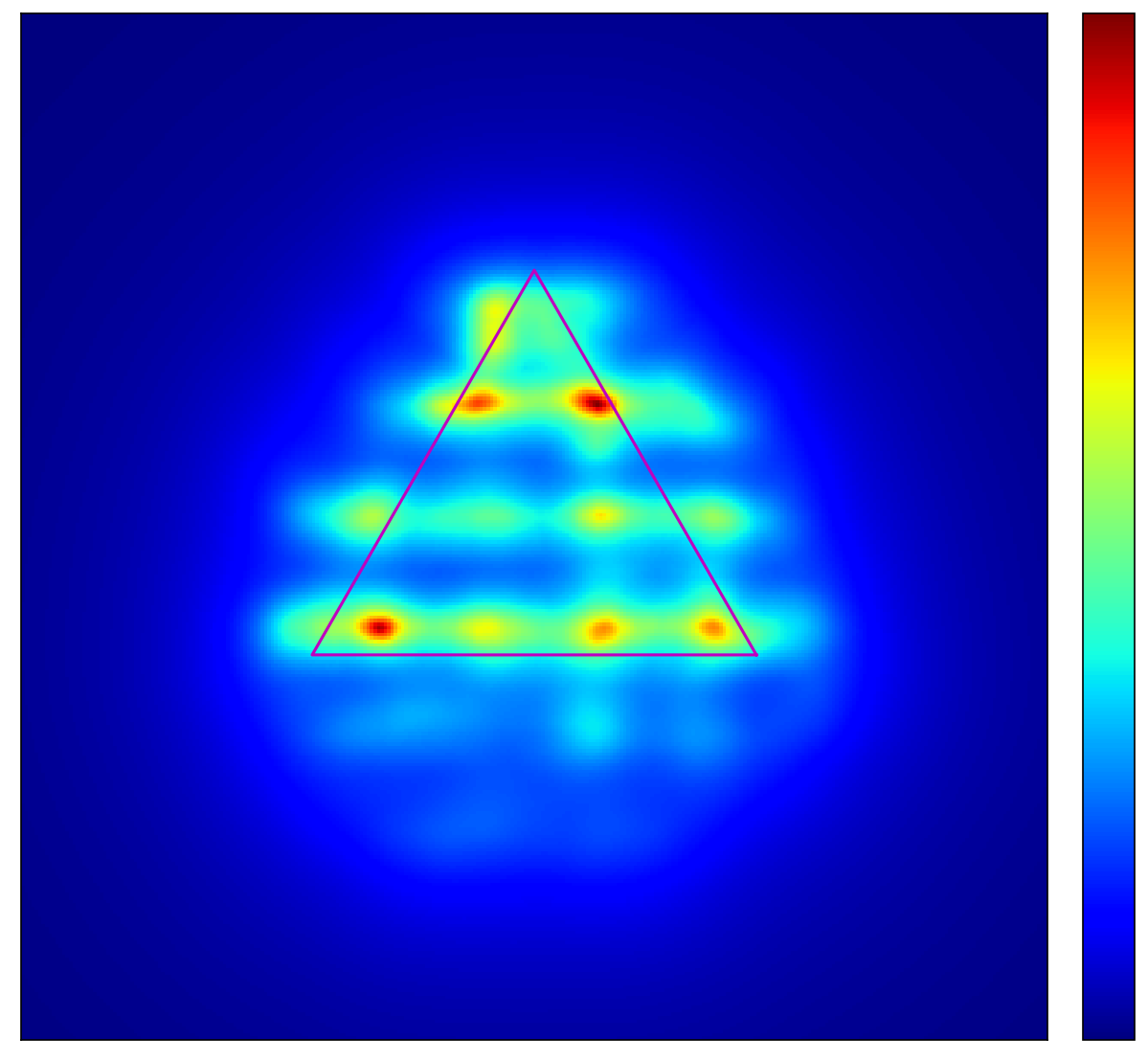}}
			\resizebox*{3cm}{!}{\includegraphics{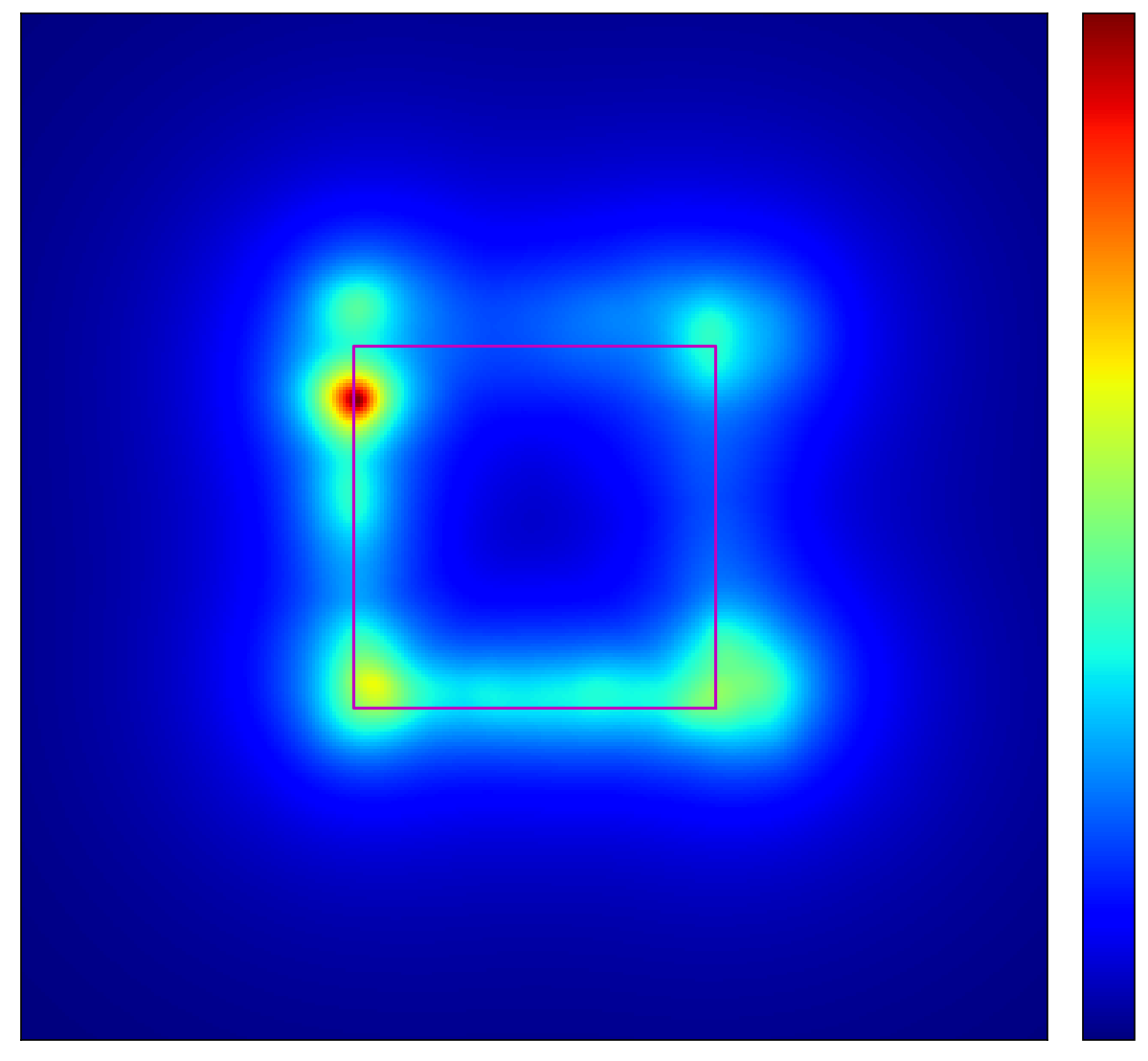}}
			\resizebox*{3cm}{!}{\includegraphics{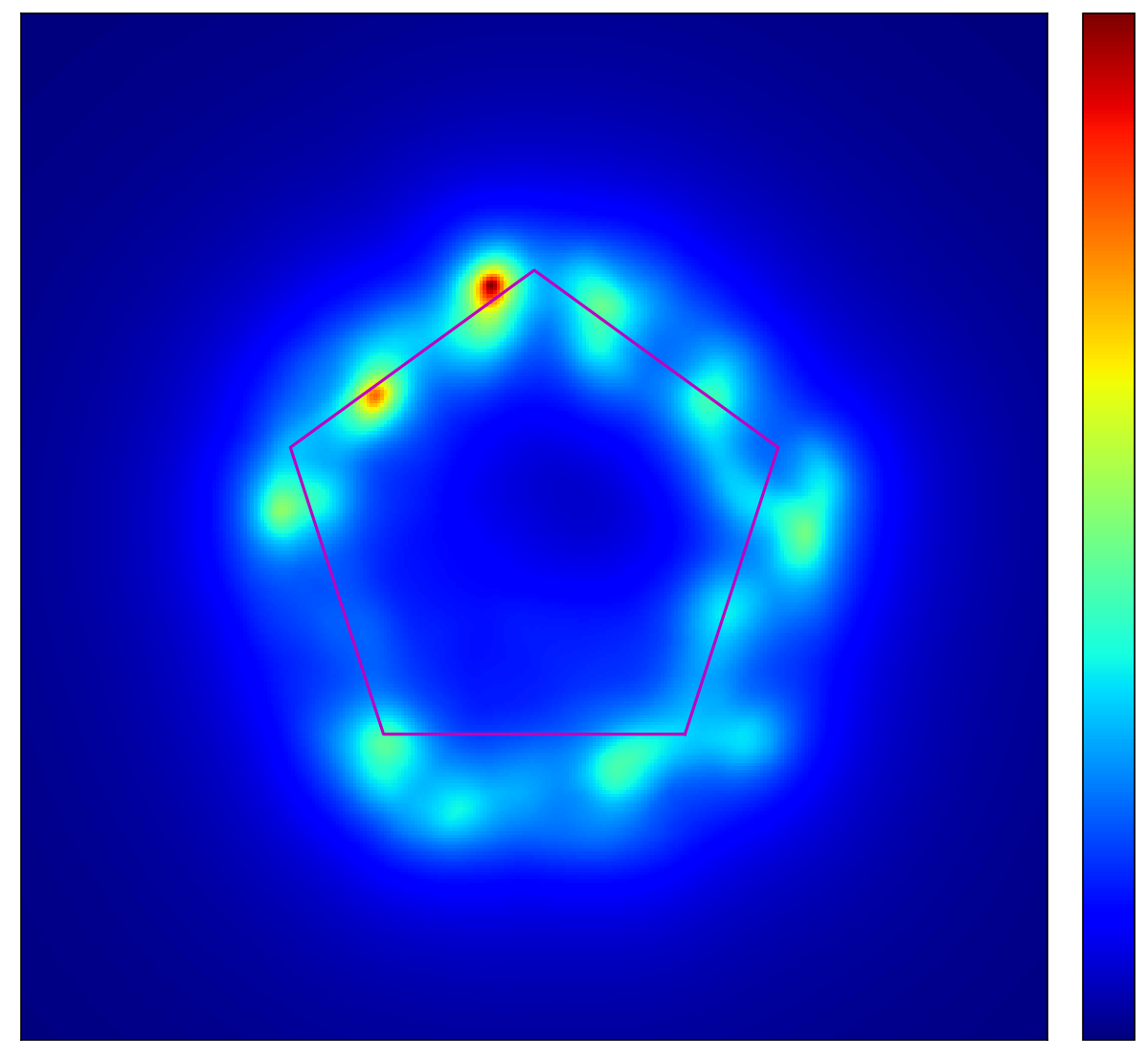}}
			\resizebox*{3cm}{!}{\includegraphics{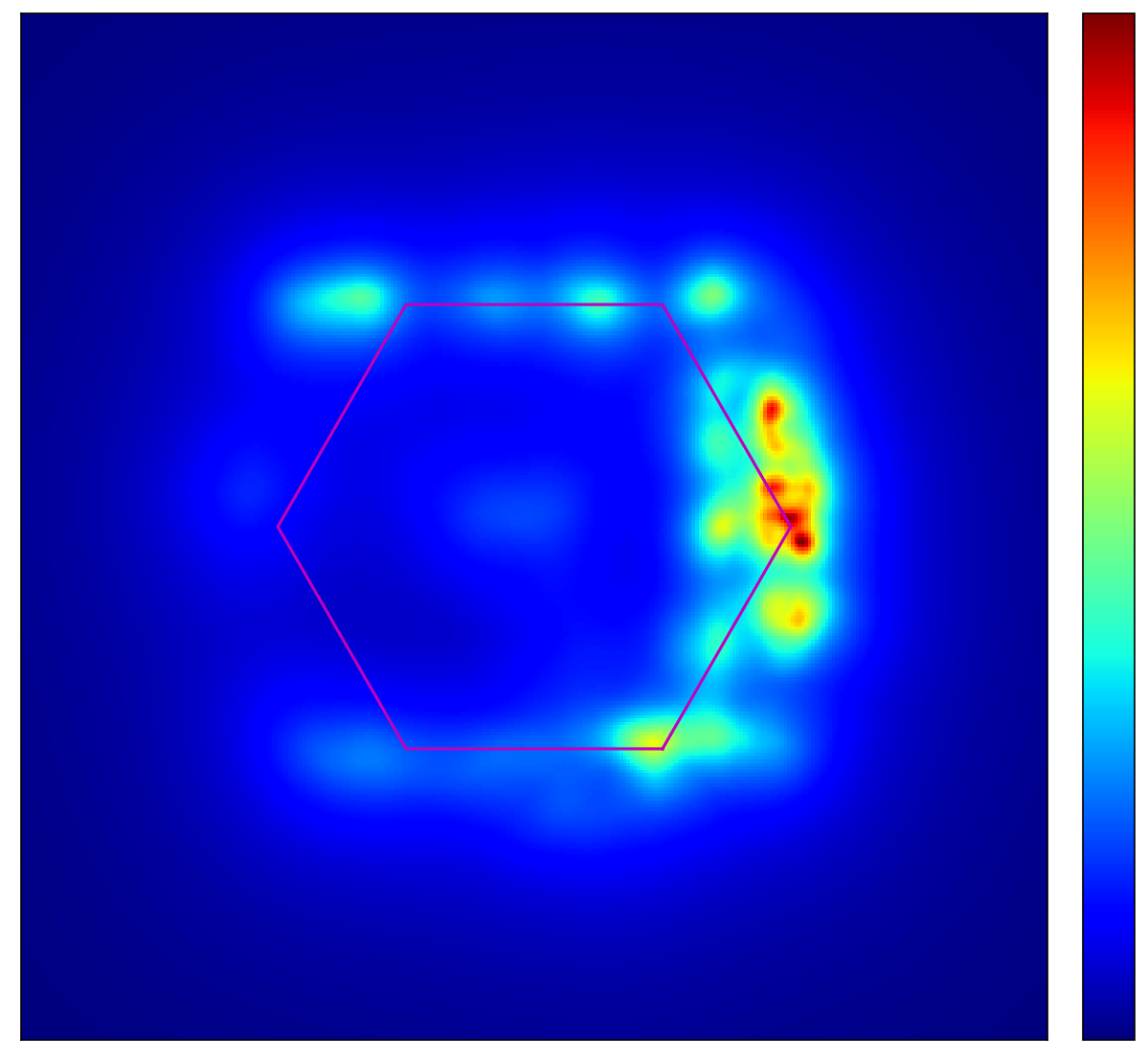}}
			\resizebox*{3cm}{!}{\includegraphics{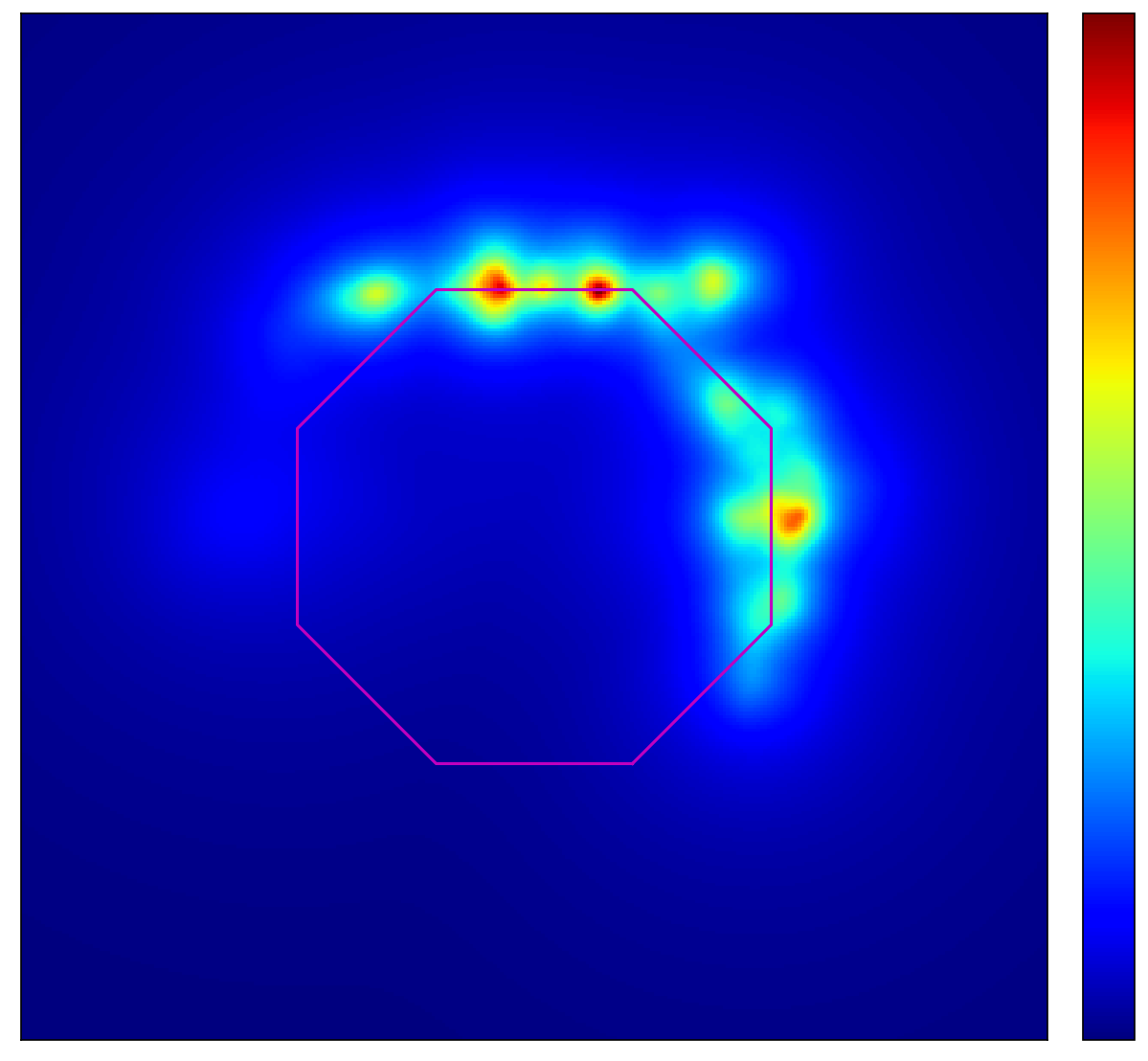}}\label{fig8c}}
		\caption{Frequency maps of randomly selected trials of participants (left to right: triangle, square, pentagon, hexagon, octagon) under three rendering conditions; (a) \ac{c1}, (b) \ac{c2}, and (c) \ac{c3}.} \label{fig8}
		
	\end{figure}

	We investigated the number of touches to the edges and corners of all shapes under all three rendering conditions. Two-way MANOVA shows a significant multivariate effect for rendering conditions $(F(2, 16)=32.522,p<0.001,\eta^2=0.803)$ and number of edges $(F(4, 32) = 142.561, p<0.001,\eta^2= 0.947)$. The interaction effect between the number of edges and rendering condition was also significant $(F(8, 64) = 4.820, p < 0.001, \eta^2 = 0.376)$. Multivariate group analysis revealed that \ac{c2} rendering condition was significantly different from the \ac{c1} and \ac{c3} rendering conditions for all shapes. Our post-hoc analyses showed that the number of touches to the corners were significantly higher for \ac{c2} rendering condition $(p_{(Edge-Inside)} < 0.001$ and $p_{(Edge-Outside)} = 0.001)$, suggesting that participants focused more on the corners in \ac{c2} rendering condition as observed from the frequency maps in Figure \ref{fig8b} and the number of touches to the edges were significantly higher for \ac{c1} and \ac{c3} rendering conditions $(p_{(Inside-Edge)} < 0.001$ and $p_{(Outside-Edge)} = 0.001)$, suggesting that they focused on the edges in those rendering conditions, as observed from Figures \ref{fig8a} and  \ref{fig8c}. 
	
	We also found that the number of active touches (i.e., the number of touches in the regions where the electrovibration was turned on) was significantly lower under the \ac{c2} rendering condition compared to the other two rendering conditions. Since there were more white spaces with no stimulus (i.e., inactive region) in \ac{c2} rendering condition, the participants spent more time identifying the shapes. These outcomes align with the results reported by \cite{gorlewicz2020design} for tactile shapes displayed by mechanical vibration.
	
	\subsubsection{Temporal Analysis}
\begin{figure}
		\centering
		\subfloat[]{%
			 {\includegraphics[width=0.68\textwidth]{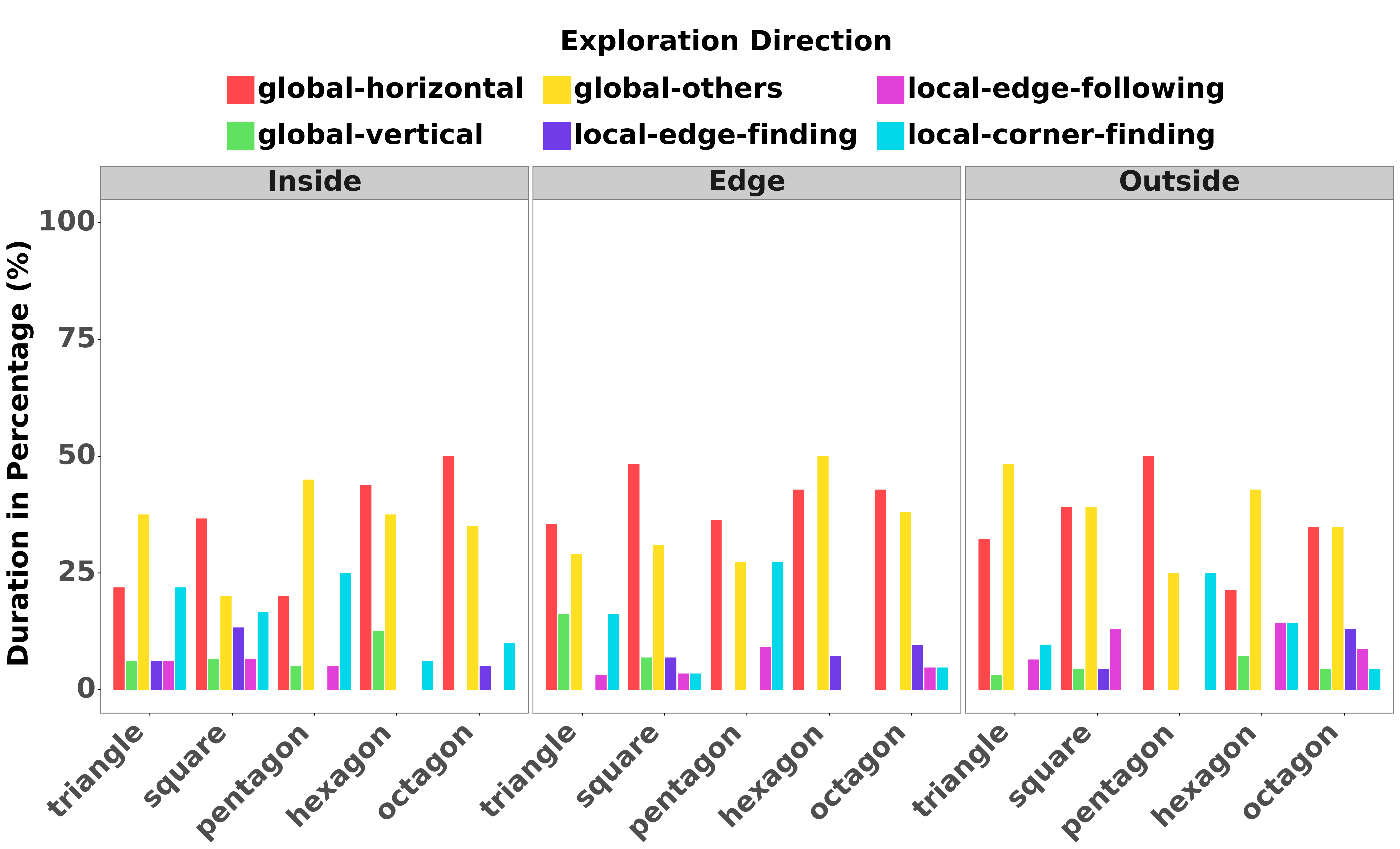}}\label{fig9a}}
			 
		\subfloat[]{%
			{\includegraphics[width=0.68\textwidth]{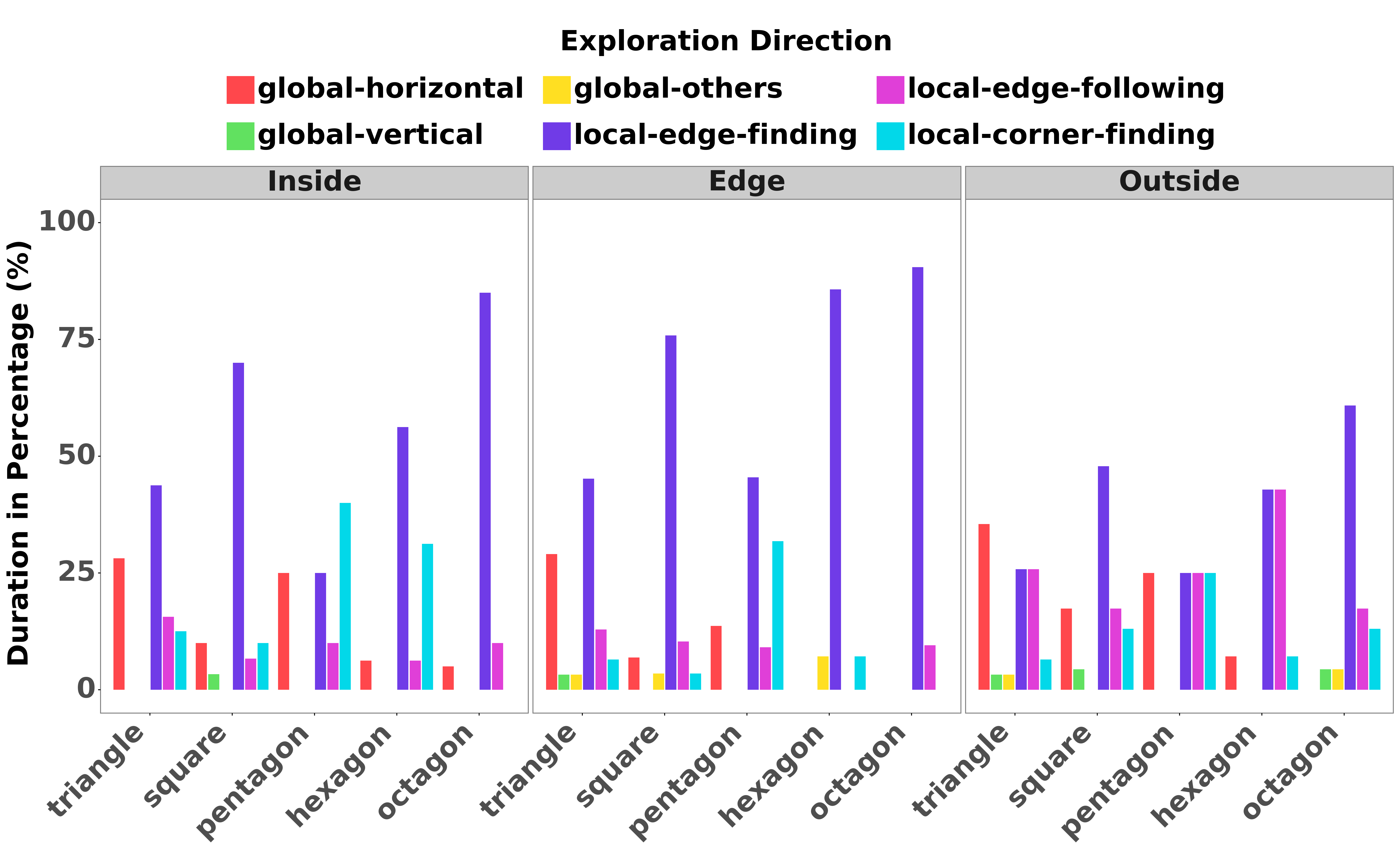}}\label{fig9b}}
			\newline
		\subfloat[]{%
			{\includegraphics[width=0.68\textwidth]{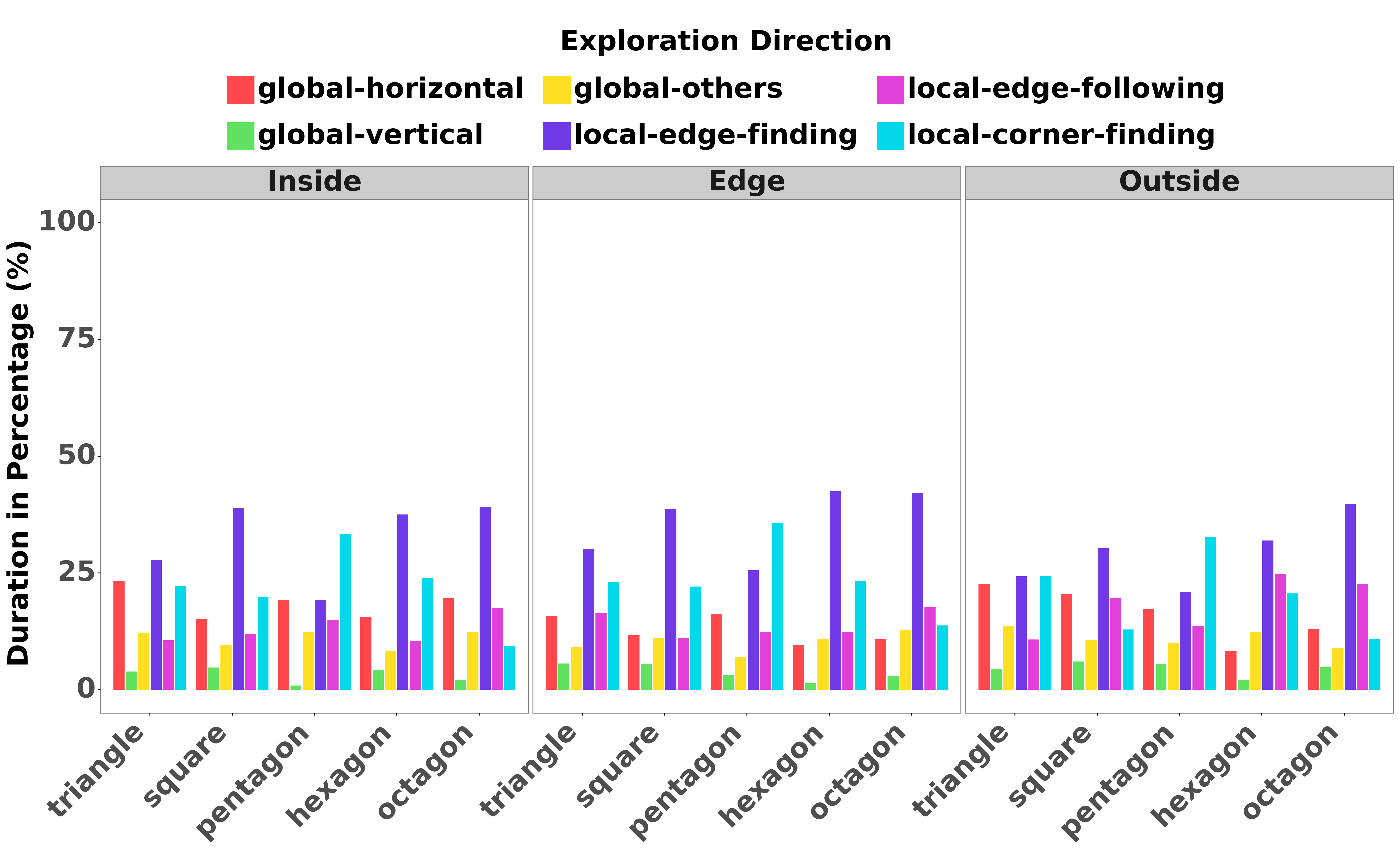}}\label{fig9c}}
			\vspace{-0.5cm}
		\caption{(a) Starting, (b) ending strategies, and (c) most/least explored strategies and directions for all participants.} \label{fig9}
	\end{figure}

	We investigated the exploration strategies used by the participants to examine the virtual shapes displayed on the touch surface. We reported the result of our analysis in terms of the starting, ending, most explored, and least explored strategies.
	\begin{itemize}
		\item The most common starting strategy adopted by the participants was global scanning with exploratory movement as horizontal (35\%) for all shapes and rendering conditions (see the red colored bars in Figure \ref{fig9a}).
		\item Most participants ended their exploration by local-edge-finding (55\%) (see the purple-colored bars in Figure \ref{fig9b}).
		\item The most explored strategy was local-edge-finding (32\%) followed by local-corner-finding (21\%) for all shapes and rendering conditions (see the purple and cyan colored bars in Figure \ref{fig9c}).
		\item The global scanning with exploratory direction vertical (3\%) was the least explored strategy for all shapes and rendering conditions (see the green colored bars in Figure \ref{fig9c}).
	\end{itemize}
	
\begin{figure}[h!]
		\centering
		\includegraphics[width=0.5\textwidth]{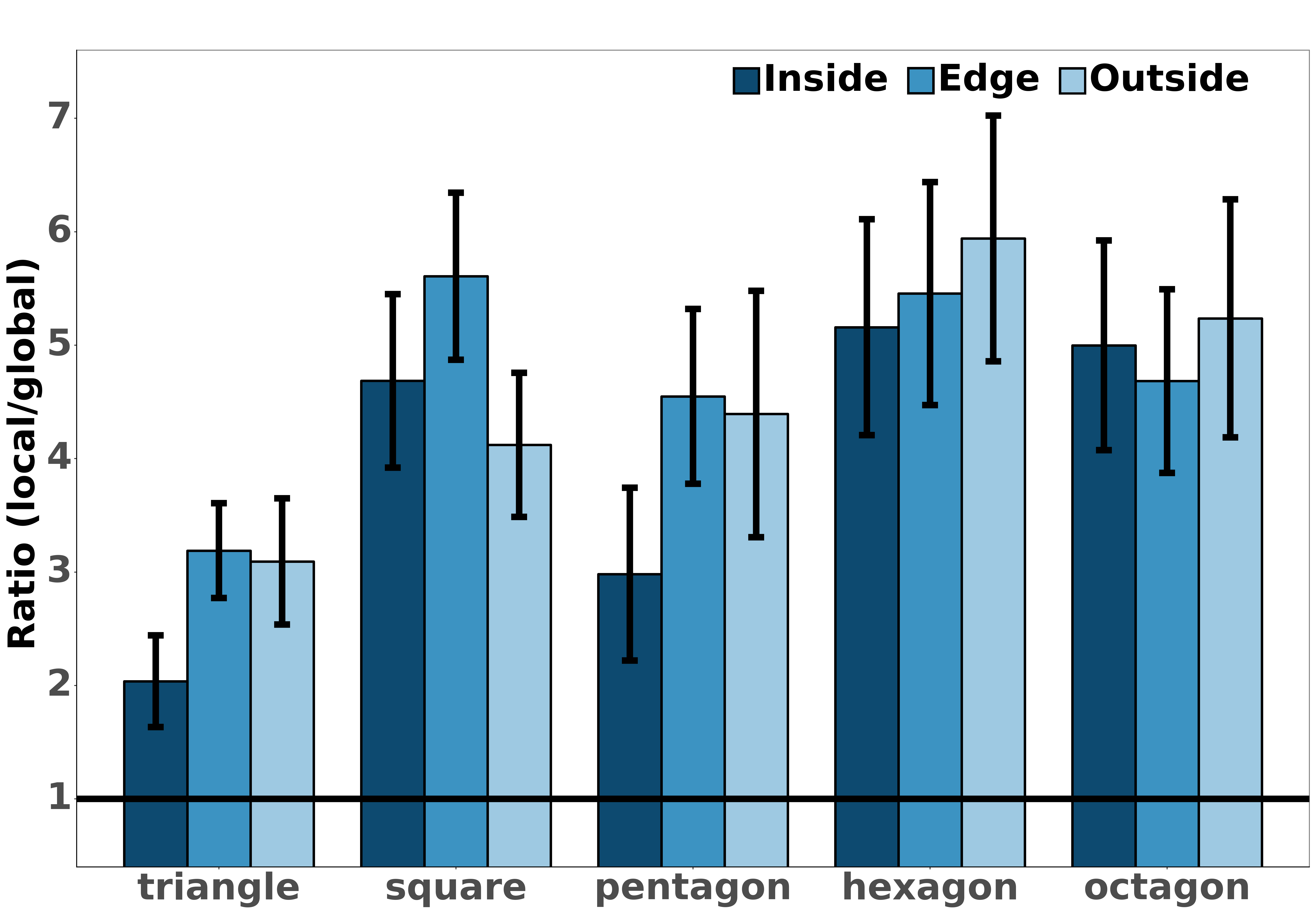}
		\caption{The ratio of local scanning time to the global scanning time with standard error of the means for all five shapes under three rendering conditions. The horizontal solid line shows the boundary when the ratio is equal to one.}\label{fig10}
	\end{figure}

	We observed that most participants followed a particular strategy while exploring the shapes. Initially, they explored the whole area inside the outer circle and then focused on the key features to identify the rendered shape. Overall, they spent less time in global scanning and more time in local scanning to identify each shape (Figure \ref{fig10}). As the shape gets complicated (i.e., the number of edges increases), the participants spent more time scanning the local features to identify the rendered shape correctly.

	\section{Experiment II}
	The results of our first experiment suggested a perceptual bias in recognition of the shapes as they were displayed to the participants in prototypical orientation. We hypothesized that this is the reason why participants achieved a higher recognition accuracy for octagon than hexagon. To test this hypothesis, we conducted a second experiment with the same shapes but applied CW and CCW angular perturbations to change their orientation (with respect to the prototypical orientation) in order to reduce the perceptual bias (see Figure \ref{fig11}). We selected \ac{c1} rendering condition for the second experiment since the recognition time for \ac{c1} rendering condition was significantly shorter than the other two rendering conditions.

	\subsection{Participants}
	
	Nine different participants (5 male, 4 female with an average age of 26 $\pm$ 6.7 years) 
	participated in the experiment. All of them were graduate students. The participants read and signed a consent form before the experiment, which was approved by the Ethical Committee for Human Participants of Ko\c{c} University.
	\subsection{Stimuli}
\begin{figure}
		\centering
		\subfloat[]{
			\resizebox*{2cm}{!}{\includegraphics{Figures/triangle.png}}
			\resizebox*{2cm}{!}{\includegraphics{Figures/square.png}}
			\resizebox*{2cm}{!}{\includegraphics{Figures/pentagon.png}}
			\resizebox*{2cm}{!}{\includegraphics{Figures/hexagon.png}}
			\resizebox*{2cm}{!}{\includegraphics{Figures/octagon.png}}\label{fig11a}}
		\newline
		\subfloat[]{
			\resizebox*{2cm}{!}{\includegraphics{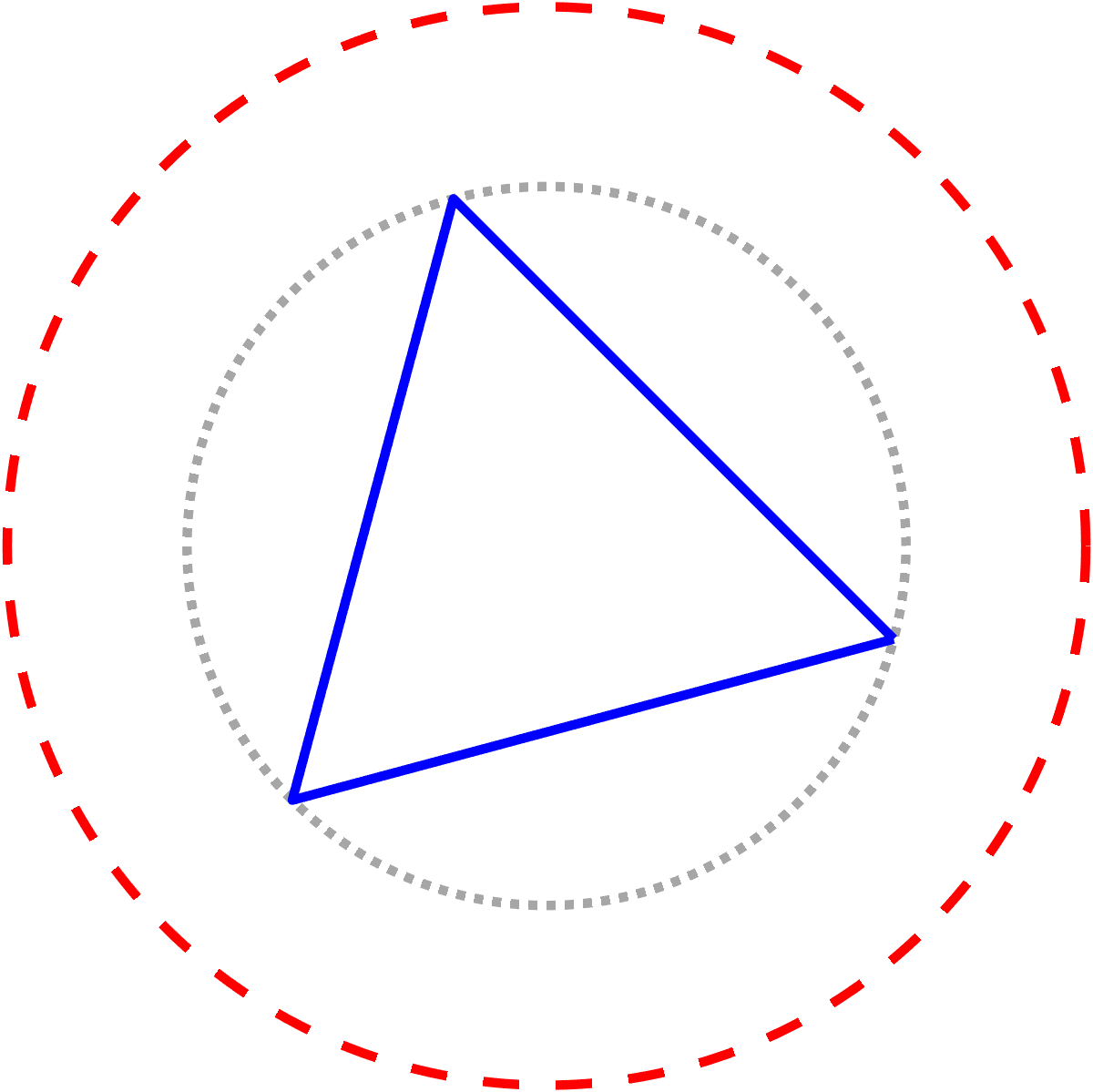}}
			\resizebox*{2cm}{!}{\includegraphics{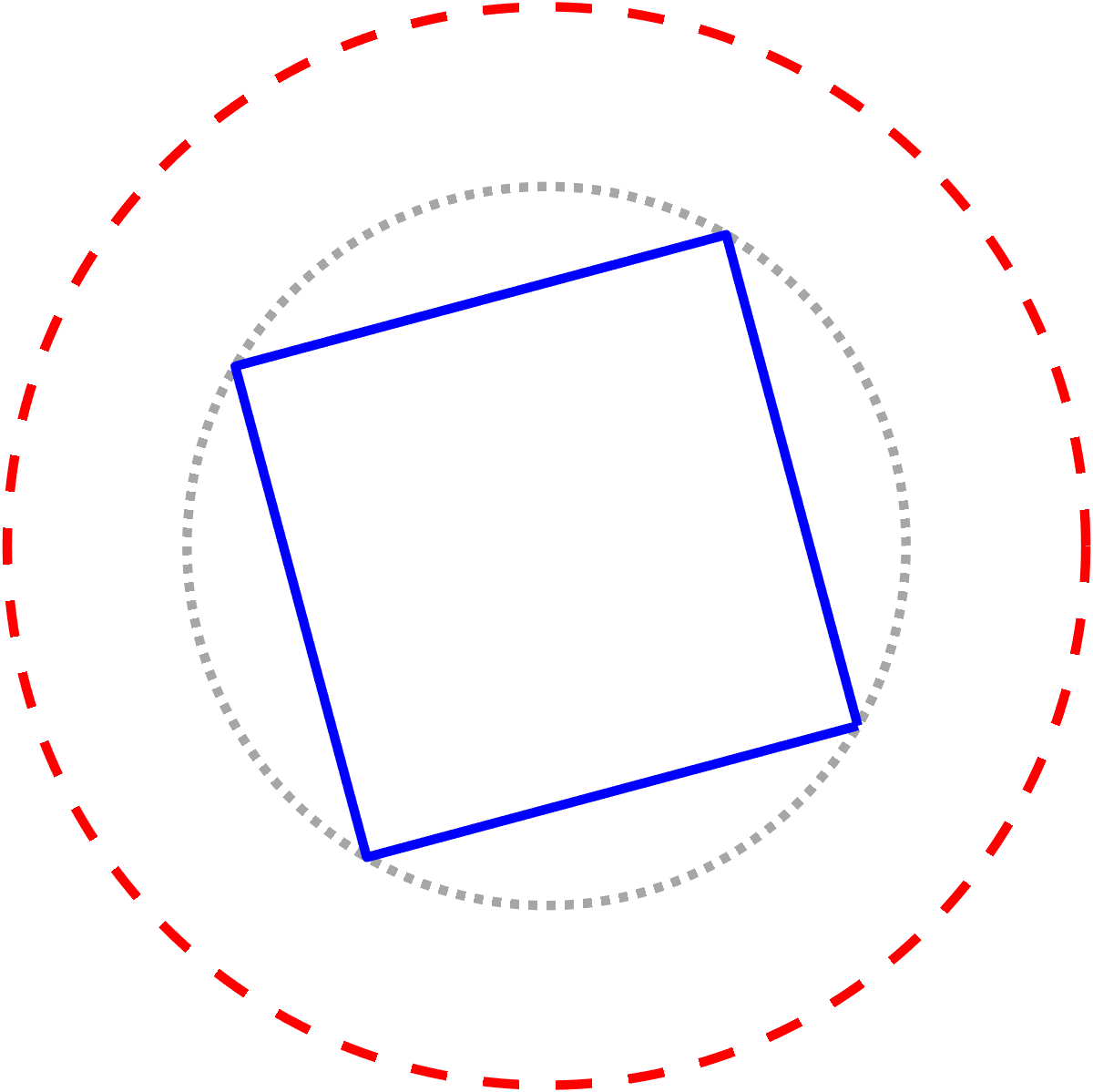}}
			\resizebox*{2cm}{!}{\includegraphics{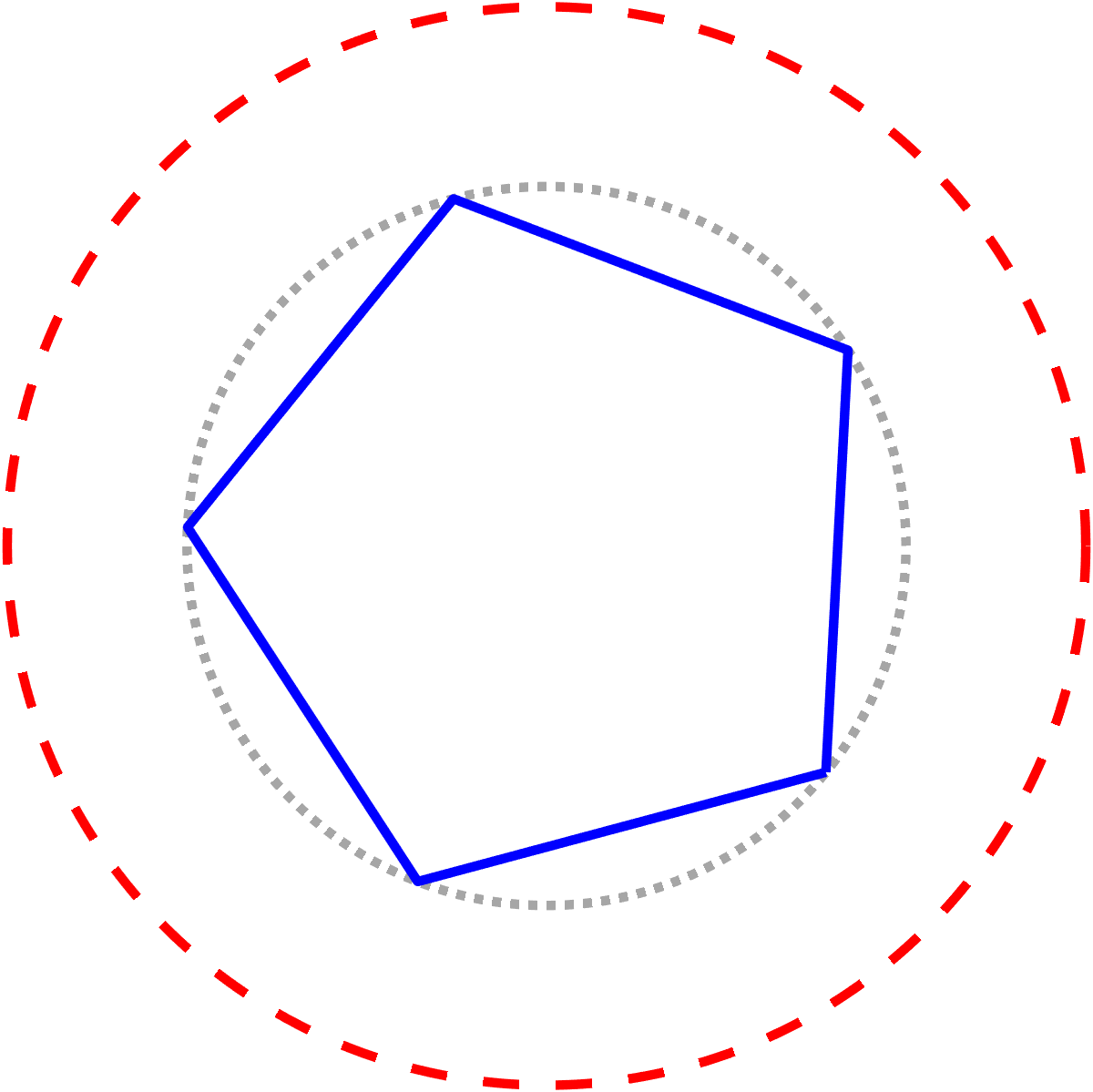}}
			\resizebox*{2cm}{!}{\includegraphics{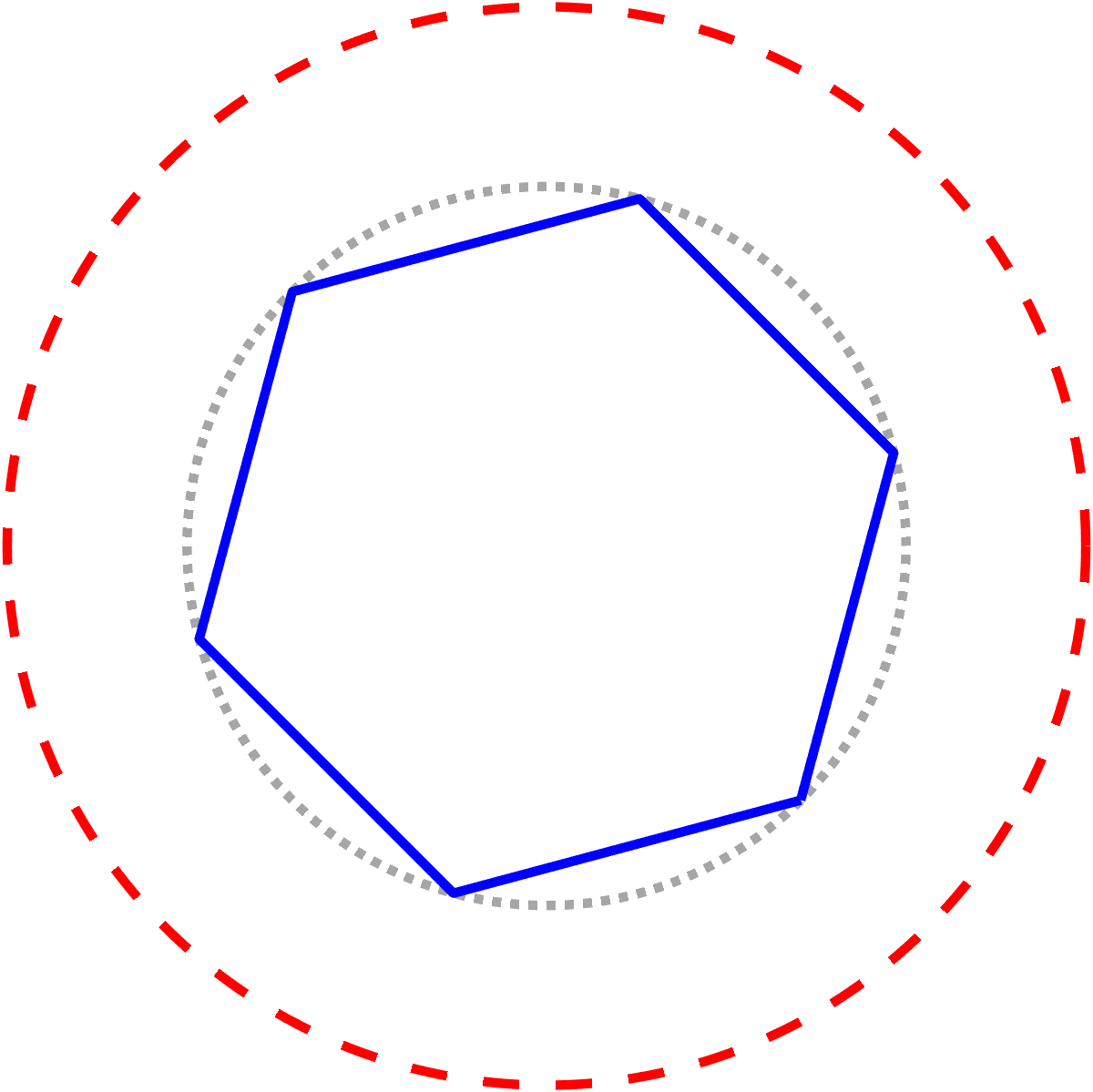}}
			\resizebox*{2cm}{!}{\includegraphics{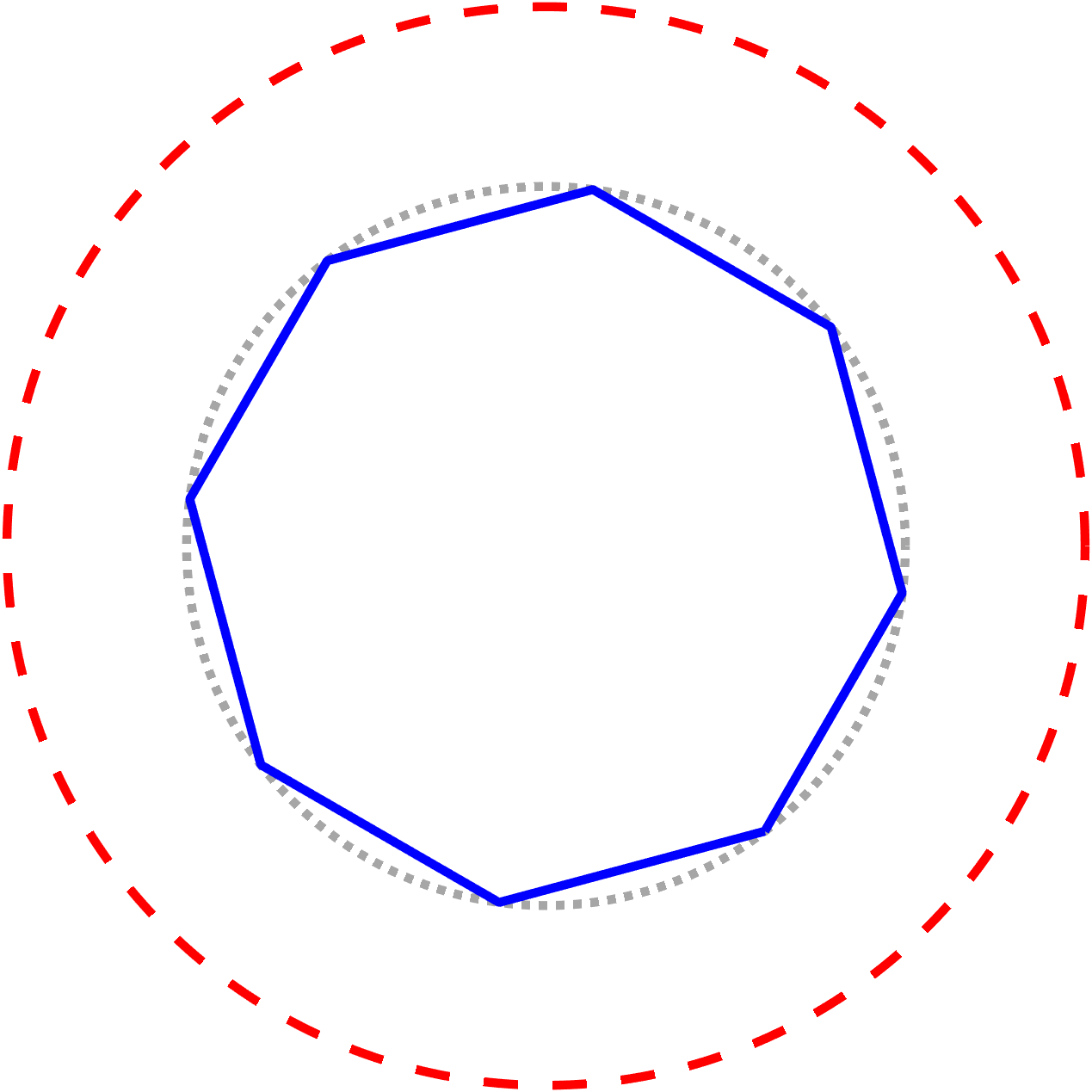}}\label{fig11b}}
		\newline
		\subfloat[]{
			\resizebox*{2cm}{!}{\includegraphics{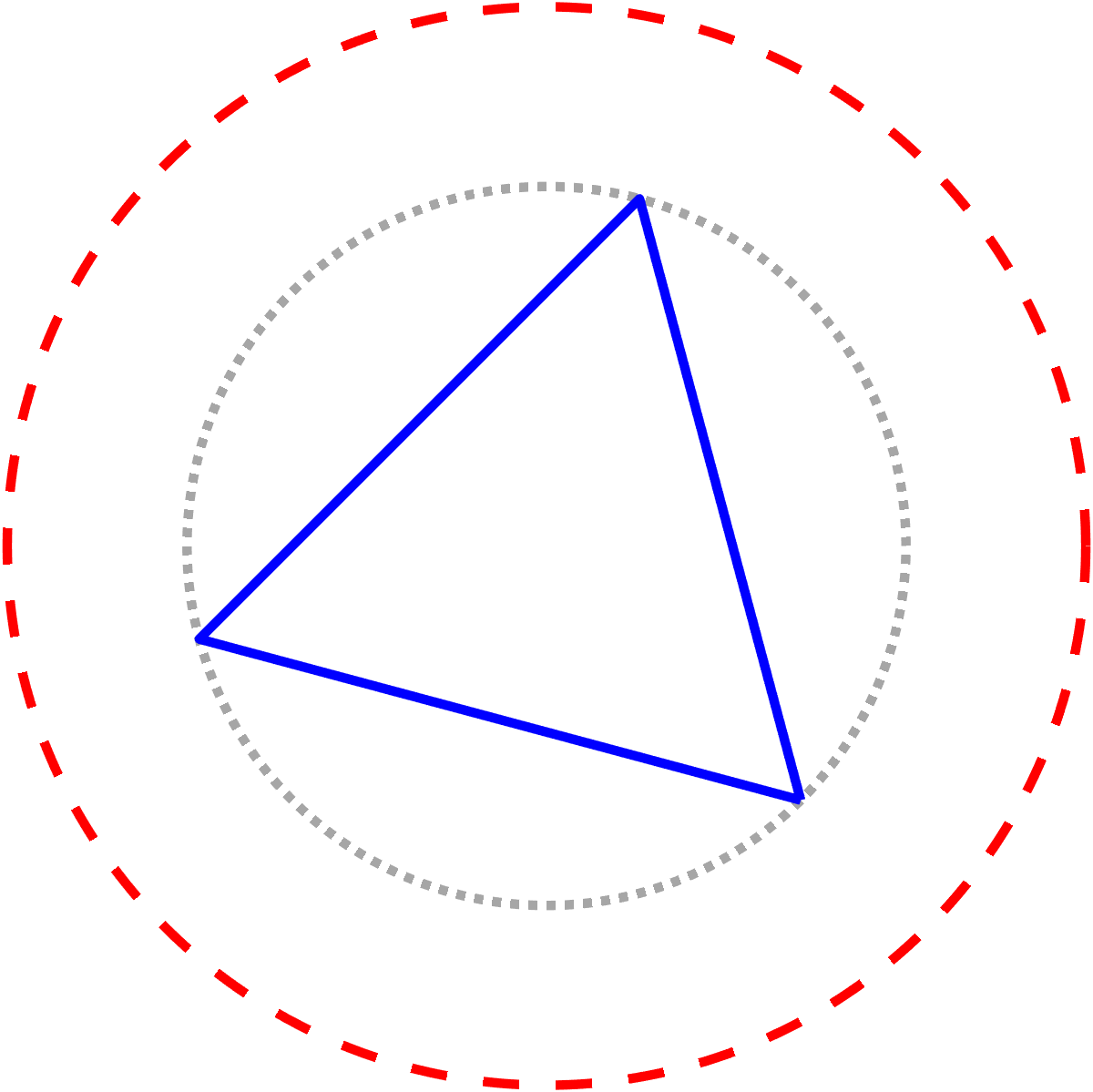}}
			\resizebox*{2cm}{!}{\includegraphics{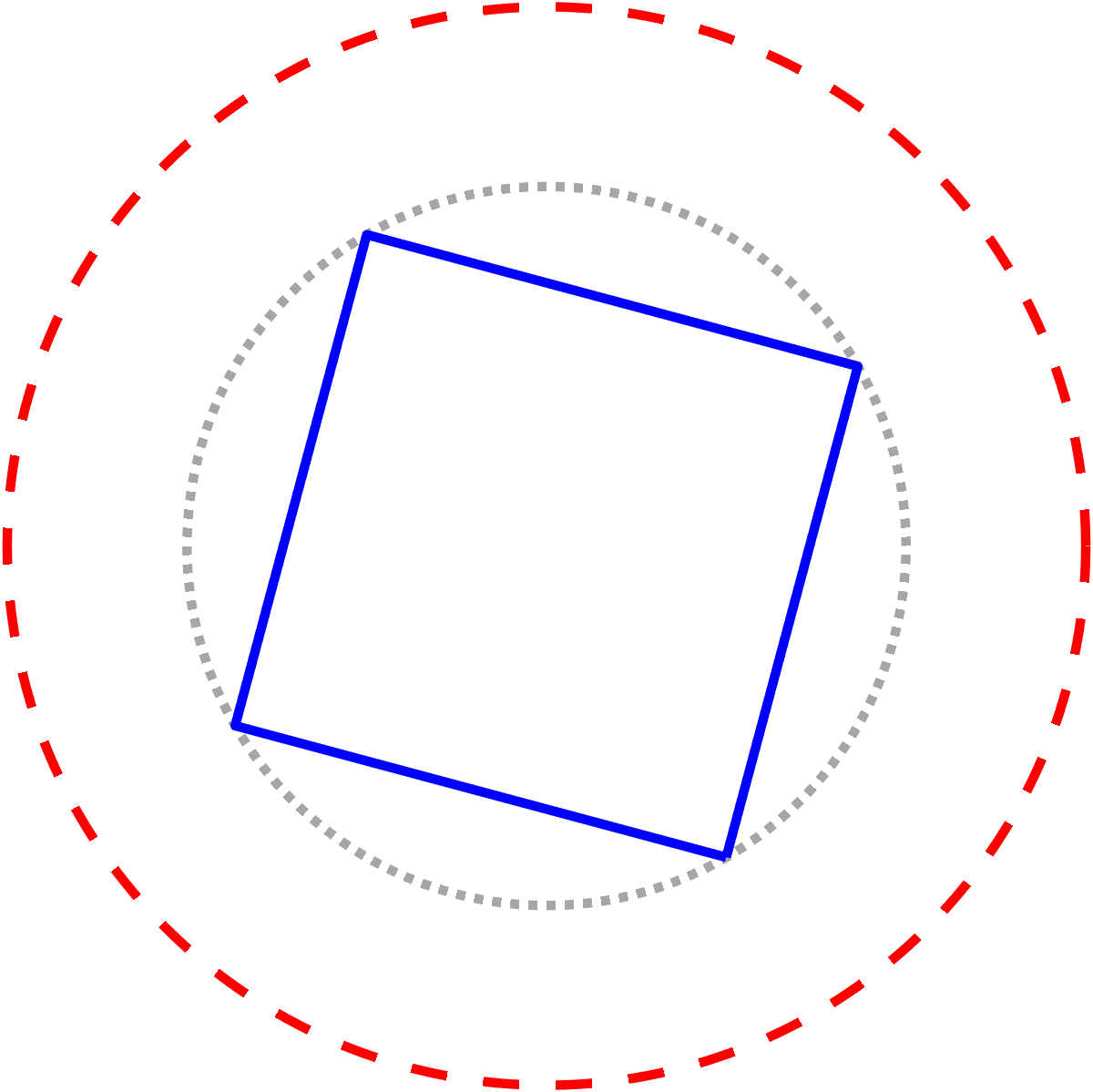}}
			\resizebox*{2cm}{!}{\includegraphics{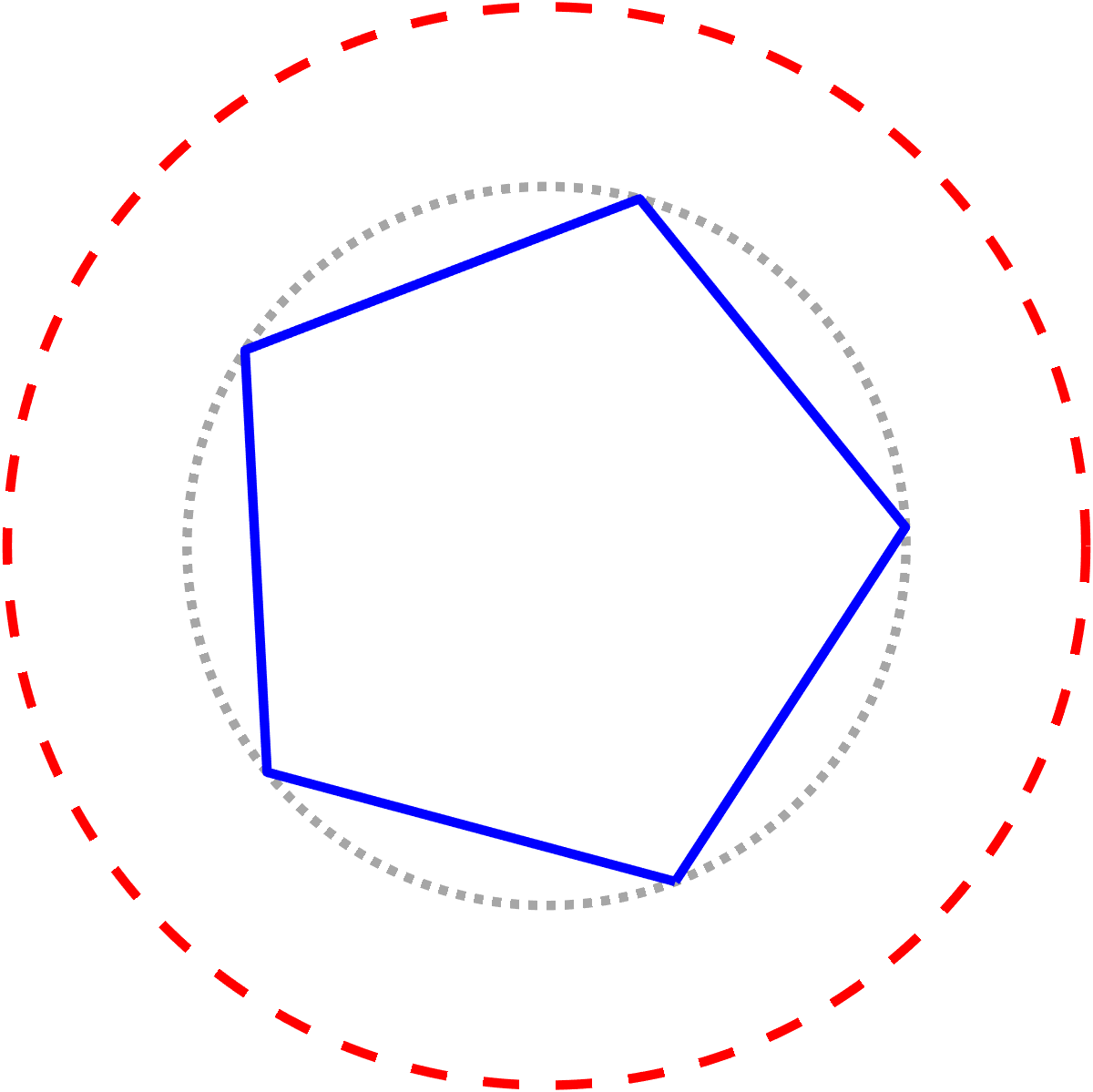}}
			\resizebox*{2cm}{!}{\includegraphics{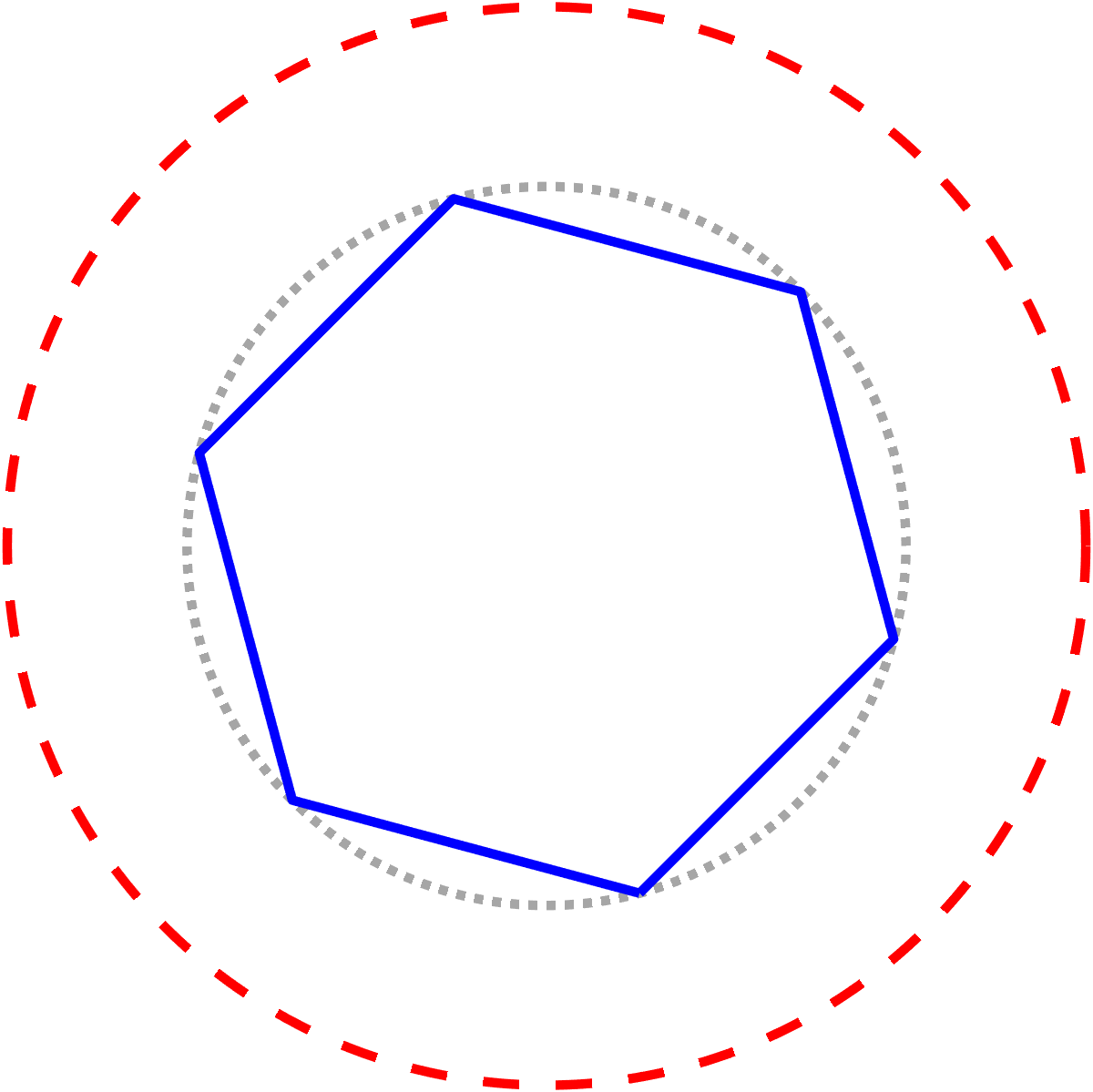}}
			\resizebox*{2cm}{!}{\includegraphics{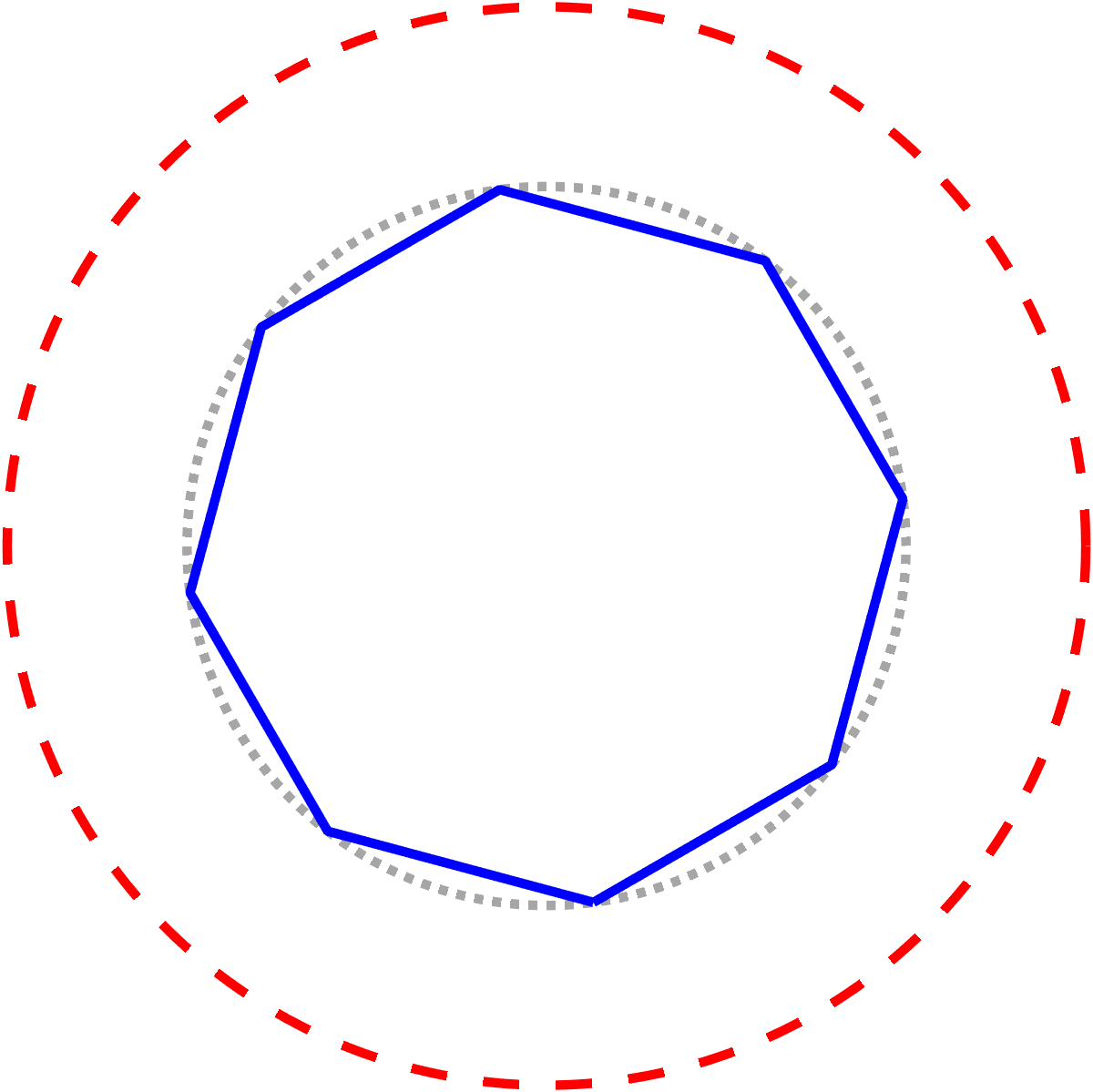}}\label{fig11c}}
		\newline
		\caption{Five tactile shapes were displayed to the participants via electrovibration during the second experiment in three different orientations: (a) prototypical orientation, (b) CCW (+15\textdegree{}), and (c) CW (-15\textdegree{}) angular perturbations were applied to the prototypical shapes in (a) to construct the shapes in non-prototypical orientations.} \label{fig11}
		
	\end{figure}

	The experimental stimuli comprised five tactile shapes (triangle, square, pentagon, hexagon, and octagon) varying in orientation as shown in Figure \ref{fig11}. As suggested by \cite{theurel2012haptic}, the shapes displayed without any angular perturbation are labeled as ``prototypical," and those displayed with angular perturbation are labeled as ``non-prototypical." The non-prototypical shapes that are rotated by 15\textdegree{} about their geometrical centers in the counter-clockwise (clockwise) direction are called positive (negative). This angle was determined such that the resultant non-prototypical shapes did not resemble their prototypical counterparts when they were rotated (Note that the maximum value of rotation angle was 22.5\textdegree{} since a CCW/CW rotation of 22.5\textdegree{} applied to a prototypical octagon results in a prototypical octagon again).
	
	\subsection{Procedure}
	The experimental procedure for this experiment was the same as the first experiment.  However, this experiment was performed in three separate sessions with one-day intervals between them. There were 30 trials in each session (5 shapes $\times$ 3 angular perturbations $\times$ 2 repetitions). Hence, the total number of trials for each participant was 90. The trials were randomized for each session, while the same randomization pattern was used for all participants.
	
	\subsection{Analysis}
	This experiment was a within-subjects design with two independent variables; angular perturbation and number of edges. The analysis procedure was the same as the first experiment.
	
	\subsection{Results}
	\subsubsection{Analysis of Recognition Rate and Time}
\begin{figure}
		\centering
		\subfloat[]{%
			\resizebox*{8cm}{!}{\includegraphics{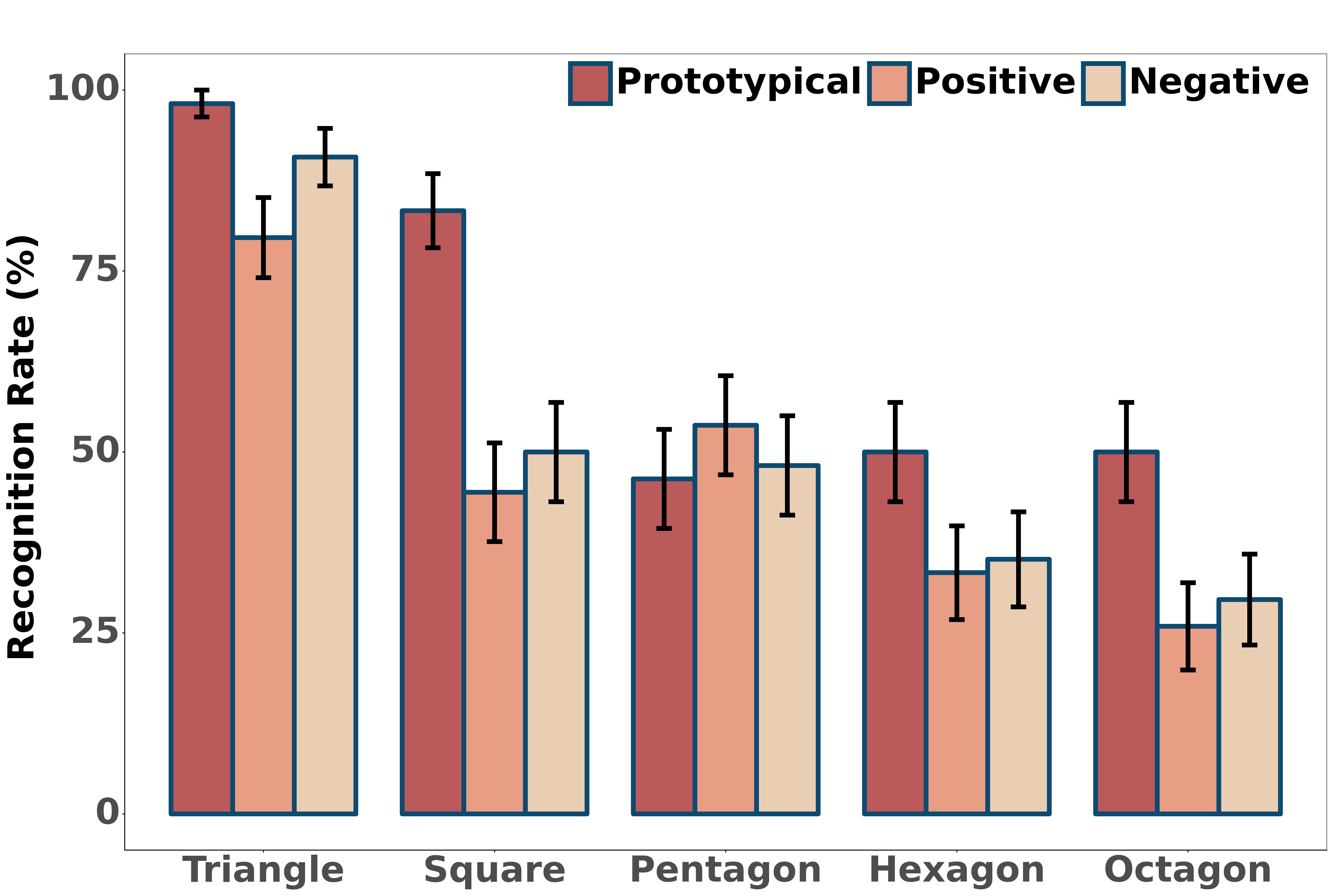}}\label{fig12a}}\hspace{5pt}
		\subfloat[]{%
			\resizebox*{8cm}{!}{\includegraphics{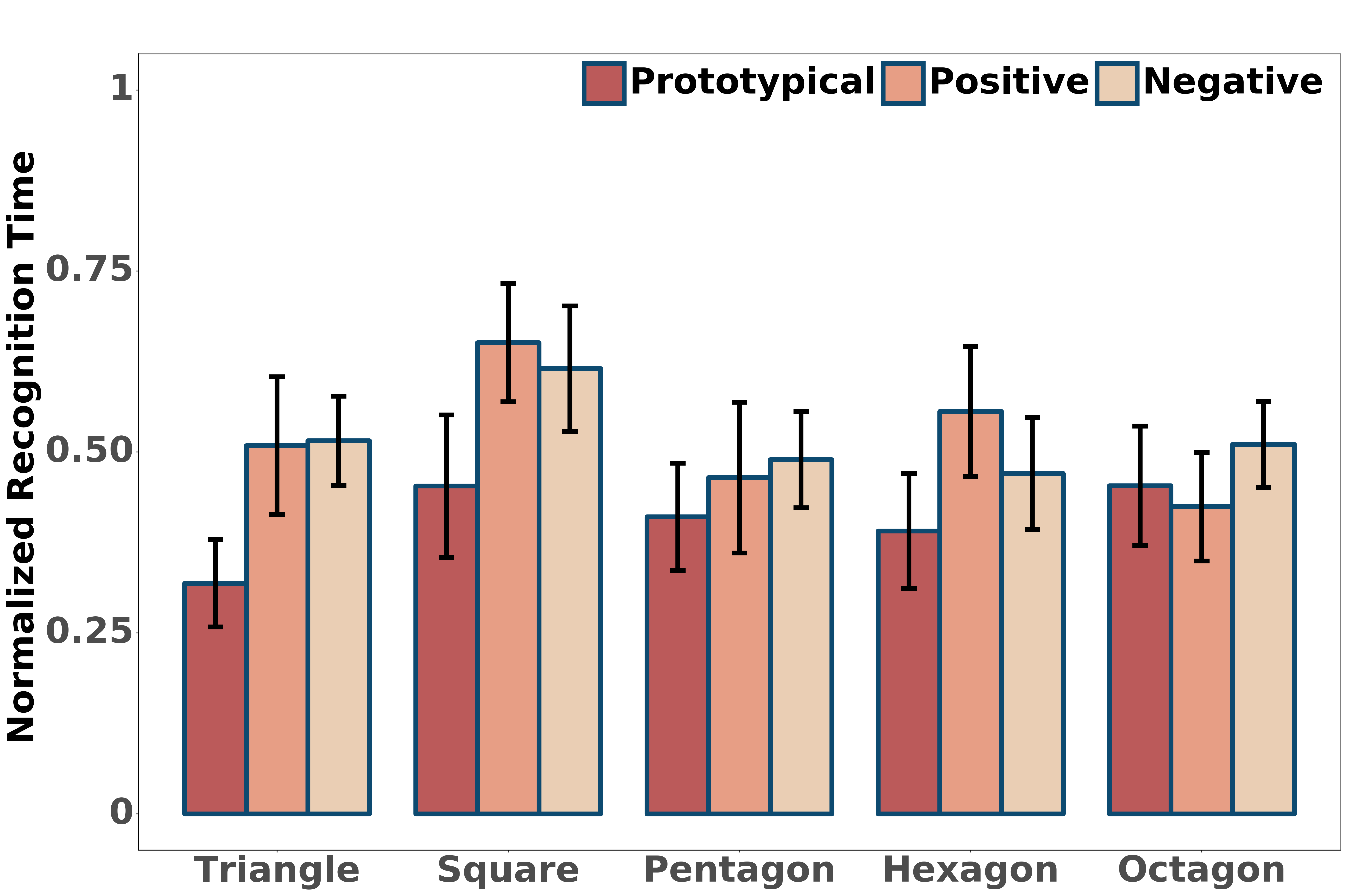}}\label{fig12b}}
		\caption{(a) Recognition rate and (b) normalized recognition time with standard error of the means for all shapes and angular orientations.}\label{fig12}
	\end{figure}

	Figure \ref{fig12} shows the average recognition rate and normalized recognition time of participants under three configurations (0\textdegree{}, +15\textdegree{}, -15\textdegree{}). The normalized recognition time of each participant for each shape was computed using her/his minimum and maximum values. The mean recognition times in seconds with the standard error of the means for all shapes and angular configuration are tabulated in Table \ref{tab2}.  Our results show that as the number of edges was increased, the recognition rate decreased significantly, more obvious for non-prototypical shapes than prototypical ones. We conducted a two-way repeated measures MANOVA with two independent variables (angular configuration: prototypical, positive, and negative; number of edges: triangle (3), square (4), pentagon (5), hexagon (6) and octagon (8)) and two dependent variables (recognition rate and recognition time). The MANOVA found a  significant main effect of angular orientation $(F(4, 32) = 4.248, p < 0.01, \eta^2 = 0.347)$ and number of edges $(F(8, 64) = 9.749, p < 0.001, \eta^2 = 0.549)$. The interaction effect between angular orientation and the number of edges was also significant $(F(16, 128) = 2.307, p < 0.01, \eta^2 = 0.224)$. Individual ANOVA found a significant main effect of angular orientation on both recognition rate $(F(2, 16) = 10.473, p < 0.001, \eta^2 = 0.567)$ and recognition time $(F(2, 16) = 8.130, p < 0.01, \eta^2 = 0.504)$. The effect of number of edges was also significant on both recognition rate $(F(4, 32) = 30.823, p < 0.001, \eta^2 = 0.794)$ and recognition time $(F(4, 32) = 3.682, p < 0.05, \eta^2 = 0.315)$. Our post-hoc analysis of recognition rate showed that the statistical difference between prototypical and non-prototypical shapes was significant $(p < 0.05)$, supporting our hypothesis that there is a perceptual bias for prototypical shapes. Also, the statistical difference between positive and negative shapes were non-significant $(p = 1.00)$.
	The recognition time was significantly smaller for the prototypical shapes as compared to the non-prototypical shapes $(p < 0.05)$. However, the statistical difference between the recognition times of positive and negative shapes was again non-significant $(p = 1.00)$. The significant interaction between the number of edges and angular orientation on the recognition rate revealed that the recognition rate dropped significantly for the non-prototypical shapes when the number of edges exceeded five $(p < 0.05)$. 
	
	\begin{table}[!ht]
    \centering
    \caption{Mean recognition times with standard error of the means for all shapes and angular configurations.}
    \label{tab2}
    \resizebox{12cm}{!}{\begin{tabular}{|p{2.5cm}|>{\centering}m{3.5cm}|>{\centering}m{4cm}|>{\centering\arraybackslash}m{4cm}|}
    \hline
    \diaghead(5,-2){\hskip4.8cm}{\textbf{Shape}}{\textbf{Angular}\\\textbf{Configuration}}
      & Prototypical & Perturbed (+15\textdegree{} CCW)  & Perturbed (-15\textdegree{} CW)          \\ \hline
    Triangle & 62.20 $\pm$ 6.30 sec & 82.10 $\pm$ 9.96 sec & 82.80 $\pm$ 6.45 sec \\ \hline
    Square   & 76.26 $\pm$ 10.29 sec  & 96.99 $\pm$ 8.56 sec & 93.23 $\pm$ 9.08 sec \\ \hline
    Pentagon & 71.81 $\pm$ 7.77 sec & 77.48 $\pm$ 10.90 sec & 80.05 $\pm$ 6.95 sec \\ \hline
    Hexagon  & 69.77 $\pm$ 8.29 sec & 87.03 $\pm$ 9.41 sec & 78.06 $\pm$ 8.09 sec \\ \hline
    Octagon  & 76.29 $\pm$ 8.62 sec & 73.27 $\pm$ 7.85 sec & 82.28 $\pm$ 6.23 sec \\ \hline
    \end{tabular}}
    \end{table}
	
\begin{figure}{!h}
		\centering
		\subfloat[]{%
			\resizebox*{5cm}{!}{\includegraphics{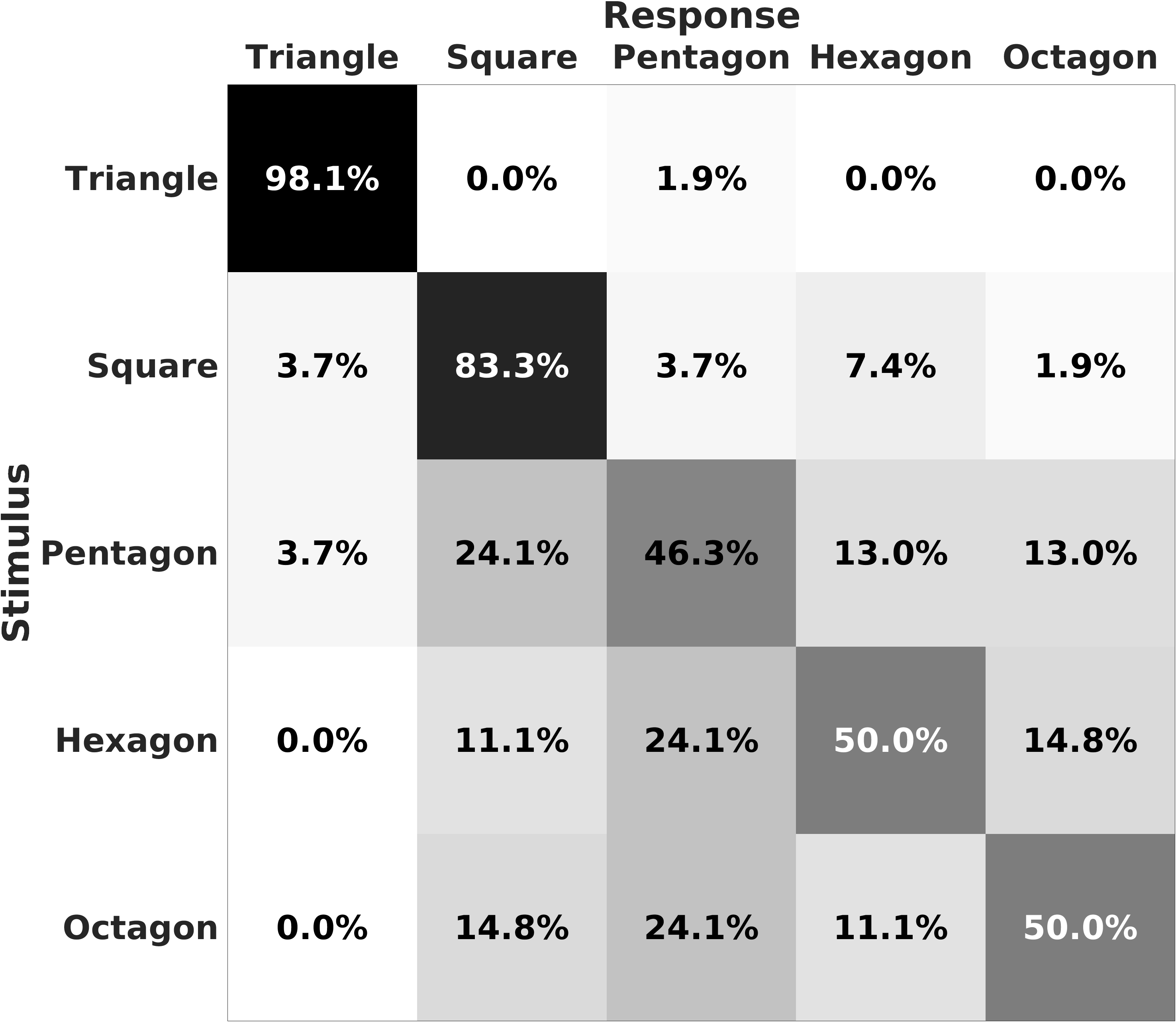}}\label{fig13a}}\hspace{5pt}
		\subfloat[]{%
			\resizebox*{5cm}{!}{\includegraphics{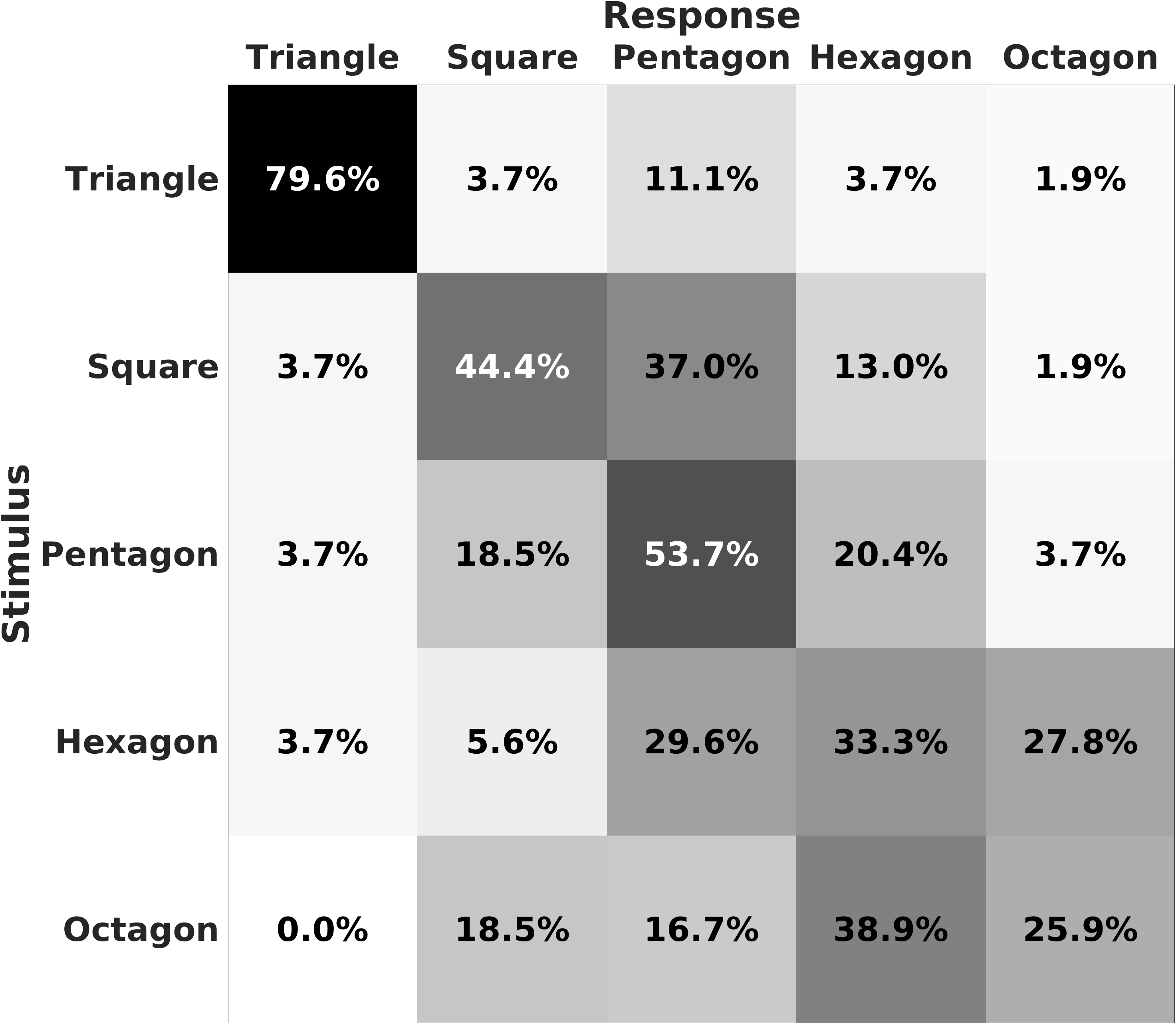}}\label{fig13b}}\hspace{5pt}
		\subfloat[]{%
			\resizebox*{5cm}{!}{\includegraphics{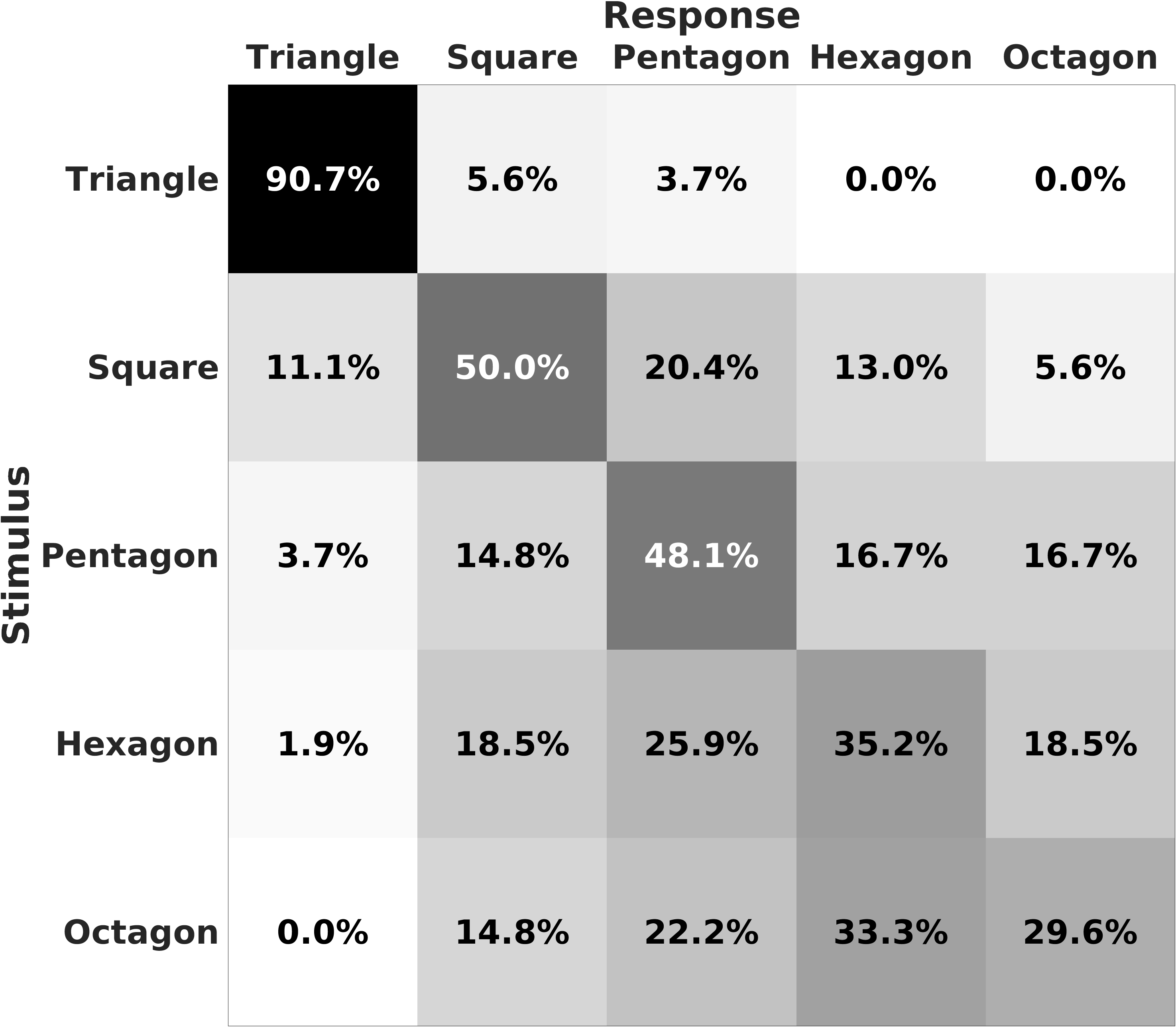}}\label{fig13c}}
		\caption{Confusion metrics of the recognition rate for (a) prototypical shapes and non-prototypical shapes; (b) positive and (c) negative.} \label{fig13}
	\end{figure}

	We also computed confusion matrices for the recognition rate (Figure \ref{fig13}). They show that as the number of edges increases, the recognition rate decreases significantly, especially for positive and negative configurations. Furthermore, as the number of edges increases, the confusion between the neighboring shapes also increases. 
	
	\subsubsection{Spatial Analysis}
	We calculated the number of times the participants touched the edges and corners of prototypical and non-prototypical shapes. A two-way MANOVA was carried out on the number of touches. We found a significant multivariate effect of angular orientation $(F(2, 16) = 11.782, p < 0.001, \eta^2=0.596)$ and number of edges $(F(4, 32) = 44.195, p<0.001, \eta^2=0.847)$ on the number of touches. The interaction effect between angular orientation and number of edges was also significant $(F(8, 64) = 4.542, p < 0.001, \eta^2=0.362)$. Our post-hoc analysis showed that the number of touches made to the corners and edges were not significantly different for all angular orientations and shapes except for square and hexagon $(p < 0.05)$. The number of touches to the corners (edges) was significantly higher (lower) for prototypical square and hexagon than their non-prototypical counterparts.

\begin{figure}[h!]
		\centering
		\subfloat[Negative Triangle]{
			\resizebox*{4cm}{!}{\includegraphics{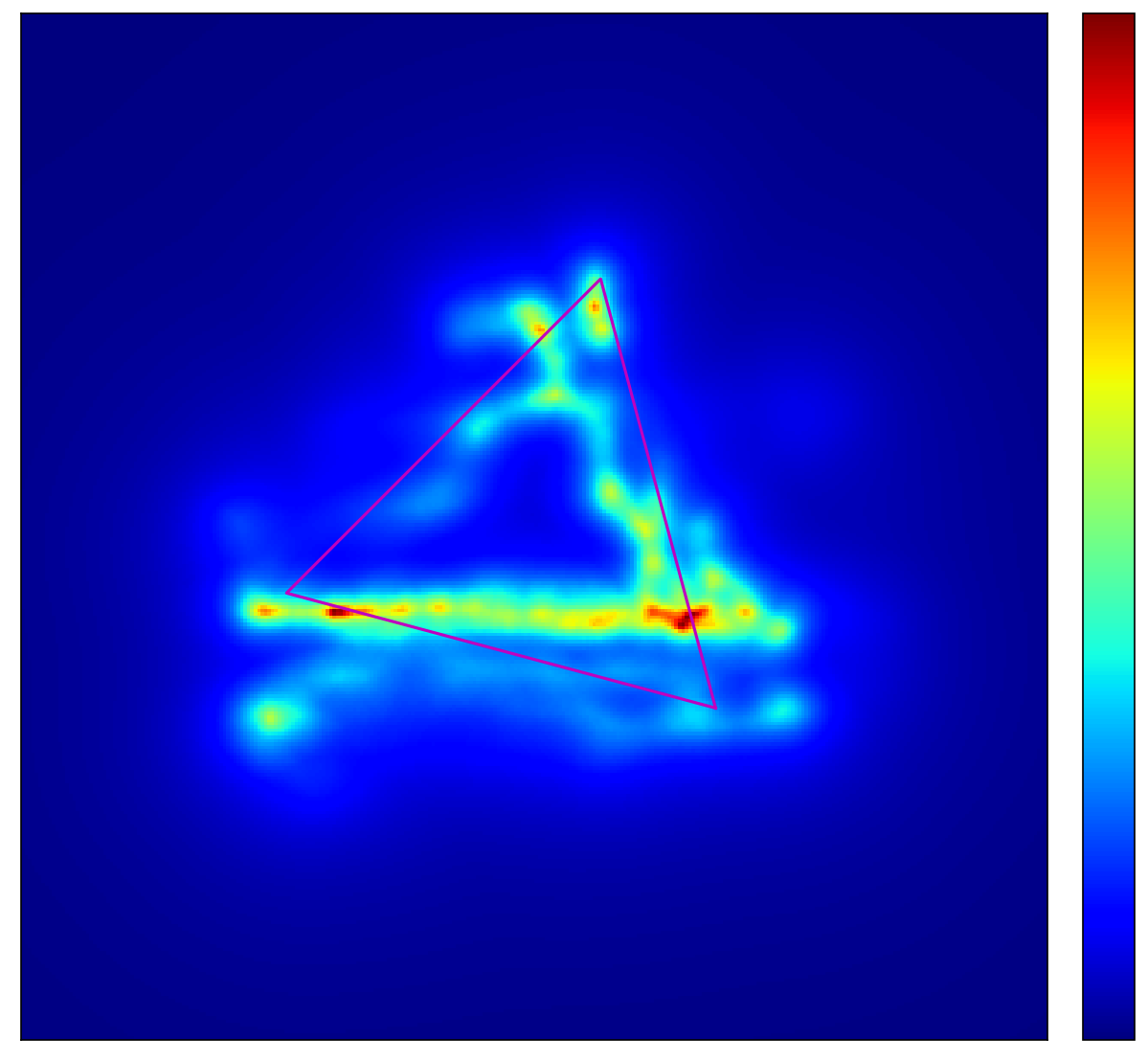}}}
		\subfloat[Positive Square]{              \resizebox*{4cm}{!}{\includegraphics{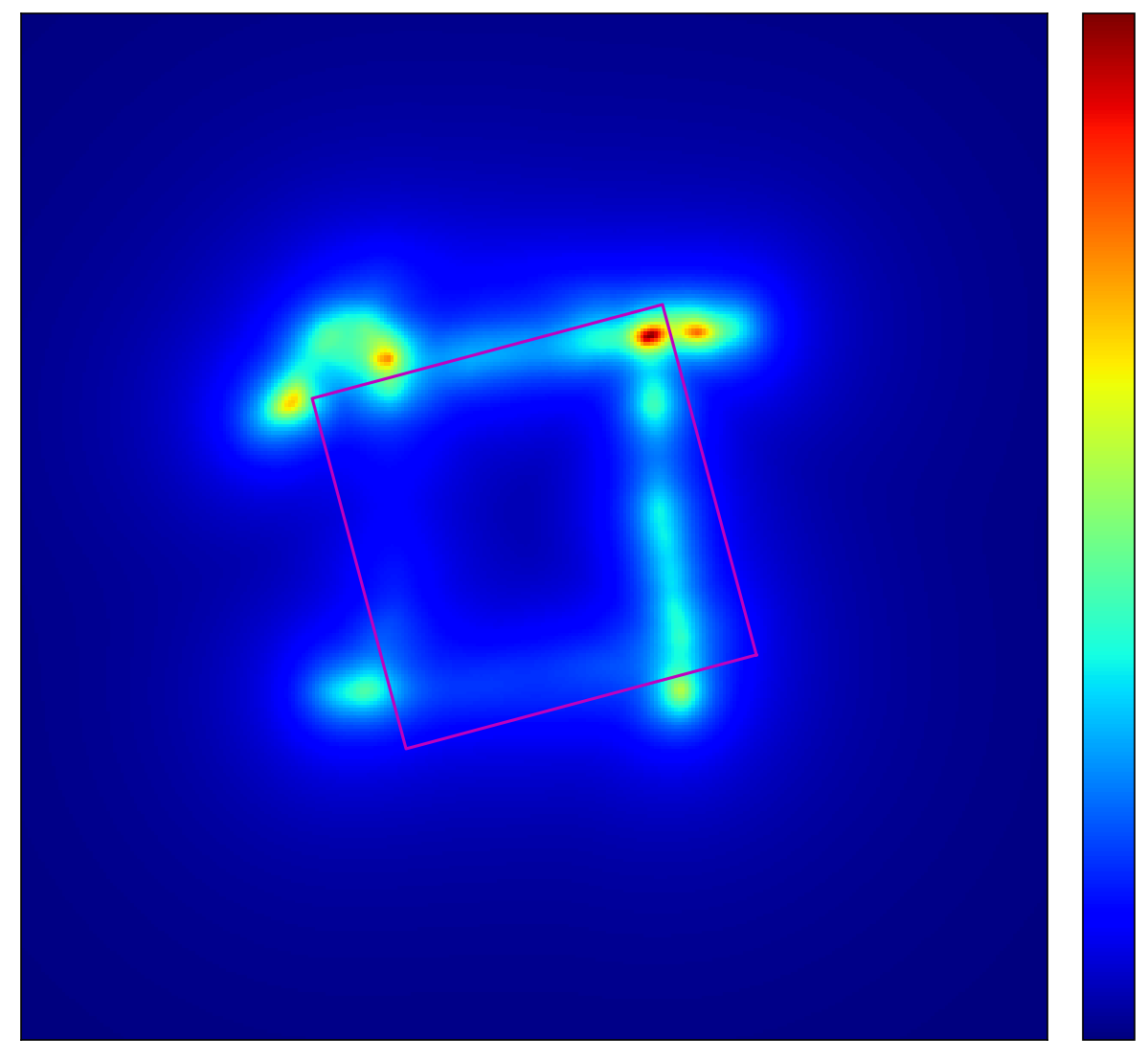}}}
		\subfloat[Negative Pentagon]{
			\resizebox*{4cm}{!}{\includegraphics{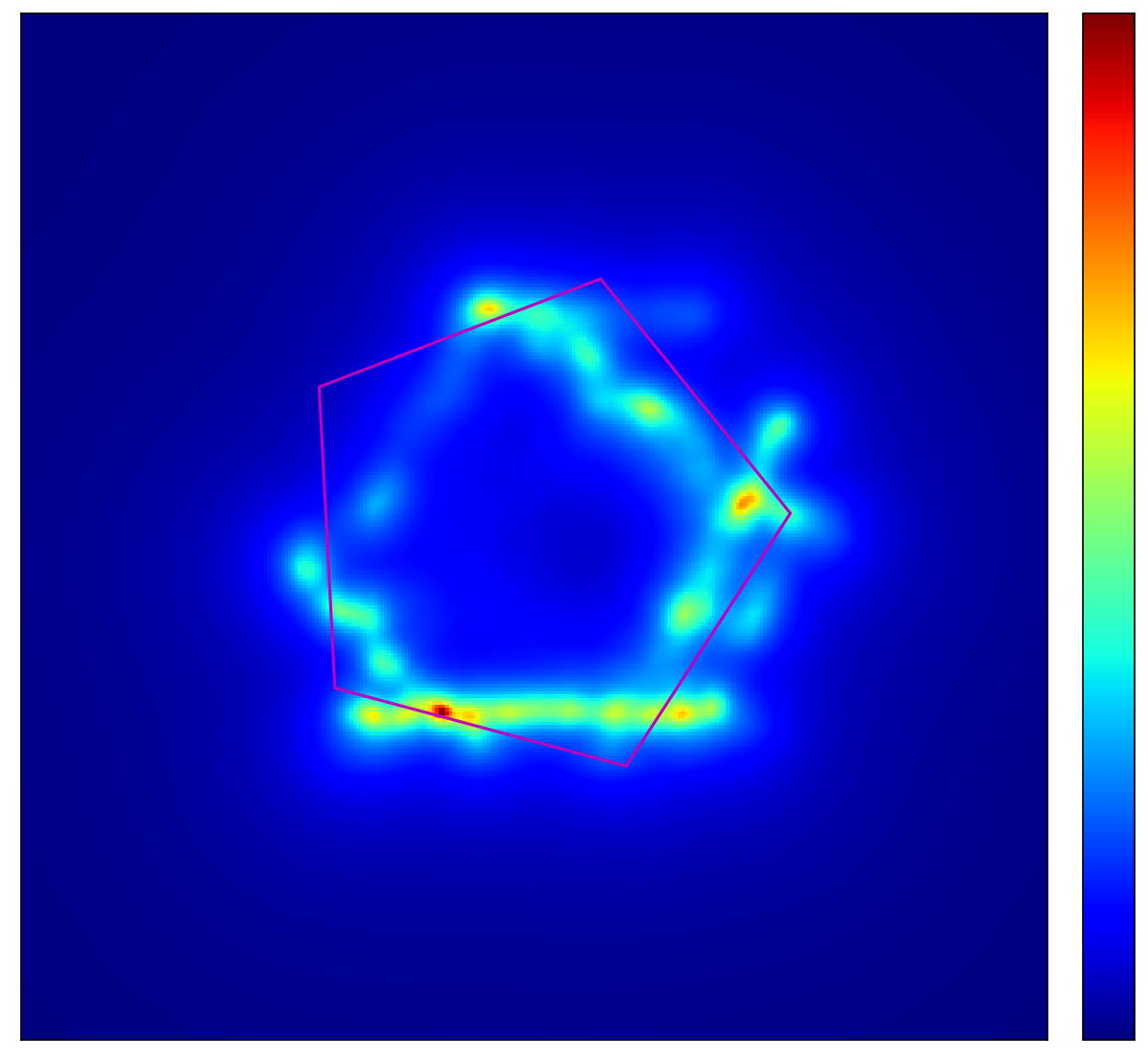}}} \newline
		\subfloat[Negative Hexagon]{
			\resizebox*{4cm}{!}{\includegraphics{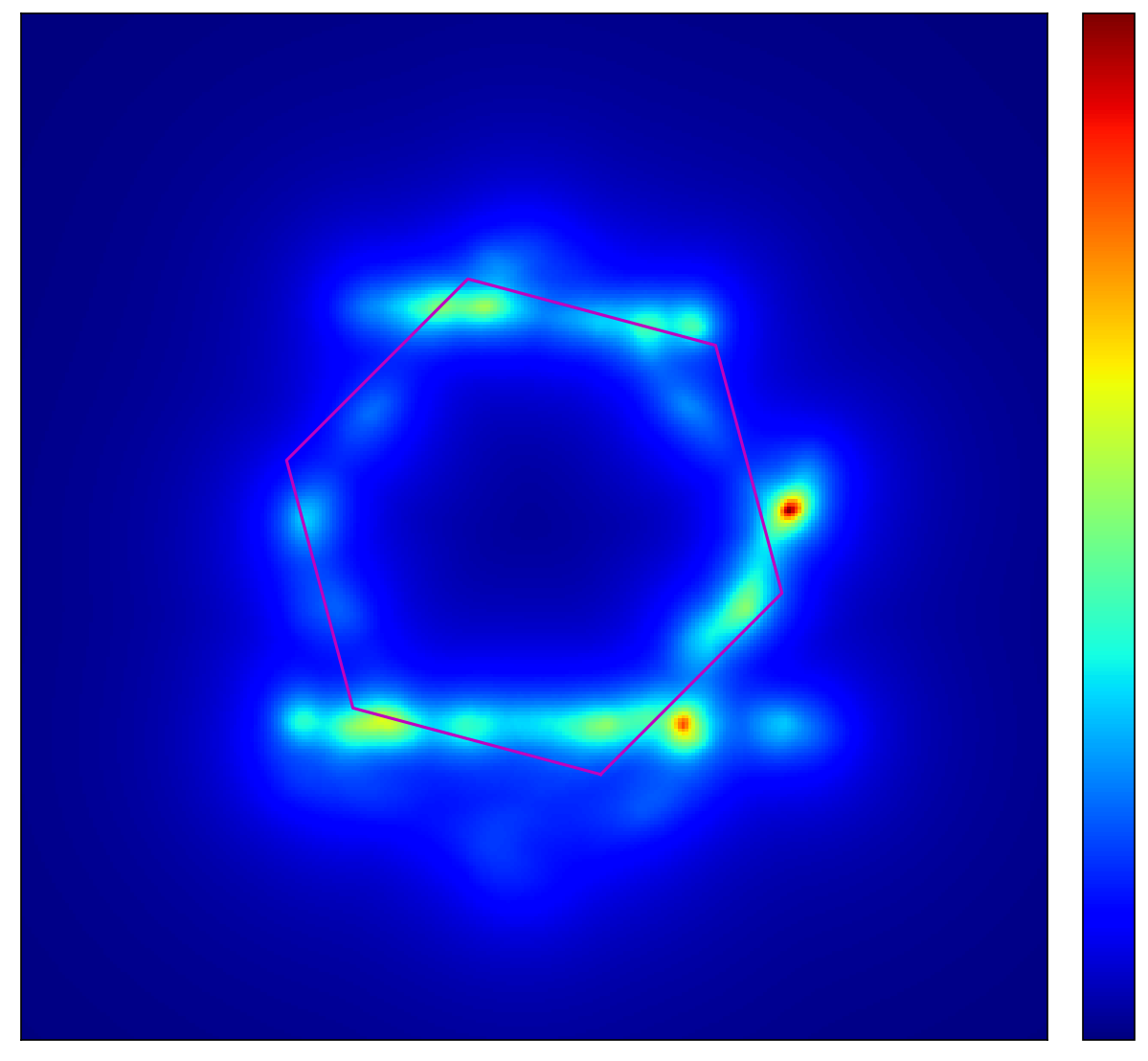}}}
		\subfloat[Positive Octagon]{
			\resizebox*{4cm}{!}{\includegraphics{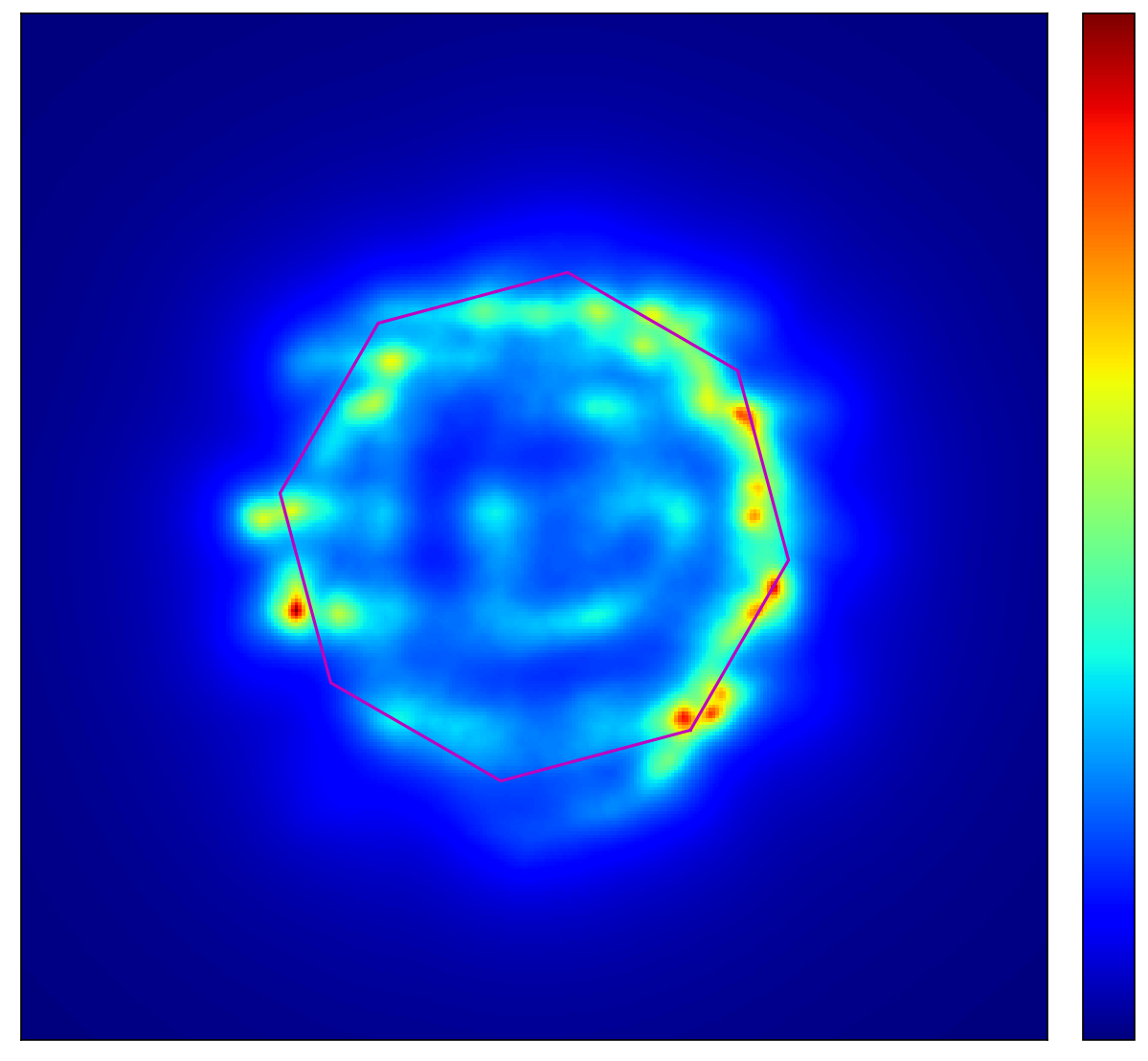}}\label{fig14a}}
		\caption{Frequency maps of randomly selected trials of participants for perturbed shapes.} \label{fig14}
		
	\end{figure}

	Figure \ref{fig14} shows the frequency maps of participants for the perturbed shapes. Interestingly, we observed that participants scanned the non-prototypical shapes as if they were scanning the prototypical shapes, demonstrating again a perceptual bias for the prototypical shapes.
	
	\subsubsection{Temporal Analysis}
	
\begin{figure}[!htbp]
		\centering
		\subfloat[]{%
		{\includegraphics[width=0.65\textwidth]{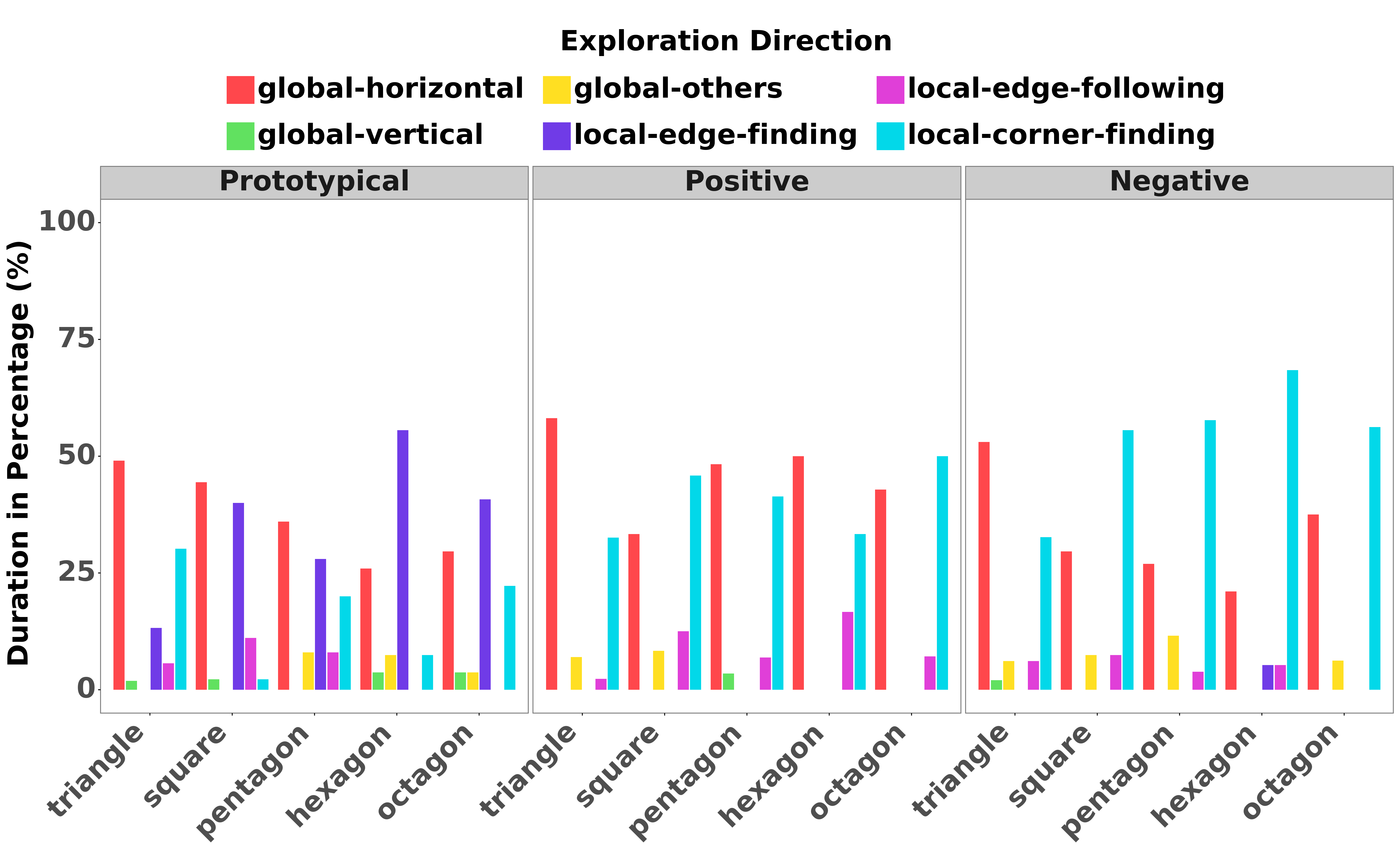}}\label{fig15a}}
			\newline
		\subfloat[]{%
			{\includegraphics[width=0.65\textwidth]{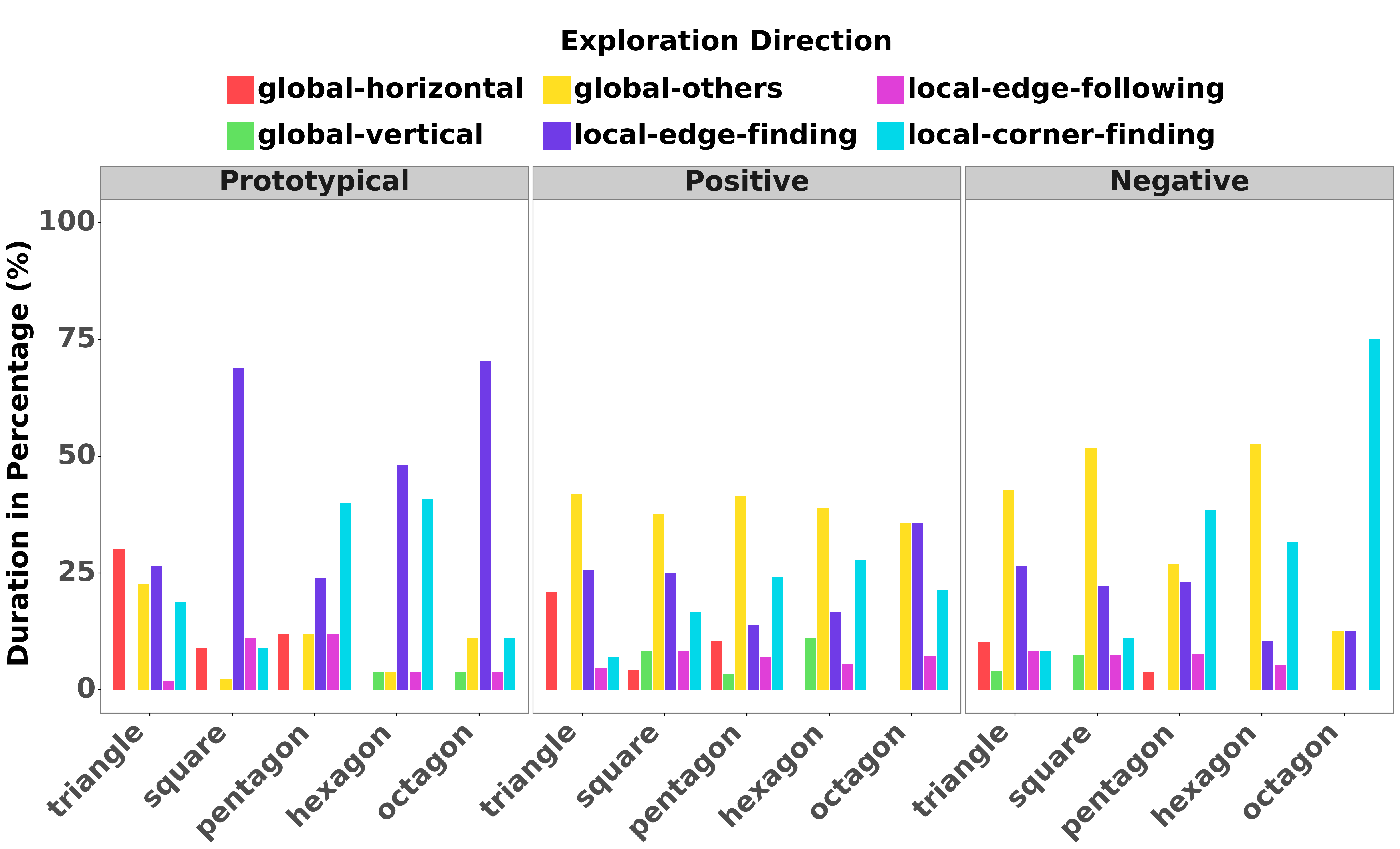}}\label{fig15b}}
			\newline
		\subfloat[]{%
			{\includegraphics[width=0.65\textwidth]{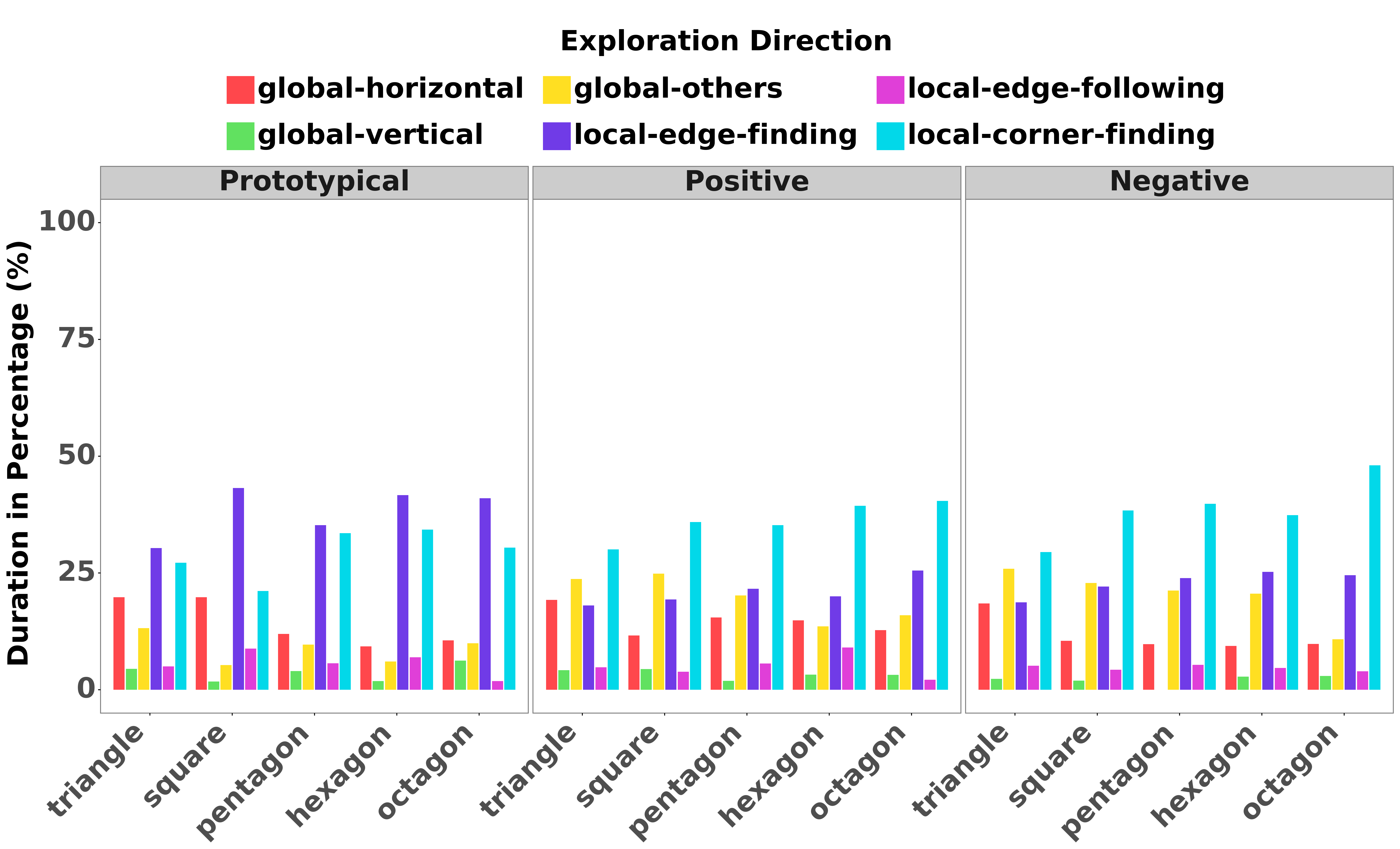}}\label{fig15c}}   
			\vspace{-0.5cm}
		\caption{(a) Starting, (b) ending, and (c) most/least explored strategies utilized for prototypical and non-prototypical shapes for all participants.} \label{fig15}
	\end{figure}

	We investigated the scanning strategies used by the participants to explore the virtual shapes displayed on the touch surface. Our results showed that the most common starting strategy adopted by the participants was global scanning with exploratory movement as horizontal for the prototypical (37\%) and the positive (46\%) shapes and local-corner-finding (54\%) for the negative shapes (Figure \ref{fig15a}). Participants have ended their exploration with local-edge-finding (47.57\%) for the prototypical shapes and global-others (39\%) for the non-prototypical shapes (Figure \ref{fig15b}). The most common exploration strategy was local-edge-finding (38.28\%) followed by local-corner-finding (29.31\%) for the prototypical shapes and 
	local-corner-finding (38\%) for the non-prototypical shapes followed by local-edge-finding (22\%) (Figure \ref{fig15c}). The least explored strategy was global-vertical for both the prototypical (3\%) and non-prototypical shapes (2\%).
	
	We observed that participants utilized global scanning first and then focused on the local features to identify the prototypical shapes.  However, for the case of non-prototypical shapes, they started with global scanning, then focused on the local features, and finally performed global scanning again to confirm their decision. 
	
\begin{figure}[!h]
		\centering
		\includegraphics[width=0.5\textwidth]{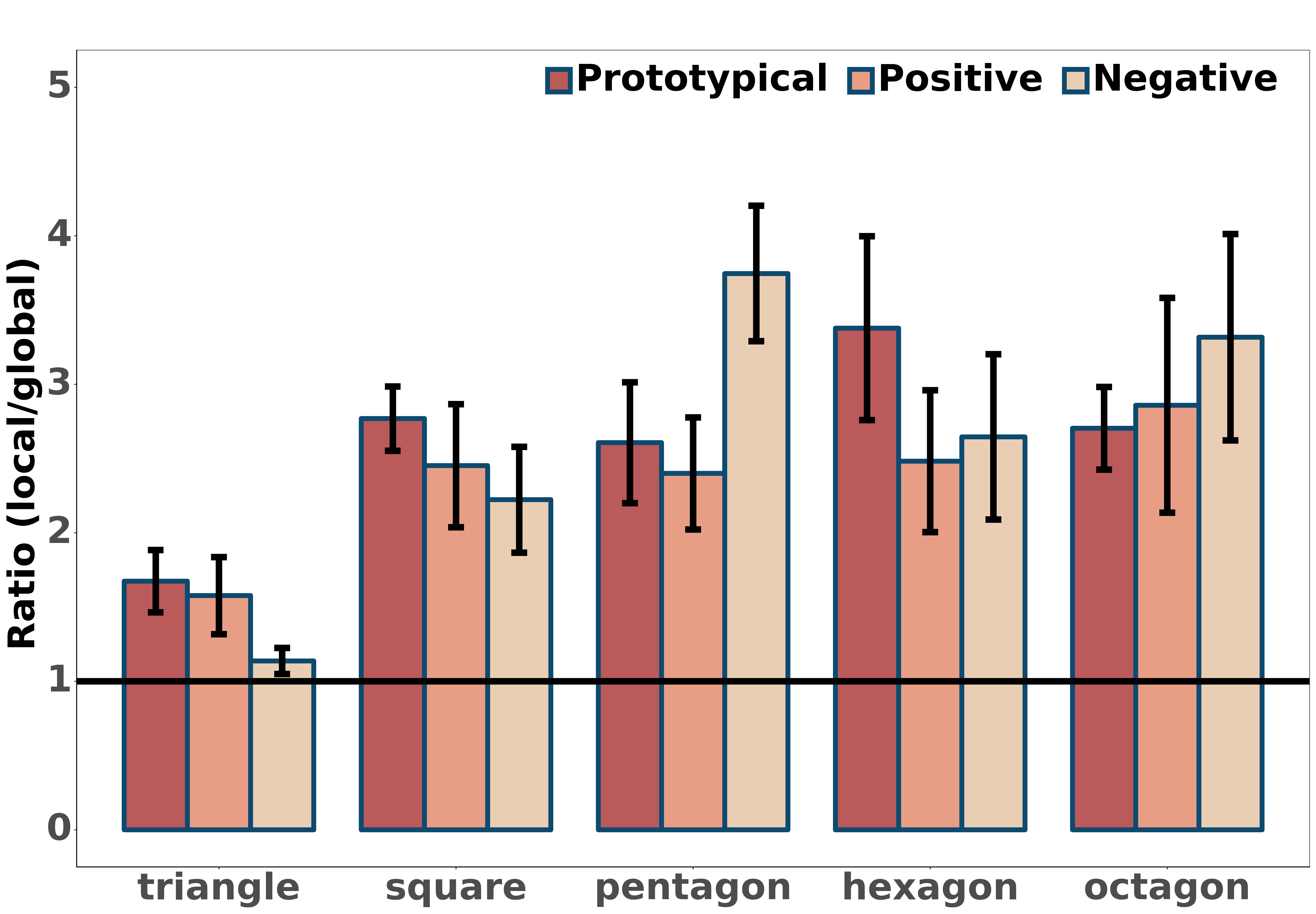}
		\caption{The ratio of exploration time spent for local scanning to global scanning  with standard error of the means for prototypical and non-prototypical shapes. }\label{fig16}
	\end{figure}

	We observed that participants spent most of their exploration time on scanning the local features to identify the shapes. This observation can be further confirmed by Figure \ref{fig16}. As the number of edges was increased, it was more difficult for the participants to identify the shapes, and hence, they relied on the local features more to distinguish them.  
	
	\section{Discussion}
	
	In this study, we first investigated the effects of (a) number of edges, (b) rendering conditions, and (c) angular orientation on recognition accuracy and time of the 2D equilateral geometrical shapes displayed by electrovibration on a touchscreen. Then, we investigated the strategies followed by the participants to explore these tactile shapes by analyzing their finger position recorded as a function of time.

	\subsection{Geometric Complexity of Shape Affects the Tactile Perception}
	
	The results of Experiment-I showed that recognition rate was inversely proportional to the number of edges (except for octagon), while recognition time was directly proportional to the number of edges. Our analysis suggested that participants utilized the same scanning patterns to explore the shapes that looked alike and therefore confused them. For example, the top and bottom edges of prototypical hexagon and octagon are similar and parallel to the horizontal axis. The main difference between these geometries is on their left and right sides. Simply speaking, the displayed shape is a hexagon (octagon), when there is a corner (edge) on each side. We observed that the participants had difficulty in differentiating shapes using such local features as the geometric complexity was increased. Despite this difficulty, they were better in recognizing octagon than hexagon, which we believe was due to some perceptual bias.

	In fact, the results of Experiment-II confirmed the perceptual bias observed in Experiment-I. After applying perturbation to the shapes to eliminate perceptual bias, the recognition rate for all non-prototypical shapes dropped significantly as the number of edges was increased. As anticipated, triangle (3) had the highest recognition rate, while octagon (8) had the lowest. Considering the confusion matrices for the non-prototypical shapes (Figure \ref{fig13b}, \ref{fig13c}), we speculate that displaying more than five edges makes the recognition task highly difficult since false negatives was greater than true positives for the shapes higher than five edges (hexagon, octagon). As the number of edges in a shape increases, it is more difficult to form a mental image of the shape due to the amount of information that has to be processed in the brain, limiting the ability to identify the shape correctly. \cite{tennison2016toward} reported that participants could successfully identify 2D shapes with up to four edges when they were rendered by vibrotactile actuators. 
	
	The participants in our second experiment showed poor performance (lower recognition rates and longer recognition time) for the non-prototypical shapes than the prototypical ones due to the mental rotation effect (\citealp{shepard1971mental}). Similar results were also reported in the literature for the mental rotation experiments conducted with different types of stimuli (\citealp{milivojevic2011turn, theurel2012haptic, zeugin2017implicit, tivadar2019mental}). \cite{theurel2012haptic} also observed a prototype effect in their experiments conducted with congenitally blind and blindfolded adolescents on the haptic recognition of geometrical shapes. The authors suggested that the early exposure of the participants to these prototypical shapes in their visual environment was the leading cause of this effect. The lower recognition rate and longer recognition time of non-prototypical shapes also linked to the fact that participants exhibited difficulty tracing edges of the tactile shapes and detecting their orientation as they deviated from the horizontal and vertical directions (\citealp{giudice2012learning, palani2014evaluation, gershon2016visual}).  
	
	\subsection{Size of Haptically Active Area Affects the Tactile Perception}
	
	The results of our first experiment showed that participants spent less time recognizing the prototypical shapes when they were rendered by using \ac{c1} condition. Frequency maps for the shapes in Figure \ref{fig8} highlighted that participants explored the corners and edges to identify the prototypical shapes. By inspection, we can say that they formed a decision about the shape that they were exploring after the third or fourth edge. They focused on exploring the edges during \ac{c1} and \ac{c3} rendering conditions and the corners in the case of \ac{c2} rendering condition to identify the rendered shape.
	
	Participants achieved the shortest recognition time under \ac{c1} rendering condition. This result suggests that a larger haptically active area (where the electrovibration is on) leads to shorter recognition time. However, a good balance between haptically active and inactive areas must be maintained as suggested by \cite{gorlewicz2020design}. They report that too much mechanical vibration on touchscreen results in confusion while too little makes the interpretation of the tactile graphics challenging. Therefore, haptics application designers should guide the users by designating haptically active and inactive areas. For example, in our experiments, we guided the participants towards the haptically active area by visually displaying a circle with a diameter of 15 cm around the rendered shape. As a result, it was more easier for the participants to explore the designated area and identify the rendered shape. All our shapes fitted into an invisible circle of 10 cm in diameter, which was large enough for the participants to recognize them correctly by tactile exploration.

	\subsection{Haptic Exploratory Procedures for Tactile Shapes}
	
	We observed differences in exploration strategies employed by the participants for the prototypical and non-prototypical shapes. Our analyses demonstrated that participants initially started their exploration with global scanning for both the prototypical and non-prototypical shapes to get an overall idea about the examined shape. They then switched to local scanning to discover its key features distinguishing the rendered shape from the others. For the prototypical shapes, they focused more on the edge-finding strategy.  Possibly, it was easy for the participants to find and count the edges of the shapes at typical orientations. However, for the non-prototypical shapes, participants employed a corner-finding strategy because it was challenging for them to follow the edges. Furthermore, we observed that participants eventually performed another global scanning at the end of exploration for the non-prototypical shapes to confirm their decision. Similar strategies have been reported by \cite{bara2014exploratory} for the exploration of tactile pictures in a real book by visually impaired children. They reported that lateral motion was the preferred strategy to obtain global information about the shapes, and contour-following was opted by the children to perceive them accurately. \cite{simonnet2012accuracy} and \cite{bardot2019investigating} referred to the initial exploration of tactile maps, where participants obtained the overall idea about the map, as the investigation or discovery phase.  When participants explored the maps in a particular way by following some specific strategies to complete the given task, they called this as the memorization phase.  \cite{tennison2016toward} and \cite{gorlewicz2020design} mentioned that users employed \textit{``circling a part of a line''} strategy to determine the number and orientation of intersecting lines. This strategy is similar to the local-corner-finding strategy used by the participants in our study.

	We observed that the least preferred strategy among participants was global scanning with the exploratory direction as vertical. A possible explanation is that users tend to explore a touch surface horizontally, i.e., from left to right, using a single finger (\citealp{withagen2013use,palani2018haptic}). Therefore, vertical scanning was utilized less in comparison with other strategies. Moreover, in horizontal scanning, the finger area that is in contact with a touch surface is larger, which provides more tactile information to the users.

	We also observed that the number of touches to the corners (edges) was significantly higher (lower) for prototypical square and hexagon than their non-prototypical counterparts. Since humans can perceive horizontal lines more easily than oriented ones (\citealp{tennison2019non}), perhaps the participants in our experiments recognized the parallel horizontal edges in prototypical square and hexagon easily and focused on the corners more to differentiate them from the others.

	Frequency maps in Figure \ref{fig14} revealed an interesting finding. The fingers of most participants in our second experiment traced the non-prototypical shapes as if they were tracing the prototypical ones. One potential reason for this is the perceptual bias of the participants for the prototypical shapes as mentioned before. Participants could have formed a spatial representation of each prototypical shape when visual feedback was provided during the training session. Although the shapes were not displayed visually during the actual experiments, participants traced the shapes using their index finger based on the information stored in their spatial memory. We believe that current limitations of the electrovibration technology also contributed to this behavior. Since SA-I receptors in fingerpad, responsible for detecting changes in static pressure, cannot be stimulated by electrovibration (\citealp{icsleyen2019tactile}), the participants had to make zig-zag movements along an edge to follow it, which was not easy especially for the non-prototypical shapes. The temporal analysis also confirmed that edge-following was the least preferred strategy among the participants for both the prototypical and non-prototypical shapes. 
	
	\subsection{General Guidelines for Designing and Displaying Tactile Graphics}
	Based on the results of our experiments, we put forth the best practices and design guidelines to display tactile graphics on touchscreens via electrovibration. Our suggestions are specific to the exploration of 2D tactile shapes using index finger.
	\begin{itemize}
	    \item As the number of edges in a shape exceeds five, it becomes more difficult for the participants to recognize it. Therefore, more sophisticated rendering algorithms and sensory augmentation with vibrotactile or auditory feedback are necessary for displaying more complex tactile shapes.
	    \item Large white spaces (haptically inactive areas) should be avoided as suggested in \cite{gorlewicz2020design}. We observed in our experiments that large white spaces led to longer recognition time (e.g., \ac{c2} rendering condition). Moreover, guiding the user towards the targeted area of exploration is helpful for reducing the recognition time. For example, we visually displayed a red-colored outer circle to help the users locate the tactile shapes more easily. Alternatively, a guiding tactile feedback could be displayed to the users when their index fingers had entered the outer circle. 
	    \item It should be preferable to display frictional tactile feedback inside the shapes rather than outside or on the edges only. In our study, the recognition time was significantly shorter for the \ac{c1} rendering condition than those for the \ac{c2} and \ac{c3} rendering conditions. 
	    \item Common tactile exploration strategies should be taken into account when guiding users on how to explore tactile graphics on a touchscreen. Our study shows that the typical strategy for exploration of tactile shapes rendered by electrovibration starts with global scanning in horizontal direction followed by local search using edge or corner finding. The participants finished their exploration by another global scanning in horizontal direction for final conformation.
	    \item It is difficult to follow the edges of a tactile shape (i.e. edge-following) using the current friction modulation displays since the user has to make zig-zag movements to follow an edge. This causes a saturation in the tactile sensation of the user and even sensory fatigue with time. Again, sensory augmentation with vibrotactile or auditory feedback can be helpful for following an edge with less fatigue. Moreover, the hardware designers of next generation friction displays should aim to generate localized tactile effects on the finger of users to make them feel the edges of shapes with minimal tangential movement, though the earlier studies show that the effect of electrovibration is only present during full slip but not before slip (\citealp{sirin2019fingerpad}).  
    \end{itemize}

	\section{Conclusion}
	Our findings can provide some guidance to haptic interface designers in developing techniques for efficient and effective interaction with tactile graphics displayed by electrovibration. In particular, we investigated the complexity of displayed geometry and the size of haptically active area on perception of tactile shapes as well as the haptic exploratory procedures enable to perceive them. In addition to measuring recognition rate and time, we performed spatial and temporal analyses on the collected finger position data in order to investigate the exploratory strategies. 
	
	We observed that the recognition rate is inversely proportional to the complexity of shape. As the number of edges was increased, it became more difficult for the participants to recognize the tactile shapes rendered by electrovibration. Our results showed that rendering 2D shapes with more than five edges causes the recognition rate drops significantly and makes it unreliable since true positives becomes less than false negatives in shape recognition.
	
	Furthermore, our results show that the size of the haptically active area also affects the perception of shape. Participants in our experiments were able to identify the displayed shapes faster when the haptically active area was larger, as in \ac{c1} rendering condition. In \ac{c2} rendering condition, there were many white spaces (haptically inactive areas), and therefore, participants spent the longest time to recognize the shapes. They mostly depended on the detection of corners for the recognition. Participants counted the corners (or the number of intersecting edges) by circling around them to recognize the shapes. 
	
	Analysis of exploration strategies exhibited that participants did not rely on a single strategy to explore the shapes. Moreover, they employed different exploration procedures for the prototypical and non-prototypical shapes. They started their exploration with global scanning to identify the shapes' global features and then switched to local scanning to focus on the shape's key features for the prototypical shapes. However, in the case of non-prototypical shapes, they performed an additional global scanning at the end of exploration to confirm their decision. We also found that participants preferred the corner-finding over the edge-finding for the non-prototypical shapes because it was challenging to detect and follow the tilted edges compared to detecting their corners. Tactile masking techniques could be utilized to increase tactile contrast and make the detection of edges easier (\citealp{vardar2018tactile, jamalzadeh2020effect}), but following them would still be challenging. Here, it should be noted again that the exploration of 2D shapes embossed on a paper easily stimulates SA (Slowly-Adapting)-I receptors, which is not possible under electrovibration. They are primarily responsible for detecting changes in static pressure in normal direction, which enables us to detect edges of real shapes and objects easily. This makes it possible to follow the edges/contours even when no visual feedback is available. However, edge-following was the least preferred local exploration strategy by the participants in our study since electrovibration modulates frictional forces in tangential direction only and hence the participants had to make zig-zag movements to follow an edge (i.e. they could not stay on the edge). 
	Finally, the current electrovibration technology is also limited by the lack of multi-touch tactile feedback, restricting the user's ability to explore tactile graphics with multiple fingers, which would lead to more accurate and faster recognition of virtual shapes. 
	
	\section*{Acknowledgments}
	We acknowledge the financial support provided by the Scientific and Technological Research Council of Turkey under Contract 117E954. We thank Pinar Onal for her technical support in the temporal analyses.

	\bibliographystyle{elsarticle-harv}
	\bibliography{MyRef}

\end{document}